\documentclass[journal,12pt,onecolumn,draftclsnofoot,]{IEEEtran}

\usepackage{times,amsmath,epsfig}
\usepackage[table]{xcolor}%
\usepackage{textcomp}
\usepackage{xcolor}

\usepackage{amssymb}
\usepackage{pifont}
\usepackage{tablefootnote}
\usepackage{booktabs, tabularx}
    \usepackage[flushleft]{threeparttable}

\usepackage{graphicx}
\usepackage{cite}
\usepackage{amsmath,amssymb,amsfonts}
\usepackage{algorithmic}
\usepackage{gensymb}
\usepackage{subfigure}
\usepackage{graphicx}
\usepackage{textcomp}
\usepackage{cases}
\usepackage{multirow}
\usepackage{multicol}
\usepackage{todonotes}
\usepackage{tablefootnote}
\usepackage{booktabs}
\usepackage{textgreek}
\usepackage{threeparttable}
\usepackage{latexsym}
\usepackage{amsfonts,amssymb,amsmath}
\usepackage{hyperref}

\usepackage{xcolor}
\newcommand{\xmark}{\ding{55}}%

\begin{document}

\title{Radar Cross Section Based Statistical Recognition of UAVs at Microwave Frequencies\\
}

\author{%
Martins Ezuma, Chethan Kumar Anjinappa, Mark Funderburk, and Ismail Guvenc\\
\vspace{-2mm}%
 \thanks{This work has been supported in part by the National Aeronautics and Space Administration (NASA) under the Federal Award ID number NNX17AJ94A. }}


\maketitle

\begin{abstract}
This paper presents a radar cross-section (RCS)-based statistical recognition system for identifying/classifying unmanned aerial vehicles (UAVs) at microwave frequencies. First, the paper presents the results of the vertical (VV) and horizontal (HH) polarization RCS measurement of six commercial UAVs at 15~GHz and 25~GHz in a compact range anechoic chamber. The measurement results show that the average RCS of the UAVs depends on shape, size, material composition of the target UAV as well as the azimuth angle, frequency, and polarization of the illuminating radar. Afterward, radar characterization of the target UAVs is achieved by fitting the RCS measurement data to 11 different statistical models.  From the model selection analysis, we observe that the lognormal, generalized extreme value, and gamma distributions are most suitable for modeling the RCS of the commercial UAVs while the Gaussian distribution performed relatively poorly. The best UAV radar statistics forms the class conditional probability densities for the proposed UAV statistical recognition system. The performance of the UAV statistical recognition system is evaluated at different signal noise ratio (SNR) with the aid of Monte Carlo analysis. At an SNR of 10 dB, the average classification accuracy of 97.43\% or better is achievable.
\end{abstract}

 \begin{IEEEkeywords}
 Akaike information criterion (AIC), automatic target recognition (ATR), Bayesian information criterion (BIC), classification, compact-range chamber, detection, RCS, UAV.
 \end{IEEEkeywords}

\maketitle

\section{Introduction}

\IEEEPARstart{C}{ivilian} unmanned aerial vehicle (UAV) technology is a disruptive technology that has the potential to transform modern society.  However, there are challenges to be overcome before UAVs can be effectively integrated into the national airspace. The most common challenge is security and privacy. For instance, recently UAVs have been used to carry out crimes and terror attacks that threaten public safety~\cite{card2018terror}. Therefore, there have been calls to investigate effective ways to detect, classify, and interdict small commercial UAVs~\cite{guvenc2018detection}.


%




Radar-based solutions have become appealing for UAV detection and recognition. This is because radars can operate effectively even in adverse weather such as fog, rain, and snow. As opposed to computer vision based approaches, radars can operate at night time. Moreover, radars can be easily deployed on different platforms on land, sea, and air. Active radars detect objects by illuminating them with electromagnetic energy and listening for the echo (back-scattered signals) from the targets. Although radar-based UAV detection is important, it is also desirable to be able to recognize and classify targets of interest.

The radar cross section (RCS) provides a measure of the target's reflectivity, and most  complex targets have unique RCS signature that can be used as the basis for target identification. The RCS signature can be used to distinguish a target from clutters. As an example, since most commercial UAVs fly at low-altitude, surveillance radars have to be tilted to low-grazing angles to search for UAVs. However, at low grazing angles, the surveillance radar could easily confuse reflections (scattering) from ground clutters and small birds with the radar return from a UAV~\cite{MartinsDrone,ezuma2019detection}. To avoid such confusion, we can create a database of RCS signature of specific UAVs of interest. This will be useful in designing automatic target recognition (ATR) and knowledge-based (KB) radar systems that use the prior knowledge of the scattering characteristics of specific targets~\cite{gini2008knowledge}.


In most of the studies on UAV RCS measurements, only the measured RCS results are provided. Extensive statistical modeling and RCS-based UAV classification of these commercial UAVs have not been investigated to our best knowledge. Motivated by the gaps in the existing literature, our contributions in this work can be summarized as follows.
\begin{enumerate}
\item
We report the results of the RCS measurement of six popular commercial UAVs in a compact range anechoic chamber at 15 GHz and 25 GHz. \textcolor{black}{The choice of 15 GHz and 25 GHz for the UAV  RCS measurement is mainly informed by the trend in the commercial counter-UAV radar industry and academic research. For example, Fortem Technologies and Ancortek Inc. are providing UAV radar solutions at 15 GHz and 25 GHz respectively~\cite{fortem_radar1, ancortek_radar1}.} We describe in detail the chamber setup and the measurement procedure. Unlike the indoor near field range~\cite{lauvcys2019investigation,guay2017measurement}, which does not satisfy the far-field condition for generating plane wave illumination for RCS measurement, the indoor compact range measurement setup uses a large parabolic reflector to ensure the target UAV is illuminated by plane waves and thus fulfilling the Fraunhofer far-field requirement.

\item
We provide extensive statistical analysis of the measured RCS data using the maximum likelihood estimator, Akaike information criterion (AIC), and the Bayesian information criterion (BIC). Using these statistical tools, we estimate the best statistical parametric models for each UAV type. For each UAV type, the best statistical model is selected from a set of 11 candidate models. Then, we rank the models according to their AIC and BIC scores with the best statistical model having the least AIC or BIC score. Also, we investigate the effect of frequency and polarization on the goodness of fit of the statistical distributions.

\item We proposed an RCS-based statistical recognition/classification for commercial UAVs. The UAV recognition system estimates the likelihood that a given test RCS data captured from an unknown UAV belongs to one of the UAV types or classes contained in the database. In the database, each UAV type is described by the most appropriate statistical distribution selected by either the AIC and BIC criteria. We analyzed the classification results with the aid of the Monte Carlo analysis. We investigated the effects of SNR, frequency, and polarization on the classification accuracy. At 10 dB SNR, using
the 15~GHz and 25~GHz HH-polarized RCS test data, the UAV recognition system achieves an average accuracy of 97.43\%
and 100\%, respectively. Similarly, for the 15~GHz VV-polarized
test data, the average classification accuracy is 99.17\% at 10~dB SNR. We provide confusion matrices to investigate the strengths and weaknesses of the recognition system.
\end{enumerate}

In an earlier work~\cite{ezuma2020compact}, we presented the VV-polarized RCS measurement data from three UAVs in a compact range anechoic chamber setup. In the current study, we provide the VV and HH-polarized RCS measurement data for six common UAVs. That way, we can investigate and compare the effects of polarization on the UAV RCS measurement results. Besides, in this study, we provide a detailed explanation of the background subtraction and calibration technique employed in our post-processing. UAV classification problem is not studied in~\cite{ezuma2020compact}, which is a major focus of the present work.

The remainder of the paper is organized as follows: Section~\ref{literature_overview} provides a brief literature overview of UAV RCS measurements and radar-based UAV detection/classification. Section~\ref{two} explains the basic principles of RCS measurement in both indoor and outdoor settings. Section~\ref{measurement_description} describes the measurement procedure while Section~\ref{statistical_analysis} describes the statistical analysis and model selection techniques employed. Section~\ref{result_DISCUSSION} presents the measurement and numerical results and Section~\ref{conclusion} concludes the paper.

\section{Literature Overview}\label{literature_overview}

\textcolor{black}{The process of classifying or identifying a radar target often requires the exploitation
of any information imprinted on the scattered signals  from the target (target echo). An example is the
use of radar micro-Doppler analysis and inverse synthetic aperture radar (ISAR) imaging for UAV
identification~\cite{MartinsDrone,rahman2019classification, rahman2018radar,ritchie2016multistatic}. It is known that complex objects like UAVs consist of several components, each with unique individual motion dynamics. For example, the rotating propellers (micromotion) have different dynamics from the mainframe. Therefore, the incoming electromagnetic wave will be modulated differently by the individual components of the UAV, thus producing different Doppler shifts. Therefore, using joint time-frequency spectral analysis on the returned signal, the features of the UAVs can be identified. In~\cite{MartinsDrone}, micro-Doppler analysis is used to distinguish a UAV from a walking man. In~\cite{rahman2019classification, rahman2018radar}, micro-Doppler analysis is used to distinguish drones from birds. In~\cite{ritchie2016multistatic}, features extracted from micro-Doppler signatures are used to discriminate between loaded and unloaded micro-drones. The study shows that the Doppler centroid and Doppler bandwidth were more suitable for drone identification as compared to the singular
value decomposition (SVD). However, a major drawback in using micro-Doppler for identifying
targets is the requirement of wide bandwidth and high computational resources~\cite{lee2016classification}. A large
bandwidth signal is needed to generate high-resolution micro-Doppler images. Also, currently,
there are no effective algorithms for decomposing a complex micro-Doppler signature into physical component-based mono-component signatures~\cite{chen2011micro}. This makes it difficult for a machine to accurately track and identify multicomponent micro-Doppler signatures without any human observer~\cite{chen2011micro}. Also, due to the requirement of the Nyquist-Shannon theory, to resolve the micro-Doppler signatures of a rotating propeller, the pulse repetition frequency (PRF) of a radar needs to be at least four times the maximum Doppler shift induced by the propellers~\cite{chen2011micro}. This requirement could pose a design challenge for surveillance radars, especially those used in airports~\cite{klaer2020investigation}. Besides, due to the small dimensions of many drones, a larger signal-to-noise ratio (SNR) is required for a radar to capture the micro-Doppler modulations of its rotating propellers~\cite{klaer2020investigation}. Furthermore, since Doppler radars can only measure radial velocity, it is impossible to perform micro-Doppler velocity if the target UAV has no radial velocity relative to the radar. Also, if there are no oscillations/rotations or vibrations, it is almost impossible to extract any micro-Doppler signature from a target. Therefore an aircraft without rotating propellers cannot be identified using micro-Doppler signature techniques.}

 \textcolor{black}{On the other hand, ISAR is a standard radar mode for high-resolution radar target identification. The ISAR images are generated by using short pulses to obtain high range resolution and exploiting the target motion to obtain high-resolution cross-range~\cite{chen2014inverse}. This requires 2D or 3D inverse Fourier transform operations. The ISAR images show the signature features of the UAV and the dominant scattering centers. In~\cite{li2016investigation,pieraccini2017rcs}, 2D ISAR images of several consumer drones were generated. However, a key challenge of ISAR imaging is that blind motion compensation is required to form a focused image of the moving UAV~\cite{li2018wide}. That is if the object has complex motion, such as non-uniform pitching, rolling, and yawing as in the case of drones, this can greatly degrade the ISAR images~\cite{chen2014inverse,jiang2016three}, Also, ISAR imaging requires wide bandwidth and high computational resources~\cite{lee2016classification}. Therefore, in this study, our focus is the 1-D radar cross-section (RCS) signature of UAVs which can be obtained with lower demand on radar memory and processing resources. RCS measurement does not require high-energy/high-resolution radar
pulse waveform, and hence considered in several works in the literature  for target identification,
association, and tracking~\cite{ehrman2010using,mertens2016ground}. RCS data can also be used jointly with other approaches to improve the detection/classification accuracy.
}

In recent times, there have been a few measurement campaigns to ascertain the RCS of some commercial UAVs. The RCS measurement can be carried out both in outdoor and indoor environments. The advantage of the outdoor measurement (outdoor range) is the ease of capturing reflections from the target UAV at far-field (plane-wave illumination). This is necessary since radar detection performance is evaluated at the far-field condition. However, in outdoor measurement scenarios, reflections from the ground, and other clutter in the environment are difficult to eliminate~\cite{galati2020characterization}. This makes it very difficult and sometimes impossible to accurately determine the actual reflections (RCS) from the target itself. Moreover, the effects of ground clutter reflections are stronger when measuring the RCS of low-altitude targets such as commercial UAVs. \textcolor{black}{Besides, the most advanced outdoor RCS measurement ranges are controlled by the military such as the secret Etcheron Valley Junction Ranch RCS Range operated by the United States Navy and the
Lockheed Martin Helendale RCS facility. These facilities are considered classified test sites and thus not available for academic research. These military-controlled facilities are often located in remote valleys or desert environments with no building, pedestrian, automobile, and poles. This is to ensure that clutters and multipath reflections are avoided. In these outdoor ranges, commercial or military grade radars are used to illuminate the target object. The scattered signals from the object are captured by the radar receiver and used to estimate the target RCS. Furthermore, RCS measurement calibration is done using standard objects with known RCS such as a sphere, cylinder, trihedral, or dihedral Corner Reflector~\cite{bradley2005investigation}. As a result of the challenges with outdoor RCS measurement, recent studies on UAV RCS measurements have focused on the controlled indoor environment (indoor ranges), which is the focus of the current study.}

\textcolor{black}{Use of RCS has also been used in the literature to distinguish small UAVs from birds. In~\cite{rahman2018flight}, the in-flight RCS of several birds (clutter) and UAVs are measured using K-band and W-band radars. The study shows that the measured RCS values of drones and birds are close. However, drones and birds could be distinguished by their RCS statistical distributions. In~\cite{gong2019interference}, an ATR system was designed to distinguish birds from UAVs using Ku-band radar echo. In~\cite{gong2019using}, the flight modes of a bird (gliding and flapping) are recognized using the wing RCS signature. In~\cite{nohara2011using}, the RCS of birds is measured around airports and this information is used to provide situation awareness for manned aircraft. In~\cite{may2017performance}, an avian radar is used for drone and bird detection. For detection, the avian radar exploits the RCS statistical model of the targets.
 We have recently been working on collecting measurement data from birds and drones, for distinguishing one from the other, using the TrueView radar from Fortem Technology~\cite{fortem_radar1}. Therefore, the techniques described in this study can be used to identify UAVs as well as distinguish between UAVs and birds. However, due to space limitations and different contexts, we will report our results with birds in our future work.
Therefore, the current study will only focus on multiple UAV identification. We will not consider the RCS-based classification of birds and other non-UAV targets.}


 \begin{table*}[t!]
\centering
\caption{Literature Review of Civilian UAV RCS Measurement and Statistical Analysis.}
\label{LITERATURE_REVIEW}
\begin{threeparttable}
\begin{tabular}{|m{0.9cm}|p{1.2cm}|c|c|c|c|c|m{1.5cm}|}
\hline
Ref. & Frequency & Measurement range  & $\#$ of UAVs & Background &  Polarization & RCS Statistical & RCS-based \\
 & (GHz) &   &   & subtraction &  & model selection & UAV Classification \\
\hline
~\cite{de2019drone} & 8.75 & Outdoor (open range) & 1 & \xmark &  & \xmark & \xmark \\
\hline

\cite{rahman2018flight} & 24, 94 & Outdoor (open range) & 3 & \xmark &  CP & \xmark & \xmark  \\
\hline

\cite{gong2019interference} & 12-18 & Outdoor (open range) & 1 & \xmark  &  & \xmark  & \xmark   \\
\hline

\cite{lauvcys2019investigation} & 9.5 & Indoor (near field range) & 1 & \xmark &  & \xmark & \xmark \\
\hline

\cite{nakamura2017characteristics} & 2.4, 24 & Indoor (near field range) & 1 &\xmark  & H-H &\xmark  & \xmark \\
\hline

 \cite{nakamura2018ultra} & 60, 79 & Indoor (near field range) & 1 & \xmark & H-H & \xmark & \xmark \\
 \hline

\cite{mizushima2020reflection} & 24/26 & Indoor/Outdoor  & 5 & \xmark & VV & \xmark & \xmark \\
\hline

\cite{li2016investigation} & 3-6, 12-15 & Indoor (near field range) & 3 &  \checkmark & VV, HH & \xmark & \xmark \\
\hline

\cite{pieraccini2017rcs} & 8, 12 & Indoor (near field range) & 2 & \checkmark & VV, HH & \checkmark  & \xmark  \\
\hline

\cite{khristenko2018magnitude} & 9 & Indoor (near field range) & 1 &  \xmark & HH  &  \xmark &  \xmark \\
 \hline

\cite{semkin2020analyzing} & 26-40 & Indoor (near field range) & 9 &  \xmark  & VV, HH & \xmark & \xmark \\
\hline

\cite{guay2017measurement} & 8–10 & Indoor (near field range) & 1 & \checkmark  & VV &  \checkmark &  \xmark \\
\hline

 \cite{ezuma2020compact} & 15, 25 & Indoor (compact range)  & 3 & \checkmark & VV & \checkmark  &  \xmark \\
 \hline
Current work & 15, 25 & Indoor (compact range) & 6 &  \checkmark & VV, HH,   & \checkmark  & \checkmark \\
 \hline
\end{tabular}
\begin{tablenotes}
    \item[1] HH, VV, and CP represents the horizontal, vertical, and circular polarization respectively.
  \end{tablenotes}
  \end{threeparttable}
\end{table*}

Table~\ref{LITERATURE_REVIEW} provides a brief survey of recent studies on the RCS measurement of commercial UAVs. In these controlled indoor measurement setups, the effect of clutters and ground reflections can be readily solved. Almost all indoor RCS measurements are carried out in an anechoic chamber where the walls, ceilings and floors are laced with high-fidelity radar absorption materials (RAM). The RAM lacing ensures that reflections from background clutter are minimized as much as possible. However, unlike outdoor environments, generating the far-field measurement conditioning in the indoor environment is not trivial, especially at microwave frequencies. This is a drawback in most of the indoor measurements surveyed in Table~\ref{LITERATURE_REVIEW}. Far-field illumination is only possible if the distance between the radar and the target is at least equal to the Fraunhofer distance~\cite{selvan2017fraunhofer}. This is difficult to achieve in almost all indoor UAV RCS measurements at microwave frequencies. Therefore, thus far, most indoor RCS measurements can be classified as near field measurement and the obtained RCS results are at best only an approximation. Moreover, background subtraction is not implemented in some of these indoor measurement setups. These omissions would affect the accuracy of the RCS measurement since antenna coupling and multipath reflections in the chamber need to be estimated and eliminated from the measurements.

\section{Principle of RCS Measurement}\label{two}
In this section, we present the basic principles and experimental techniques for RCS measurements in both outdoor and indoor settings.

\subsection{Theoretical Background}
The RCS is a measure of how much power is scattered by a target in a given direction when illuminated by a radar. The far-field RCS of the target is given by~\cite{knott2004radar}:
\begin{equation} \label{RCSequation_1}
\sigma = \lim_{R\to\infty} 4\pi R^2\frac{|\boldsymbol{\rm{E_s}}|^2}{|\boldsymbol{\rm{E_i}}|^2} = \lim_{R\to\infty} 4\pi R^2\frac{|\boldsymbol{\rm{H_s}}|^2}{|\boldsymbol{\rm{H_i}}|^2},
\end{equation}
where $\sigma$ $(\rm{m}^2)$ is the RCS of the target while $\boldsymbol{\rm{E_s}}$ (V/m) and $\boldsymbol{\rm{E_i}}$ (V/m) are the far field scattered and incident electric field intensities respectively as seen at a distance $R$. Also, $\boldsymbol{\rm{H_s}}$ (A/m) and $\boldsymbol{\rm{H_i}}$ (A/m) are the far field scattered and incident magnetic field intensities respectively as seen at a distance $R$. In terms of the horizontal ($\rm H$) and vertical ($\rm{V}$) polarization components of $\boldsymbol{\rm{E_s}}$ and $\boldsymbol{\rm{E_i}}$, the response of the target to the incident wave is described by a $2\times 2$ complex backscattered matrix equation:


\begin{align}
 \begin{bmatrix} \boldsymbol{\rm{E}}_{s,H}  \\ \boldsymbol{\rm{E}}_{\rm s,V} \end{bmatrix}
  & =\frac{1}{4\pi R^2}\begin{bmatrix}
   \sqrt{\sigma_{\rm HH}} &  \sqrt{\sigma_{\rm VH}}  \\\nonumber
  \sqrt{\sigma_{\rm HV}} &  \sqrt{\sigma_{\rm VV}}
   \end{bmatrix}
   \begin{bmatrix} \boldsymbol{\rm{E}}_{\rm {i,H}}  \\ \boldsymbol{\rm{E}}_{\rm{i,V}} \end{bmatrix},
\end{align}
where $\sigma_{\rm HH}$ and $\sigma_{\rm VV}$ are co-polarized RCS of the object while $\sigma_{\rm HV}$ and $\sigma_{\rm VH}$ are the cross-polarized RCS. In the case of backscattering (monostatic and quasi-monostatic), $\sigma_{\rm HV}=\sigma_{\rm VH}$.

An RCS measurement setup can be broadly classified into two main groups: outdoor-range and indoor-range (near-field and compact-range). These ranges will be explained in the next two subsections.

\begin{figure}[t!]
 \center
 \includegraphics[scale=0.3]{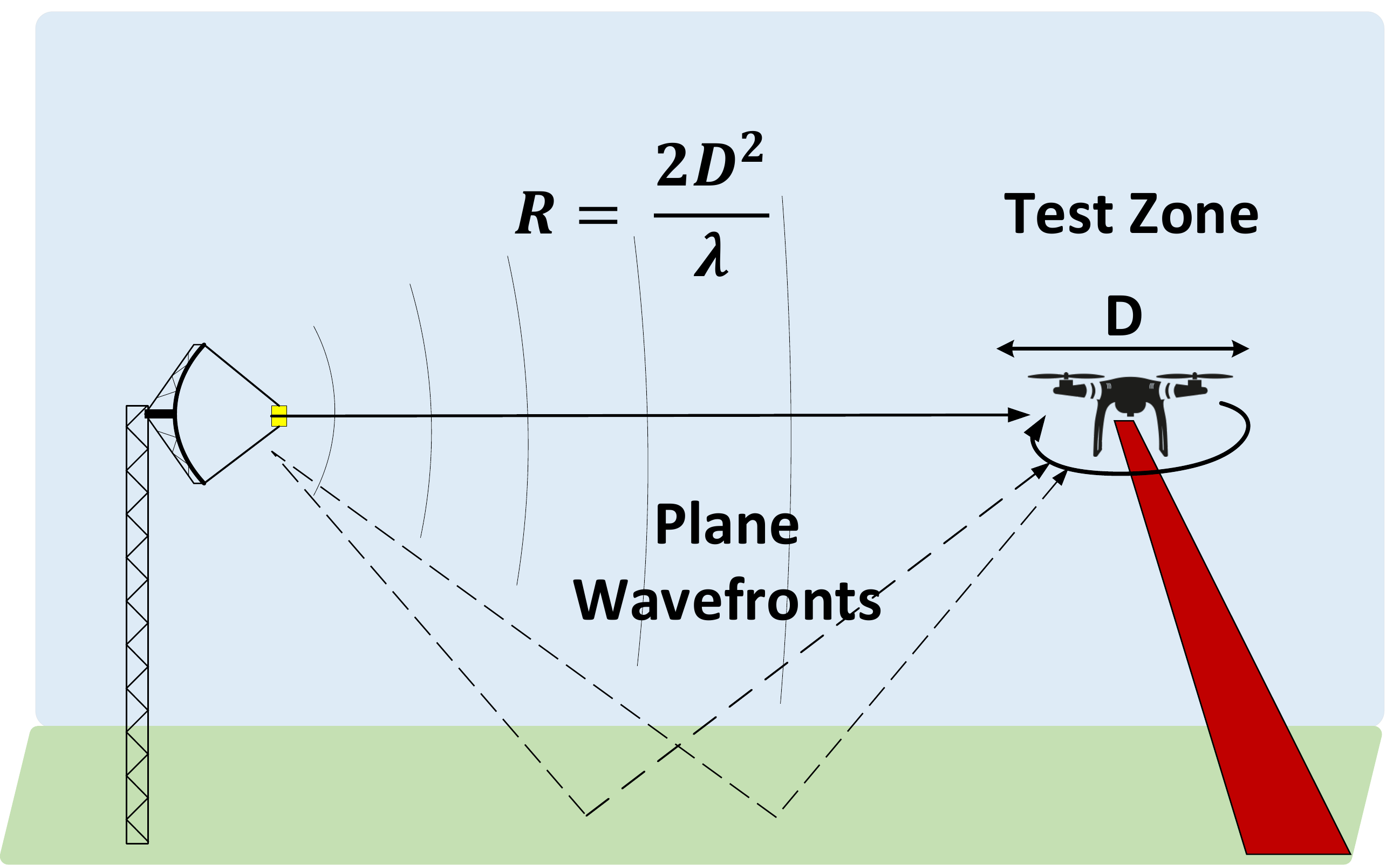}
\caption{Outdoor RCS measurement scenario of a small UAV. Atmospheric condition and inability to isolate the target are issues.}
\label{Fig:OUTDOOR}
\end{figure}

\subsection{Outdoor RCS Measurement}
The RCS definition in~(\ref{RCSequation_1}) requires the physical distance of $R$ between the radar and the target UAV to be sufficiently large. This is called the far-field condition and is necessary to generate plane wave illumination for the target UAV. Besides, the far-field requirements eliminate any distance dependency in the RCS signature of the target~\cite{knott2004radar}. Mathematically, the far-field distance (Fraunhofer distance) $R$ is given by:
\begin{equation}
  R\geq\frac{2D^2}{\lambda},
   \label{RCS_farfield_condition}
\end{equation}
where $D$ is the transverse length (or diameter) of the target and $\lambda$ is the wavelength of the radar signal~\cite{knott2004radar}. At this distance $R$, the spherical wavefronts emitted by the radar will have approximately equal phases across any measurement area (plane wave fronts). For instance, a DJI Matrice 600 UAV, a popular commercial grade UAV, have diameter of about 1.133 m~\cite{DJI_M600Spec}. Therefore, using a 25 GHz radar, we need a separation distance $R$ of at least 213.95~m to accurately measure the RCS of the UAV in the far-field. This is almost twice the length of a standard football field. However, in outdoor measurement environment, we can achieve the desired separation distance in free space. This is the advantage of the outdoor range experiment. During the measurement, the UAV is placed on a Pylon stand or allowed to hover in the radar's field of view. Fig.~\ref{Fig:OUTDOOR} shows a typical outdoor range RCS measurement scenario. However, as shown in Fig.~\ref{Fig:OUTDOOR}, the ground clutter reflection is an unavoidable drawback in outdoor range experiment. Besides, environmental clutters like buildings will generate multipath scattering which will affect the accuracy of the measurement.

\begin{figure}[t!]
 \center
 \includegraphics[scale=0.25]{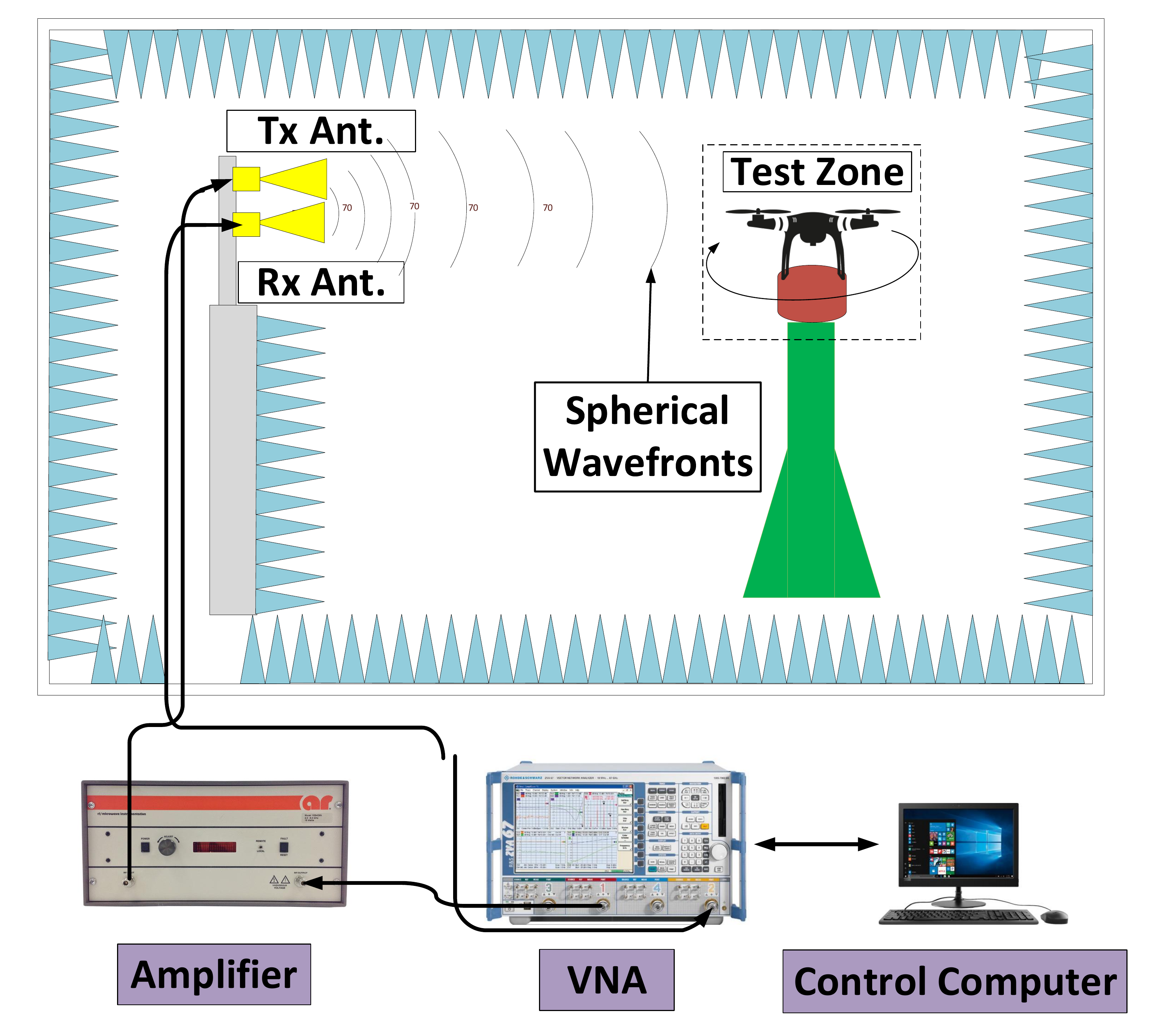}\caption{Indoor near field RCS measurement.  Spherical waves are used to illuminate the target.}
\label{Fig:indoor_far-field}
\end{figure}

\subsection{Indoor RCS Measurement}
Due to the challenge of ground reflection, environmental clutter, and atmospheric conditions associated with outdoor range, RCS measurement of commercial UAVs can be carried out in an indoor environment. In most cases, anechoic chambers are used as a controlled environment for indoor RCS measurement. Fig.~\ref{Fig:indoor_far-field} shows a typical indoor measurement setup. Here, the radar system is implemented using a combination of vector network analyzer (VNA), antennas, and power amplifiers. The consumer UAV is placed on a turntable which is controlled by computer. However, for high-frequency measurement, say 25~GHz, the far-field distance requirement is difficult to achieve within an indoor experimental setup. It will be too expensive to design an indoor anechoic chamber that is the size of a football field. Therefore, the indoor setup shown in Fig.~\ref{Fig:indoor_far-field} is considered as a near-field RCS measurement setup~\cite{semkin2020analyzing}. The illuminating signals emitted by the radar have spherical wavefronts. Therefore, the measured RCS is only an approximation of the true value.

\begin{figure}[]
\center{
\begin{subfigure}[Offset feed]{\includegraphics[width=0.3\linewidth]{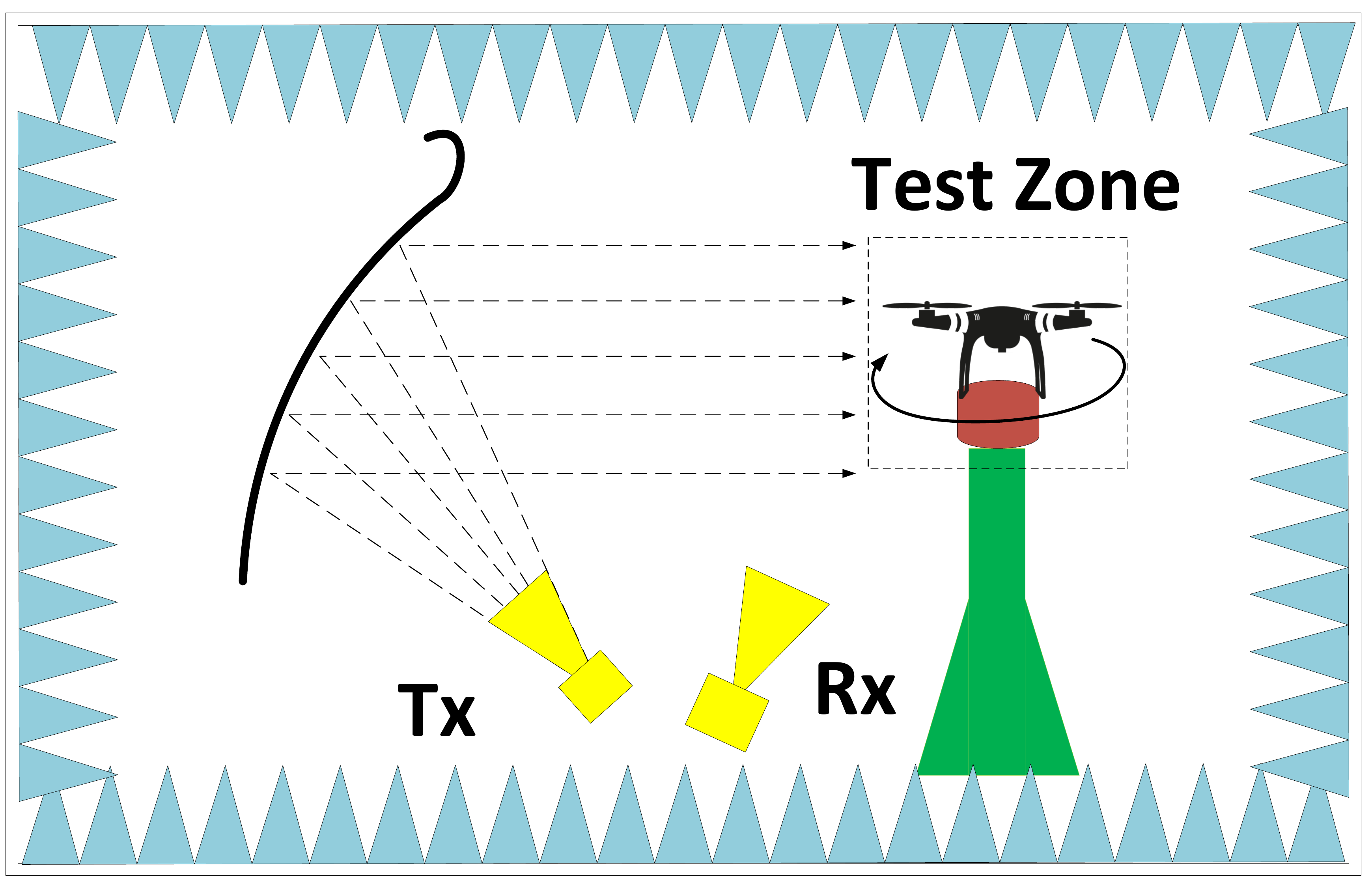}\label{trimbler_RCS_UAV}}
\end{subfigure}
\begin{subfigure}[Cassegrain dual reflector]{\includegraphics[width=0.3\linewidth]{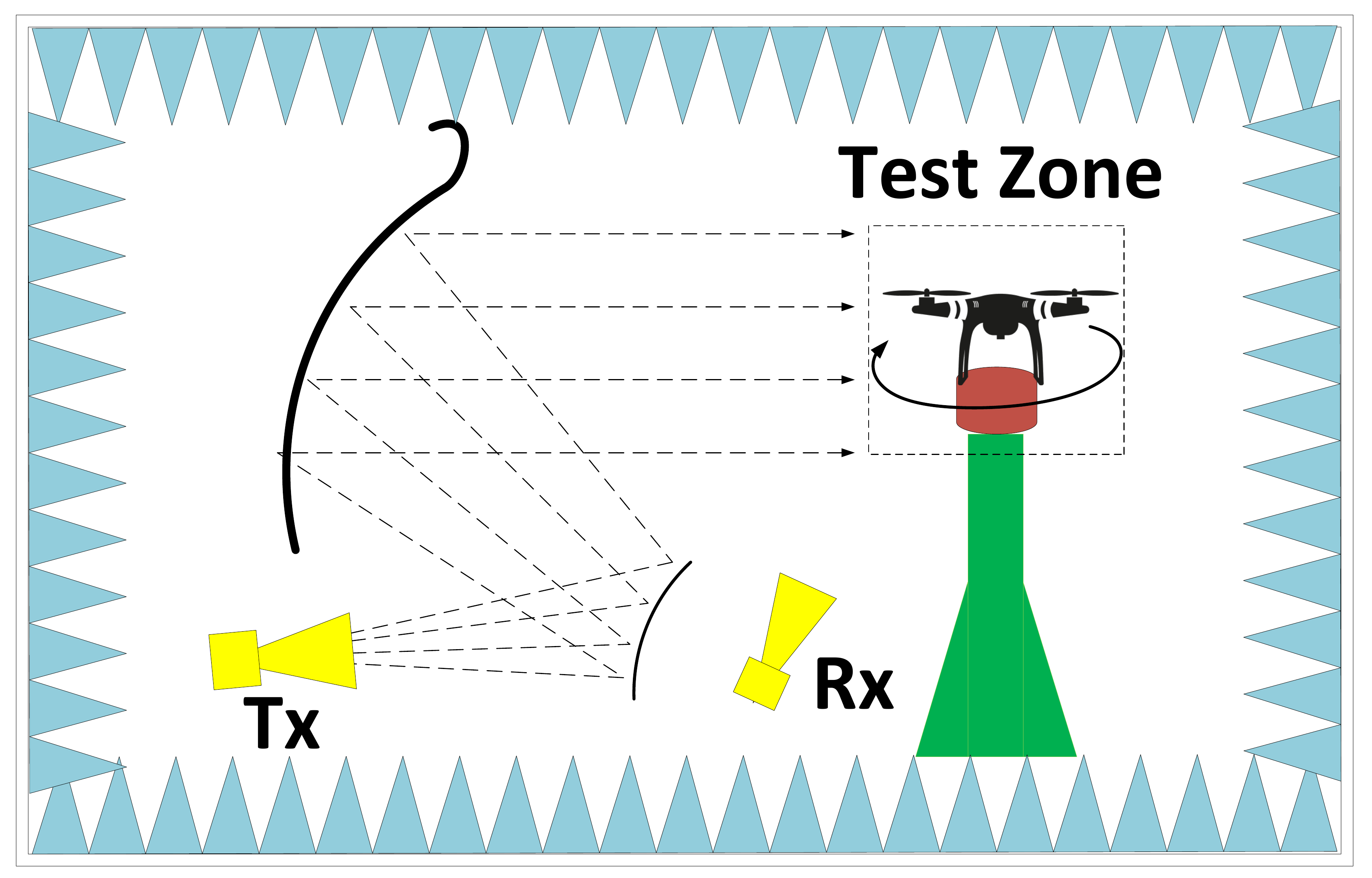}\label{inspire_RCS_UAV}}
\end{subfigure}\\
\begin{subfigure}[Gregorian dual reflector]{\includegraphics[width=0.3\linewidth]{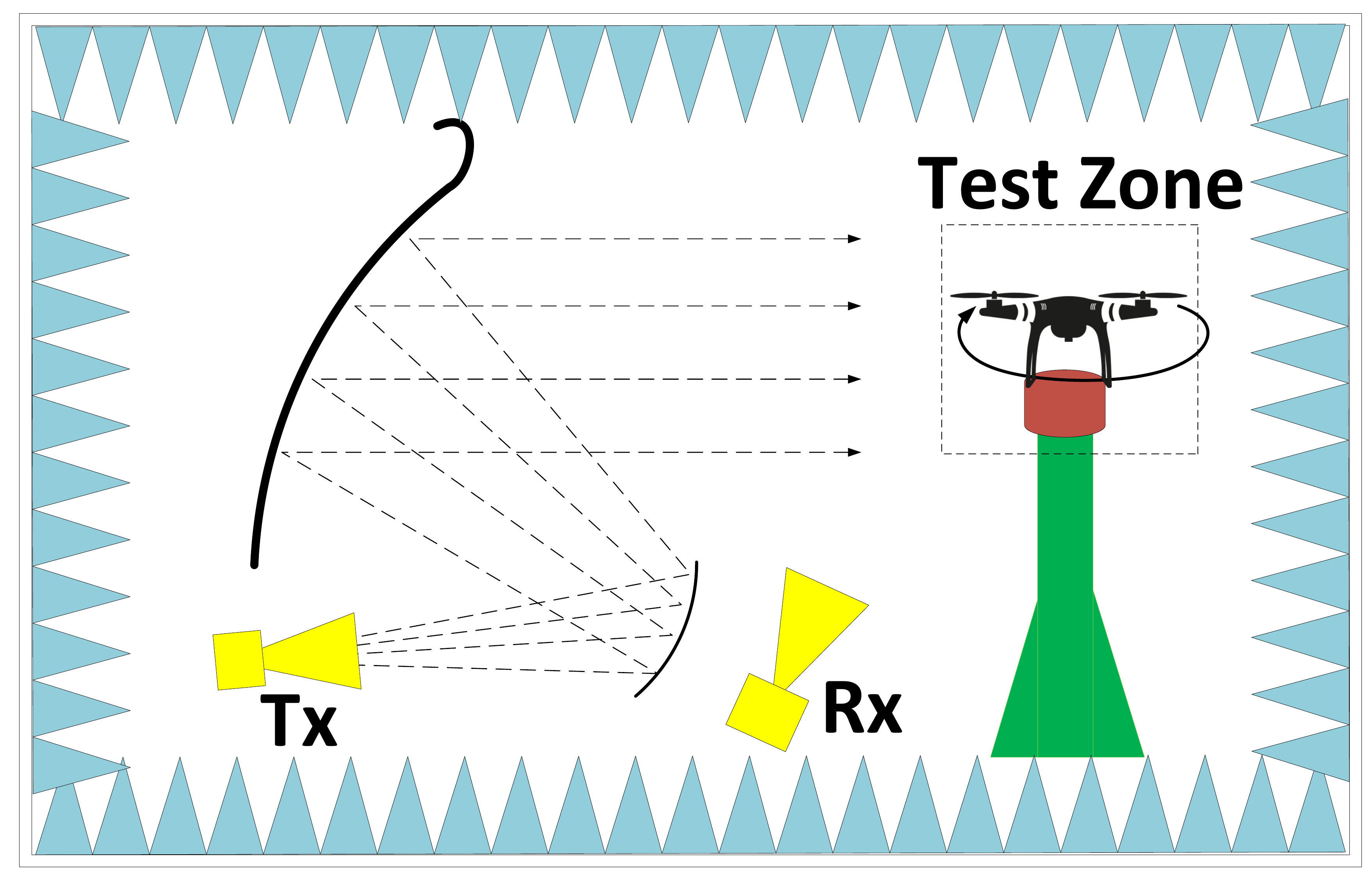}\label{inspire_RCS_UAV}}
\end{subfigure}
\begin{subfigure}[Dual cylindrical reflector]{\includegraphics[width=0.3\linewidth]{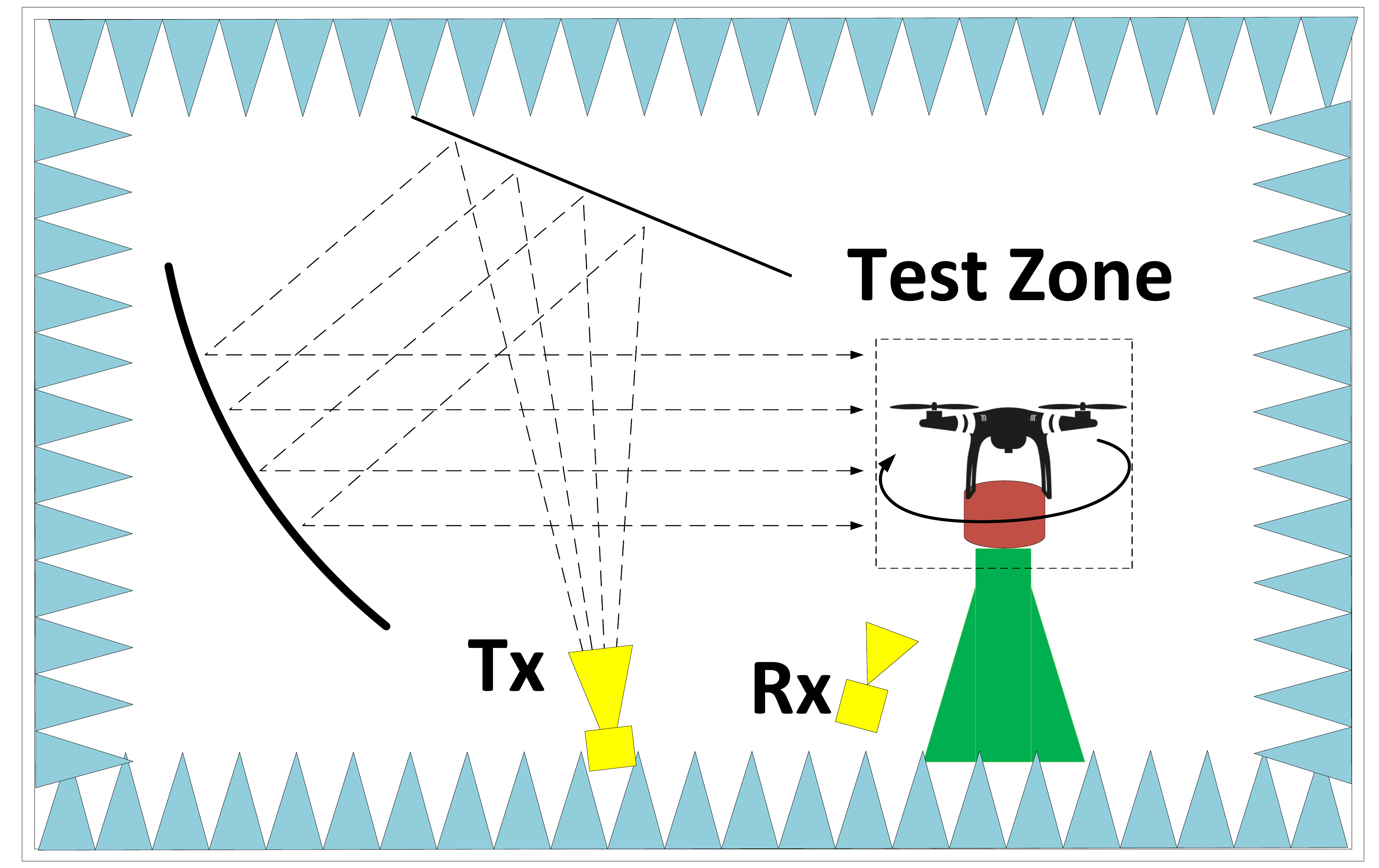}\label{inspire_RCS_UAV}}
\end{subfigure}\\
\begin{subfigure}[Dielectric lens reflector]{\includegraphics[width=0.3\linewidth]{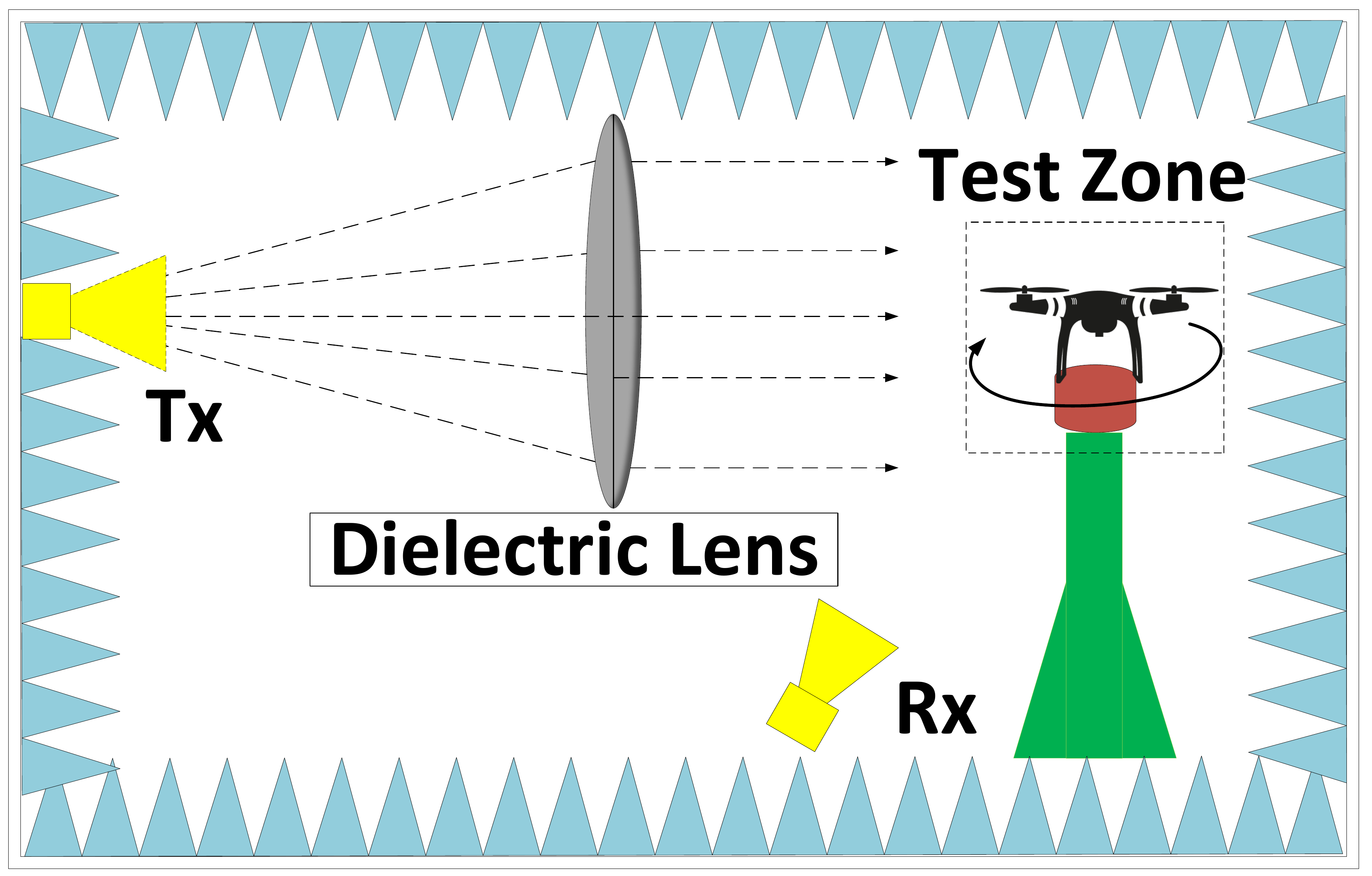}\label{inspire_RCS_UAV}}
\end{subfigure}
\begin{subfigure}[Hologram-based compact range]{\includegraphics[width=0.3\linewidth]{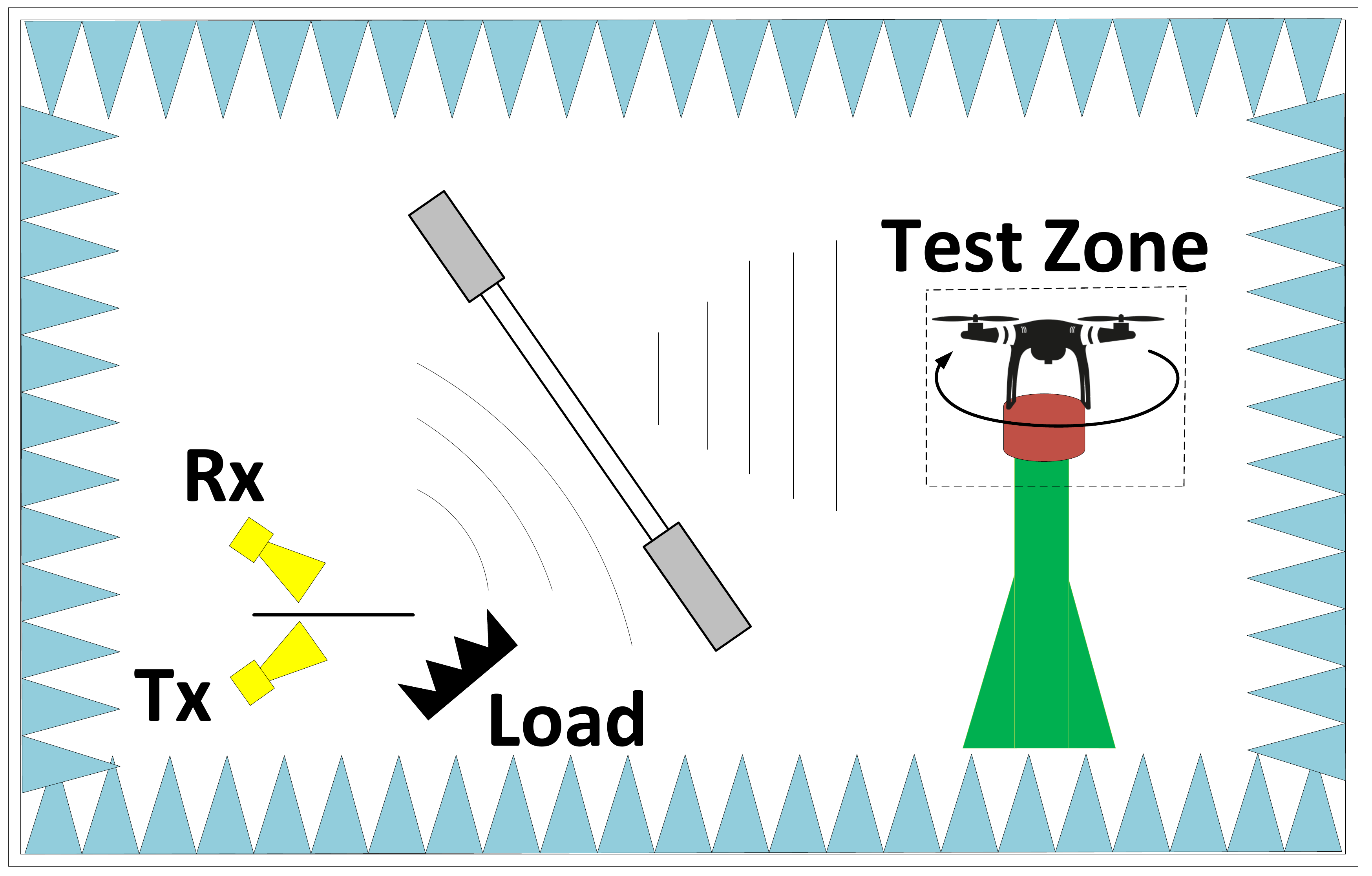}\label{phantom_RCS_UAV}}
\end{subfigure}
\caption{{Different compact range configurations that could be used for measuring the RCS of a small  UAV. Each  configuration uses a different approach to collimate the transmitted wave from the radar~\cite{IEEE_standard,lonnqvist2006phase}.}\label{compact_range_configurations}}}
 \vspace{-4mm}
 \end{figure}

 In order to accurately measure the RCS of a UAV in an indoor environment, we need to find a way to generate plane wave illumination (far-field condition) within a short distance. This is achievable in a compact range configuration. The basic principle of a compact range chamber is to use either an optical lense, parabolic reflector(s), or dielectric lense to collimate the spherical wavefronts emitted by the radar into a planar wave in a
relatively short distance, thus the term “compact range”. Fig.~\ref{compact_range_configurations} shows different possible compact configurations. However, due to the stringent requirements and cost associated with designing a large perfect collimator, many researchers have not been able investigate this techniques for the RCS measurement of commercial UAVs. The next section describes how we use an offset feed  compact range anechoic chamber to accurately measure the RCS of commercial UAVs.

\section{UAV RCS Measurements and Calibration in an Offset-Fed Compact-Range Anechoic Chamber}\label{measurement_description}
\begin{figure}[t!]
 \center
 \includegraphics[scale=0.75]{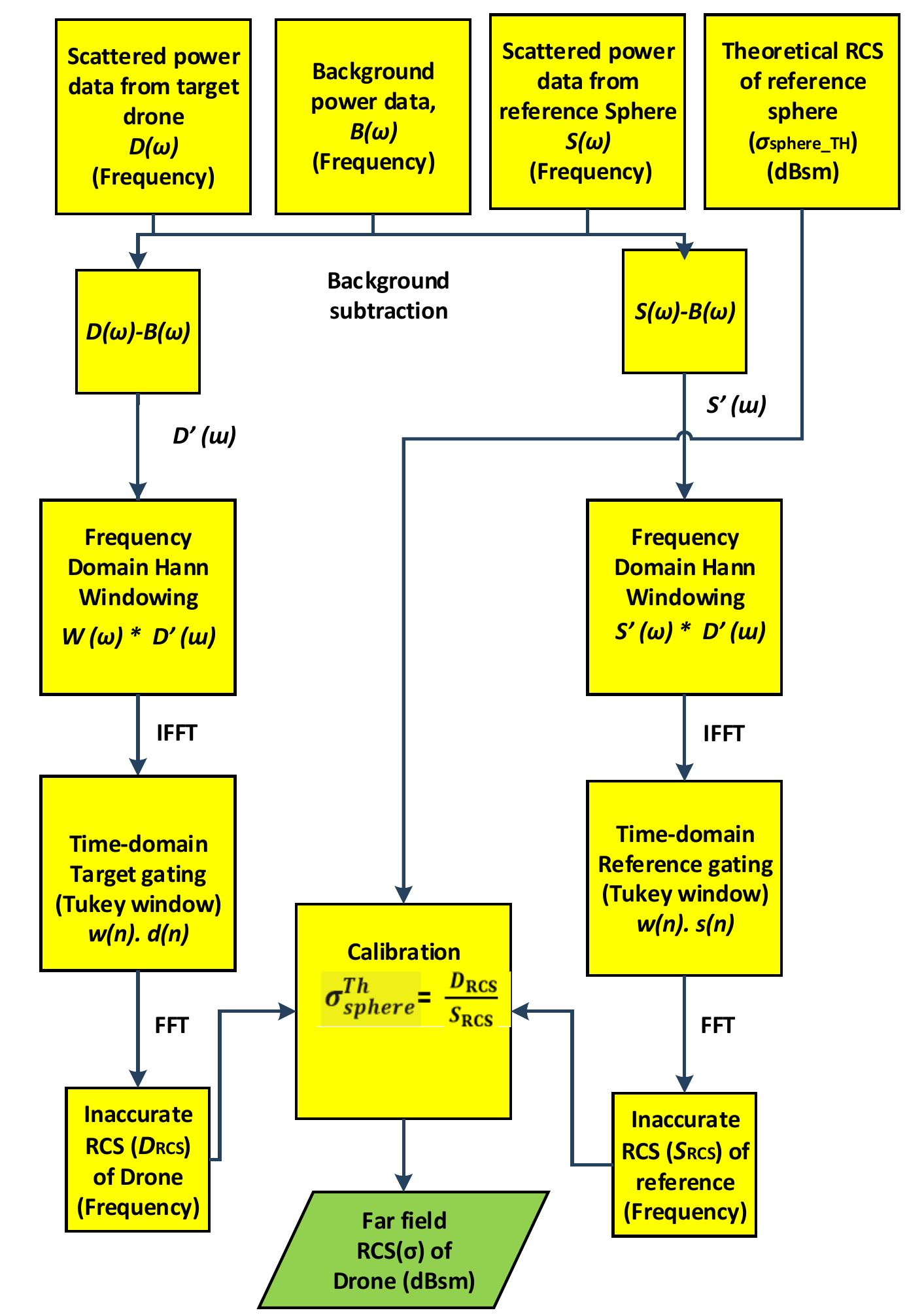}
\caption{Flowchart of the Far-field (plane wave) RCS measurement of the small UAVs in compact range anechoic chamber.}
\label{Fig:RCS_FLOWCHART}
\end{figure}
\begin{figure}[t!]
\center{
 \begin{subfigure}[]{\includegraphics[width=0.445\linewidth]{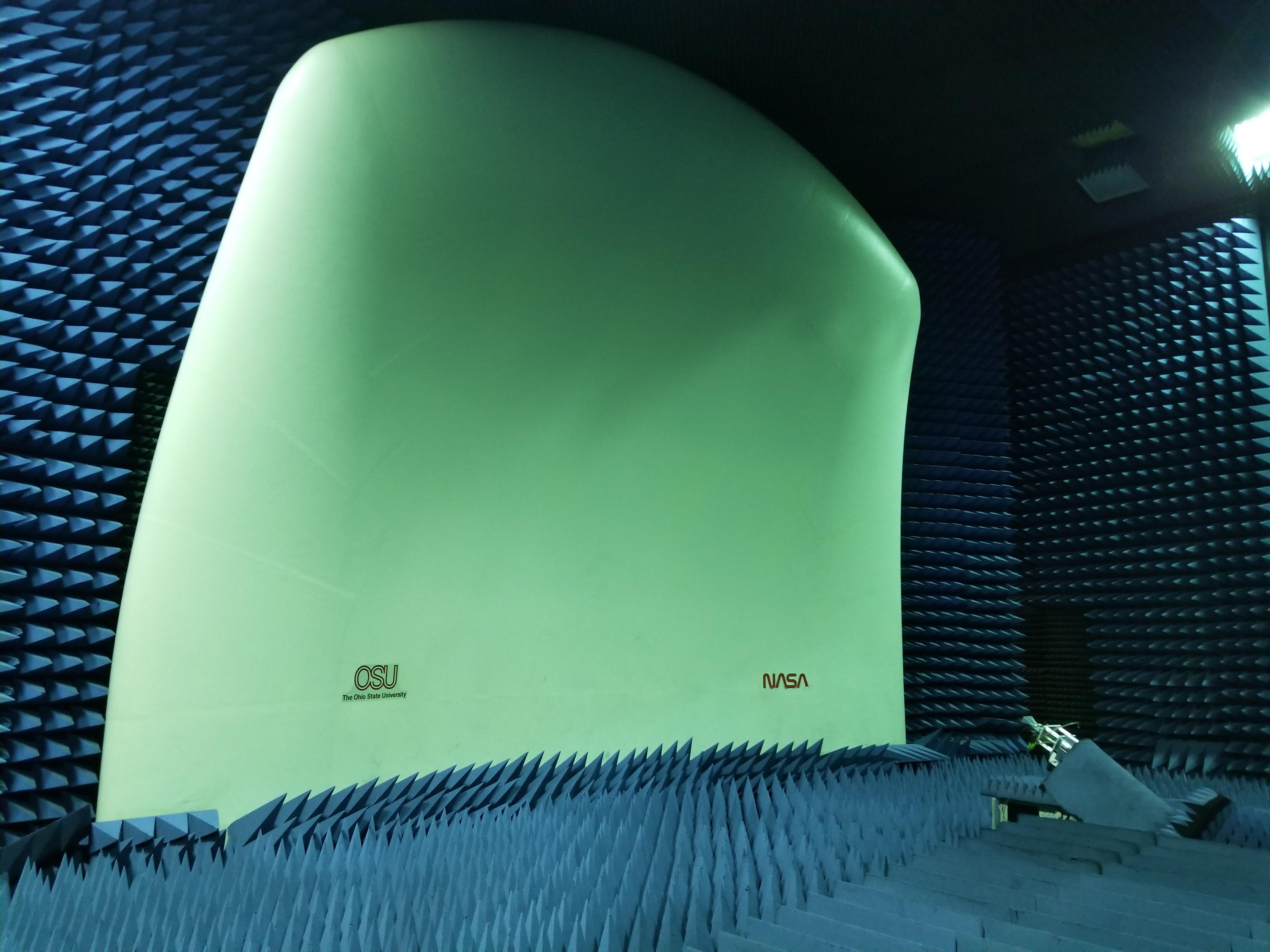}
 \label{uav_fig_A}}
\end{subfigure}
 \begin{subfigure}[]{\includegraphics[width=0.45\linewidth]{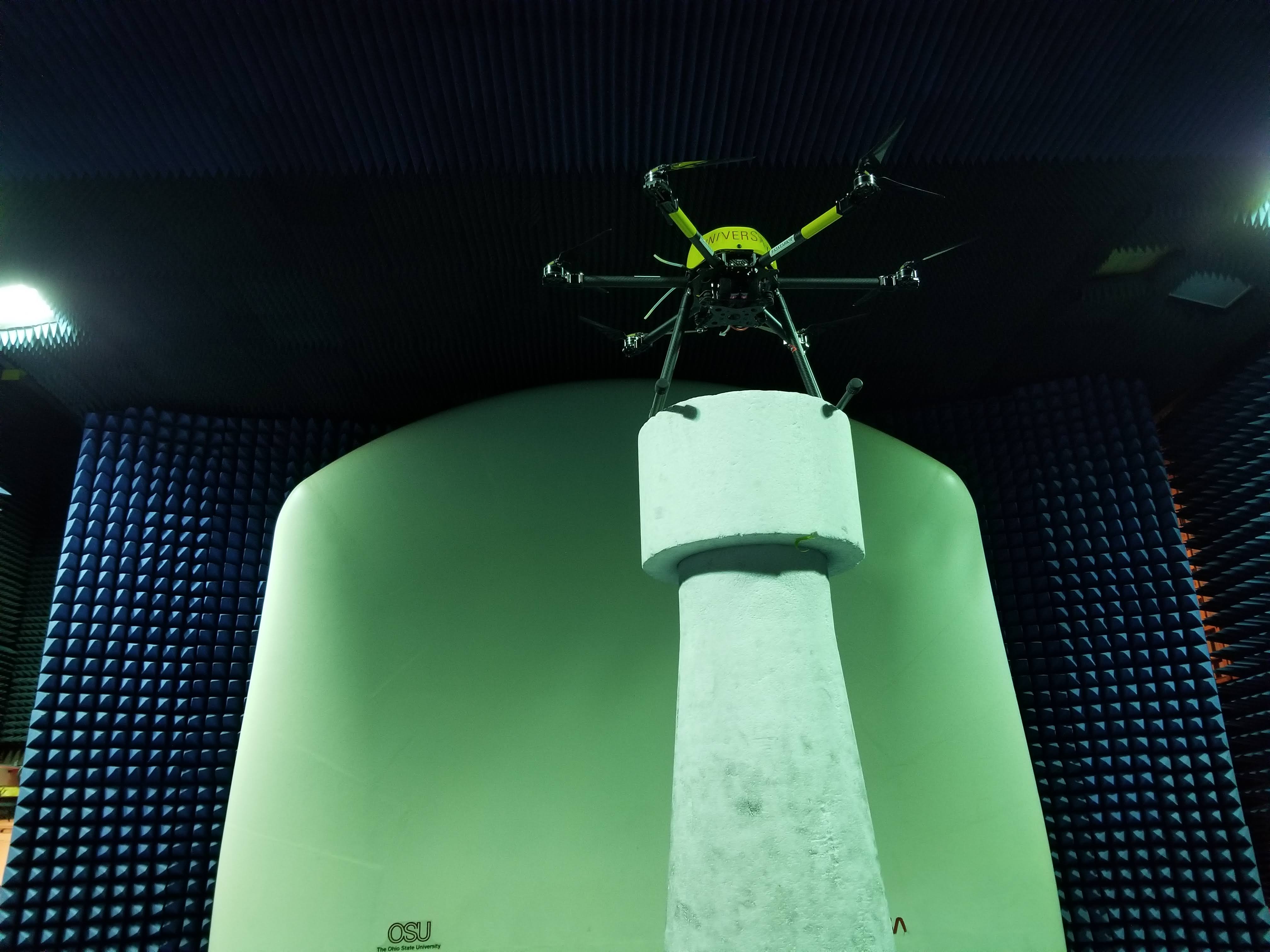}\label{uav_fig_b}}
\end{subfigure}
\caption{{(a) Large offset feed parabolic reflector, and (b) UAV RCS measurement scenario. }\label{compact_range}}
}
 \end{figure}

In this section, we describe our UAV RCS measurement procedure using an offset feed compact-range anechoic chamber. Fig.~\ref{Fig:RCS_FLOWCHART} is a flowchart that graphically describes the process and techniques employed in this study. As shown in Fig.~\ref{Fig:RCS_FLOWCHART}, the first step in measuring the RCS of a target UAV is to capture scattered radar data from the target UAV during a controlled experiment. \textcolor{black}{The transmit power of the radar is 5 dBm. However, the measured RCS does not depend on the transmit power or the bandwidth of the radar receiver. Although the bandwidth will determine the noise floor (noise level) of the radar receiver and thus the signal to noise ratio (SNR) of the received signal, it does not determine the RCS or radar signature of the target}. Also, we probe the background with the radar signals to obtain a background response. The scattered data measured from both the target and background are processed and calibrated to obtain the accurate RCS of the target. The measurement setup is described next.

\subsection{Measurement Setup and Procedure}
\textcolor{black}{To obtain a plane wave radar illumination within a limited indoor environment, we use the 20 foot high collimating parabolic reflector shown in Fig.~\ref{uav_fig_A}}. During the measurement, the chamber uses the Keysight E8362B programmable VNA to generate continuous-wave radar signals, centered at the test frequency. Using a pair of horn antennas, the radar signals are transmitted and the backscattered signals from the target UAV are received. \textcolor{black}{The transmit horn antenna (Tx) is connected to Port 1 of the VNA and the receiver antenna (Rx) is connected to Port 2. The VNA measures the scattering transmission coefficient $S_{21}$ which is proportional to the ratio of the reflected (scattered) power at Port 2 to the input power at Port 1. Therefore, using the VNA to implement a radar system, the magnitude of the measured $S_{21}$ is proportional to the RCS of the target UAV in the anechoic chamber~\cite{grace2011measurement}.} For the 15~GHz RCS measurement, we use the Cobham H-1498 broadband transverse electromagnetic (TEM) horn antennas in the TX and receive (RX) antennas. And for the 25 GHz RCS measurement, we use a Narda 638 Standard Gain Horn at both the TX and the RX. These antennas are placed at the offset/focus of the parabolic reflector.

The effect of the parabolic reflector is two-fold. First, the reflector changes the curvature of the transmitted wave from spherical to planar. The curvature and smoothness of the parabolic reflector ensure that the reflected waves are collimated to simulate far-field conditions at a relatively short distance.
\textcolor{black}{Second, from physics, we know that one of the main properties of radio waves is reflection. By reflection, a surface (a reflector) changes the original direction of an incoming or incident wave. However, if the reflector is a parabolic surface and the wave source/feed horn (TX) is located at its focus, then the incident wave from TX is reflected from the parabolic surface as plane waves as shown in Fig.3 (a). From Fig. 3(a), we see that the reflected plane waves are now in the direction of the UAV. On reaching the UAV, the plane waves are scattered/reflected once again. The scattered electromagnetic signals are captured by the receiver antenna (RX), which is connected to Port 2 of the VNA.
Therefore, the parabolic reflector has performed the dual task of plane wave generation and direction reversal of the incident wave propagating from TX.}

The scattered signal power ($P_{rcv}^{sc}$) is processed to measure the RCS of the UAV. Mathematically, the offset feed compact range is governed by the following expression~\cite{hess2008introduction}:
\begin{equation}
  \frac{P_{\rm rcv}^{\rm sc}}{P_o}=\frac{\sigma}{\lambda^2}\frac{1}{(4\pi)^3}
\left(\frac{\lambda}{R_o}
\right)^4 G_{Tx} G_{Rx} = \sigma\cdot k,
   \label{RCS_eq3}
\end{equation}
where $P_{\rm o}$ is the transmitted power by the feed horn, $G_f$ characterizes the gain of the feed horn antennas, $R_o$ is the distance from focus (focal point) and the reflector along with the principal ray intersection point, and $k=\frac{\lambda^2 G_{Tx} G_{Rx}}{(4\pi)^3 R_o^4}$ is a constant. \textcolor{black}{Therefore, in terms of the scattered and transmit power, the $S_{21}$ measured by the VNA is given in~\cite{grace2011measurement} as}:
\begin{equation}
  S_{21} = 10\cdot \log_{10} \frac{P_{\rm rcv}^{\rm sc}}{P_o},
   \label{RCS_eq31}
\end{equation}

From  geometry of conic sections, the value of $R_o$ can be estimated with respect to the focal length ($f_L$) and outside distance ($K$) of the parabolic reflector as~\cite{johnson1979conceptual}:
\begin{equation}
  R_o=f_L+\frac{1}{16}\frac{K^2}{f_L}.
   \label{RCS_eq4}
\end{equation}

During the measurement, the UAV is placed on the Styrofoam turntable which is 6 foot away from the antennas, see Fig.~\ref{uav_fig_b}. The turntable, controlled by a stepper motor, rotates the target through the azimuth plane $\phi\in[0\degree, 360\degree]$ with a 2$\degree$ increment. For each look angle, continuous wave signals, centered at the test frequency, are generated in the VNA and transmitted through the transmit horn antenna (TX). To reduce the effects of unwanted reflections (clutter and multipath reflections) in the chamber, radar absorption materials (RAM) are used. The chamber used in this study employs pyramidal and wedge-shaped extra high performance (EHP) RAM absorbers. The ETS-Lindgren's type EHP (Extra High Performance) microwave pyramidal absorbers are used in the front of the feed and the back wall while EHP-18EGCL microwave wedge absorbers are used behind the feed on the side walls, floor, and ceiling. These absorbers reduce multipath reflections from the walls, floor, and ceiling of the anechoic chamber. Typically, the performance of any RAM absorber depends on its thickness in wavelength~\cite{knott2004radar}.


 After the RCS measurement, the raw data (scattered power in dB) is post-processed to obtain the accurate RCS of the target UAV in dBsm.

\subsection{Post Processing}
Post processing is done in MATLAB. Four major post processing operation, shown in he flowchart in Fig.~\ref{Fig:RCS_FLOWCHART}, are performed on the captured scattered data. These operations can be summarized as follows.
\begin{itemize}
 \item{Step 1:} Perform background subtraction on the captured raw data in frequency domain.
 \item{Step 2:} Band limiting and side lobe reduction using Hann window.
 \item{Step 3:} Transform the resultant data from frequency to time domain using inverse Fourier transform (IFFT).
 \item{Step 4:} Perform time (range) gating on the target zone.
 \end{itemize}

 \begin{figure}[t]
 \center
 \includegraphics[scale=0.65]{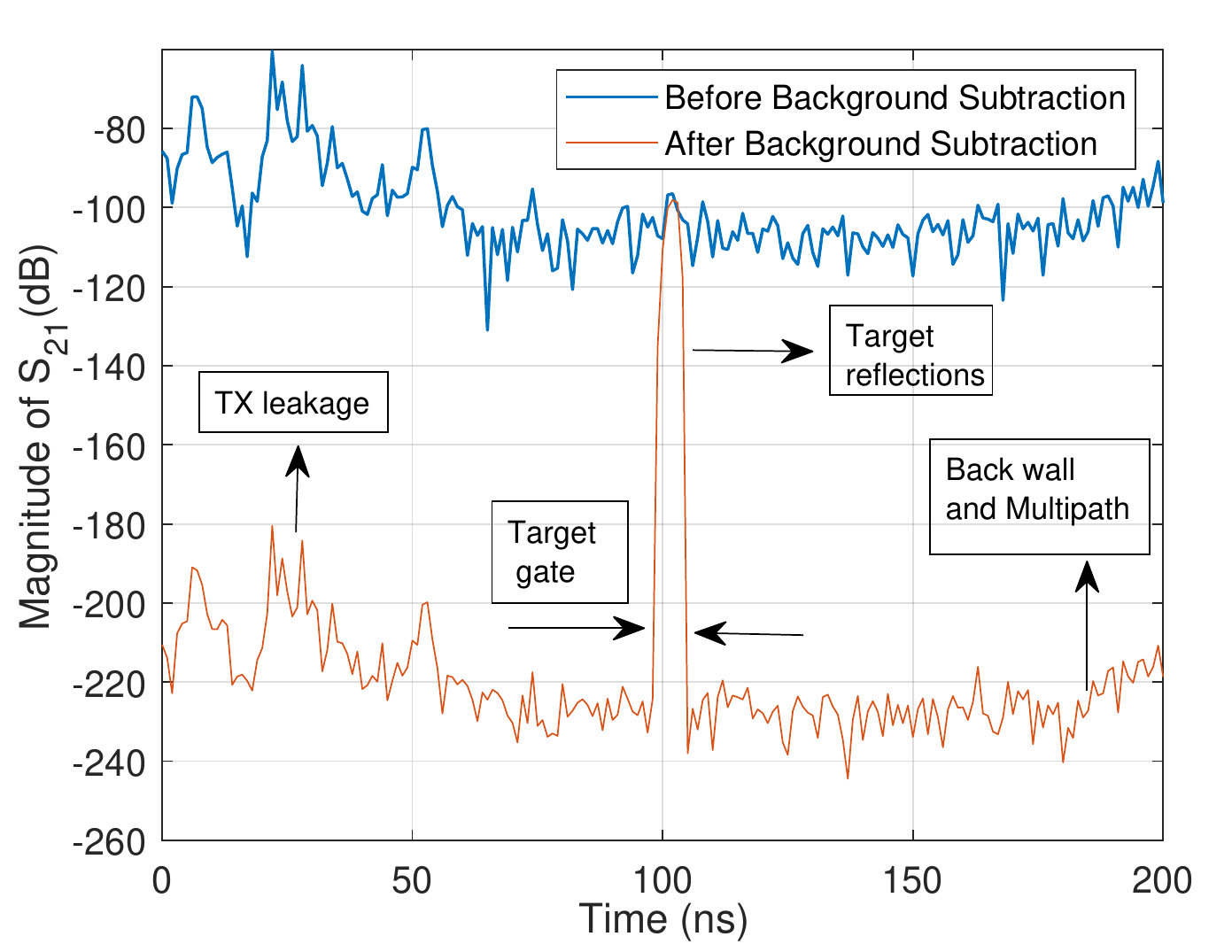}
\caption{Time domain response of the measurement data showing the effects of background subtraction. The target reflections can be isolated from the clutters using a target gate.}
\label{Fig:Target_gate}
\end{figure}


The first step in post-processing the raw data obtained during the target measurement is the background subtraction. The background measurement characterizes the frequency response of the chamber. To implement background subtraction, we measure the chamber without the target UAV mounted on the rotating stand. This measurement is then subtracted from all the UAV RCS measurements (vectorial subtraction). The purpose of background subtraction is to eliminate signals generated by clutters in the environment. Moreover, background subtraction also helps us to remove unwanted spurious signals such as leakage in the transceiver system and coupling between the TX and RX antennas. Since all the UAV RCS measurements were carried out over a wide bandwidth and completed within a short time duration (due to the automation of the measurement process), the background characteristics of the chamber is assumed to be unchanged or very slowly changing  during the entire measurement period. Therefore, we do not have to remeasure the background every time a different UAV is placed on the rotating stand. It is important to note that background subtraction may not completely remove all the clutter and leakage signals from the measurement data. Therefore, to accurately isolate the intrinsic scattering data from the target, we perform time domain transformation using the IFFT operation.

 After performing the IFFT operation on the resulting data, we obtain the time domain characteristics of measurement data. This could be considered as the intrinsic response of the target in the time domain. \textcolor{black}{Besides, we apply the frequency-domain Hann window before the IFFT operations to band limit the frequency domain data captured by the VNA~\cite{gaberson2005applying,beyene2003accurate}. This is necessary to cut-off side high-frequency noise and ripples/side lobes which are present when data are captured in the frequency domain. Frequency domain ripples tend to asymmetrically approach non-zero constants as frequency increases~\cite{beyene2003accurate}. That is, the high-frequency ripples/noise do not terminate at zero. As a result, if we perform an IFFT operation, without band-limiting the spectrum, there will be unwanted time-domain rings. Also, the Hann window is better at band-limiting a frequency spectrum than flattop windows. This is because Hann windows can easily resolve multiple frequency peaks and by so doing separate the signals of interests from noise/ripples/sidelobes~\cite{schaldenbrand2019window}. Also, unlike the Hamming window, the endpoints of the Hann window smoothly transition to zero making the latter better suitable as a band-limiting filter for frequency-domain signals}.

 Fig.~\ref{Fig:Target_gate} shows the impulse response of the scattered data. We can see the scattering from the target as well as every remaining clutter and leakage. \textcolor{black}{Therefore, to obtain the intrinsic scattering from the target object, we perform time (or range) gating using a windowing function in the time-domain. This operation is sometimes called software gating~\cite{borkar2010radar}. The time-domain software gating operation is equivalent to the multiplication of the processed data/signal with a windowing function. The Tukey window is suitable for the software gating operation. This is because, in comparison with other time-windowing functions such as the Hann and Hamming windows, the time-domain Tukey window (tapered cosine function) is less likely to distort the amplitude of the time-domain transient (time-domain impulse response) which measures the RCS of the target. Therefore, amplitude/FFT gain correction is not necessary when we use a Tukey window in the time-domain just before an FFT operation.~\cite{schaldenbrand2019window}}.

 The software gating (or target gating) is used to filter out unwanted time domain responses from the captured signal. That is, signals are accepted or rejected according to the time (range) gates. Signals outside the target time window (target zone) are gated out. This includes multipath and clutter returns from the chamber. The former arises from scatterers that are further away from the target zone while the latter (clutter sources) arises from targets that are located at, or near the target~\cite{knott2004radar}. The resulting signal is the RCS of the target UAV in time domain. This is converted into the frequency domain by means of the FFT operation. The resulting frequency domain RCS data ($D_{RCS}$) is fairly inaccurate. To correct the measurement error, we perform calibrations using standard objects with known RCS.

\begin{figure}[t!]
\center{
 \begin{subfigure}[]{\includegraphics[scale=0.44]{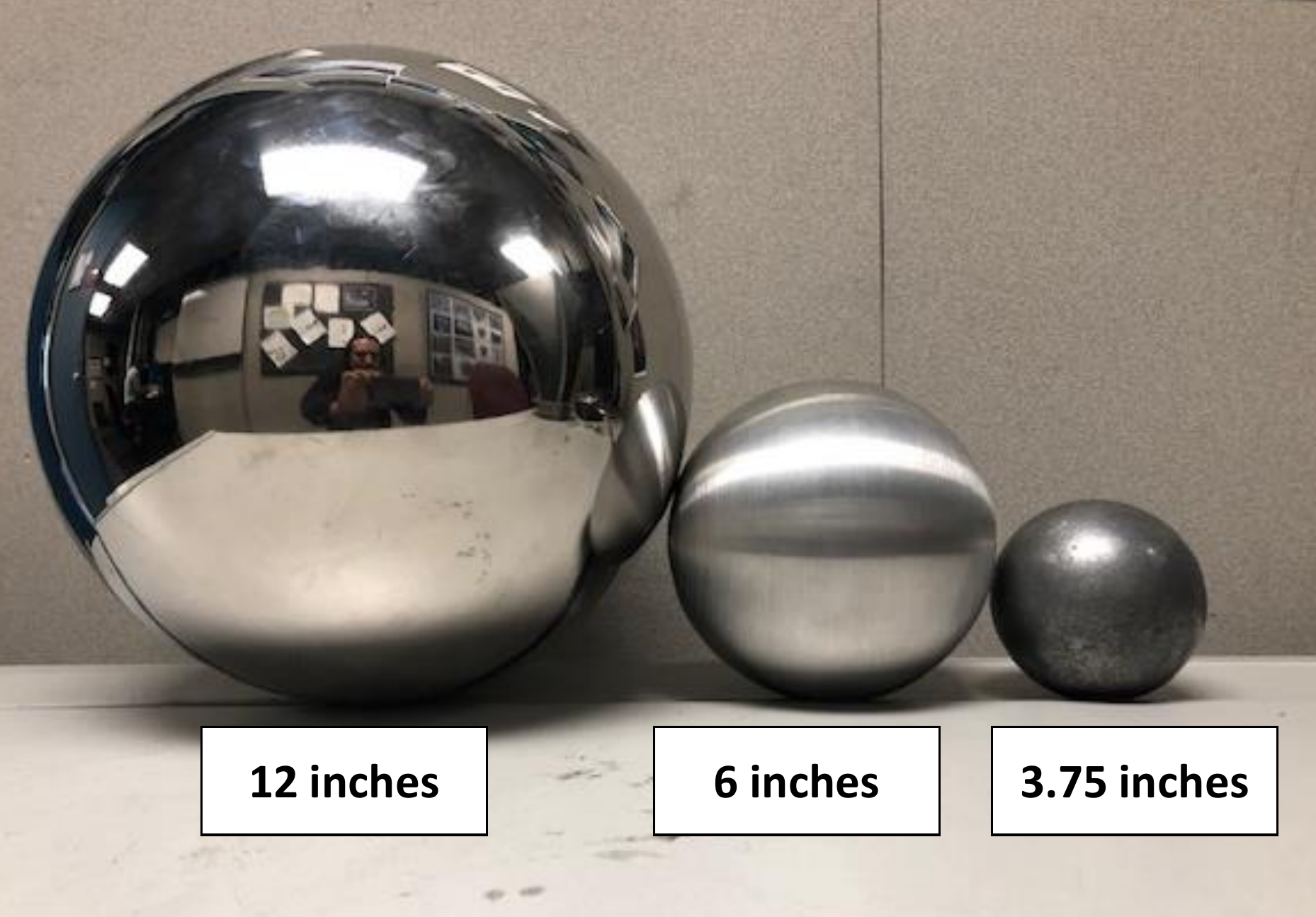}
 \label{Fig:standard_sphere}}
\end{subfigure}
 \begin{subfigure}[]{\includegraphics[scale=0.48]{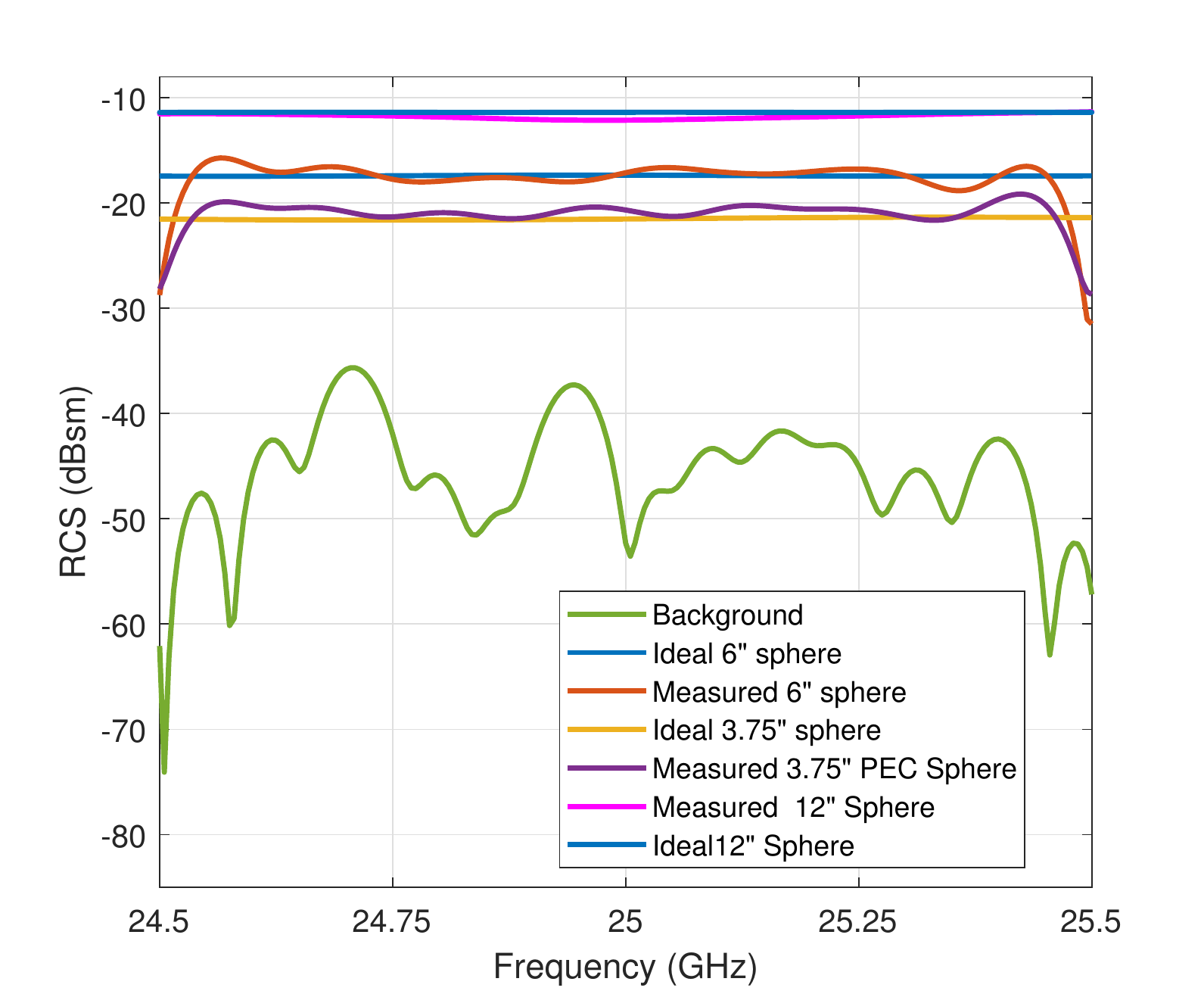}\label{Fig:calibration_measurement}}
\end{subfigure}
\caption{{(a) Three standard calibration PEC spheres: 12", 6", and 3.75". The theoretical RCS of each of these spheres can be estimated from their radius, and (b) the ideal and measured RCS (dBsm) of the three standard versus frequency.}\label{compact_range}}
}
 \end{figure}


\subsection{RCS Measurement Calibration}
In this study, we use perfectly electrical conducting (PEC) spheres with known theoretical RCS value for measurement calibration.
Besides, due to spherical symmetry, the RCS of the PEC sphere is independent of aspect angle (nonfluctuating or Marcum objects). This makes the PEC sphere a perfect calibration object. Fig.~\ref{Fig:standard_sphere} shows three standard calibration spheres used in this experimental study. Only one sphere can be used at a time. However, having measurement from multiple PEC calibration spheres allows us to verify the reliability and repeatability of our calibration procedure.

First, we measure the RCS ($S_{\rm{RCS}}$) of a PEC calibration sphere in the compact range chamber. Next, we compute the theoretical or exact RCS ($\sigma_{\rm{sphere}}^{Th}$) of the calibration sphere. In the far-field, the RCS of a PEC sphere has a closed form analytical expression given by~\cite{balanis1999advanced}
\begin{equation}
\label{RCS_sphere_equ}
     \sigma_{\rm{sphere}}^{Th} =\frac{\lambda^2}{4\pi}\Bigg|\sum_{n=1}^{\infty}\frac{(-1)^n(2n+1)}{\hat{H}_n^{(2)'}(ka){\hat{H}_n^{(2)}(ka)}}\Bigg|^2~,
\end{equation}
which can be approximated in two different regions as
\begin{equation}
   \sigma_{\rm{sphere}}^{Th} \approx\begin{cases}
    \frac{9\lambda^2}{4\pi}(ka)^6 & 2\pi a \ll \lambda~~(\text{Rayleigh region})\\
    \pi a^2 & a>2\lambda~~~~~(\text{Optical region})
    \end{cases}~,\\
\end{equation}
where $k=\frac{2\pi}{\lambda}$ is the wavenumber, $a$ is the radius of the sphere, ${\hat{H}_n^{(2)}(ka)}$ and ${\hat{H}_n^{(2)'}}$ are the spherical Hankel function of the second kind of order $n$ and its derivative, respectively.

From~(\ref{RCS_sphere_equ}), we can show that the backscattered RCS of a PEC sphere is a function of its circumference measured in wavelength ($\frac{2\pi a}{\lambda}$). In the Rayleigh region, the size of the PEC sphere is small relative to the wavelength of the signal transmitted by the radar ($\frac{2\pi a}{\lambda}\ll1$). On the other hand, in the optical region, the size of the PEC sphere is far larger than the wavelength of the sphere ($a>2\lambda$).  Moreover, between the Rayleigh and the optical regions is the Mie (or resonance) region where the radar wavelength is comparable to the size of the sphere ($\frac{2\pi a}{\lambda}\approx1$). In the Mie region, the RCS of the PEC sphere is dependent on frequency and continuously perturbed. However, in the optical region, the RCS of the PEC sphere is a constant that is independent of frequency. For this reason, we use standard PEC spheres with known theoretical RCS to calibrate the UAV RCS measurement.

Fig.~\ref{Fig:calibration_measurement} shows the measured and theoretical (ideal) RCS of the three PEC spheres considered in this study. As shown in this figure, the measured RCS varies slightly, while the ideal or theoretical RCS is a constant. For each calibration sphere, the ratio between the measured ($S_{\rm{RCS}}$) and theoretical RCS ($\sigma_{\rm{sphere}}^{Th}$) of the PEC sphere is used to calibrate the measured RCS of the target UAV according to:
\begin{equation}
   \sigma_{\rm{UAV}}=\frac{D_{\rm{RCS}}}{S_{\rm{RCS}}}\cdot \sigma_{\rm{sphere}}^{Th}.
\label{RCS_caliberation}
\end{equation}

After measuring the RCS, the obtained data can be used for further analysis such as UAV RCS statistical modeling and target classification. This will be discussed in the next section.

\section {RCS Statistical Analysis and UAV Classification}\label{statistical_analysis}
In this section, we describe the UAV statistical recognition system and the RCS statistical model selection techniques employed in this study.

\subsection{RCS-based UAV Classification}
 Complex radar targets can be modeled as consisting of large numbers of individual scattering centers that are randomly distributed. If the individual scatterers are assumed to have the same RCS, then the echo power from the target object, target RCS returns, can be modeled using an exponential distribution and ground clutter reflections can be modeled using a Weibull distribution~\cite{richards2010principles}. Using these statistical models,~\cite{MartinsDrone} developed a likelihood ratio test for a single UAV target detection in the presence of ground clutter interference.

 However, in practice, many complex radar targets cannot be modeled as an ensemble of equal-strength scatterers. Besides, for such targets, the radar echo power varies strongly with the aspect angle, frequency, and polarization of the transmitter and receiver of the radar. For such targets, the best RCS statistical model can only be obtained by empirically fitting the measured RCS data to a range of possible distributions. This is particularly helpful when we want to discriminate between multiple targets. That is, if we know the statistical models that best describe different UAV types, we can use this knowledge as the basis for target classification (or target discrimination). In such scenarios, the Bayesian-based maximum aposteriori probability (MAP) decision rule is optimal for target classification~\cite{du2008radar,hou2010new}.

 Given a test RCS data  $\boldsymbol{y}=(y_1,\cdots,y_n)$ that has been recorded from an unknown UAV, its RCS follows a specific parametric distribution. \textcolor{black}{Suppose, we have $M$ possible UAV classes, then the UAV classification problem becomes an M-ary Bayesian hypothesis testing problem,
with the corresponding hypothesis $H_1, H_2, \cdots H_M$}. The likelihood that the test RCS data $\boldsymbol{y}$ belongs to the $z^\text{th}$ UAV class (i.e., the likelihood that the $H_z$ hypothesis is correct) is given by the posterior probability $P({C}=z|\boldsymbol{y})$. According to Bayes theory, $P({C}=z|\boldsymbol{y})$ is given by:
\begin{equation}
 P({C}=z|\boldsymbol{y})= \frac{P(\boldsymbol{y}| {C}=z) P({C}=z)}{\sum_{k=1}^{M} P(\boldsymbol{y}| {C}=z) P({C}=z)}~,
\end{equation}
where $P({C}=z)$ is the prior probability of the $z$th class and $\sum_{z=1}^{M} P(\boldsymbol{y}| {C}=z) P({C}=z) = P(\boldsymbol{y})$ is the evidence. Therefore, given the RCS data obtained from an unknown UAV, the RCS-based UAV statistical recognition system determines the class membership of the UAV by computing the posterior probabilities of all class membership $P({C}=z), z=1,\cdots M$. Therefore, the MAP decision rule for the UAV classification problem can be written as:
\begin{align}
\label{decision_rule_1}
\widehat{C}= \arg\max_C {\ln P({C}=z|\boldsymbol{y})}.
\end{align}

In practice, since the evidence is not a function of $C$, it can be ignored. Besides, if we have equal number of RCS test datasets for each UAV class, then we can assume all the classes are equi-probable. That is, $P(C)=\frac{1}{M}$, and the decision rule in (\ref{decision_rule_1}) becomes:
\begin{align}
\label{decision_rule_2}
   \widehat{C}= \arg\max_C {\ln P(\boldsymbol{y}| {C=z})}.
\end{align}


For example, suppose the $z$th UAV class is described by a gamma distribution model $P(\boldsymbol{y}| {C=z}) = P(\boldsymbol{y}|\beta,\gamma)$ with scale parameter $\beta$ and shape parameter $\gamma$ that can be estimated by fitting the training data to the statistical model of the $z$th UAV class~\cite{minka2002estimating}. The training data $\boldsymbol{\sigma}=(\sigma_1,\cdots,\sigma_n)$ is obtained from measuring the RCS of the $z$th UAV in the compact range anechoic chamber as described in Section~\ref{measurement_description}. The model parameters $\beta$ and $\gamma$ can be estimated using the maximum likelihood estimation (MLE) technique. The MLE technique is tractable if we assume the components of the training data $\boldsymbol{\sigma}$ are independent~\cite{anderson2004model, du2008radar,hou2010new}, then the $z$th UAV class is described by the class model:
\begin{eqnarray}
\begin{aligned}
  \ln P(\boldsymbol{\sigma}|\beta,\gamma) &= \prod_{i=1}^{n} \ln P(\sigma_i|\beta,\gamma)\\
  &=\sum_{i=1}^{n}\ln \frac{1}{\beta^\gamma\Gamma(\gamma)}\sigma_{i}^{\gamma-1} e^{\frac{-\sigma_{i}}{\beta}}\\
  &= -n\gamma\ln\beta-n\ln{\Gamma(\gamma})+\\
  &(\gamma-1)\sum_{i=1}^{n}\ln(n)-\frac{1}{\beta} \sum_{i=1}^{n}\sigma_i.
\end{aligned}
\end{eqnarray}

Using the MLE, we solve the differential equations $\frac{\partial \ln P(\boldsymbol{\sigma}|\beta,\gamma)}{\partial \beta}=0$ and $\frac{\partial \ln P(\boldsymbol{\sigma}|\beta,\gamma)}{\partial \gamma}=0$, and we obtain
\begin{align}
\hat{\beta}&=\frac{\bar{\sigma}}{\gamma}~,\label{beta_equation}\\
 \ln{\hat{\beta}}+ \psi(\hat{\gamma})&= \frac{1}{n}\sum_{i=1}^n \ln\sigma_i~, \label{beta_expression}
\end{align}
where $\bar{\sigma}$ is the sample mean of the training data, $\hat{\beta}$ and $\hat{\gamma}$ are the MLE parameter estimate of the $z$th UAV class, and $\psi(\hat{\gamma})=\frac{\partial   ln\Gamma(\hat{\gamma})}{\partial \hat{\gamma}}$ is the diagamma function. Asymptotically, $\psi(\hat{\gamma})$ can be expanded as a function of either the Riemann zeta function $\zeta$ or Bernoulli number $B$ as follows:
\begin{eqnarray}
\begin{aligned}
\psi(\hat{\gamma}) & \approx \ln{\hat{\gamma}}-\frac{1}{2\hat{\gamma}}+\sum_{g=1}^\infty	 \frac{\zeta (1-2g)}{\hat{\gamma}^{2g}}~,\\
&=\ln{\hat{\gamma}}-\frac{1}{2\hat{\gamma}}-\sum_{g=1}^\infty	 \frac{B_{2g}}{2g\hat{\gamma}^{2g}}.
\label{diagmma_expansion}
\end{aligned}
\end{eqnarray}

Approximating $\psi(\hat{\gamma})$ using the first two terms in the expansion given in (\ref{diagmma_expansion}) and substituting (\ref{beta_equation}) in (\ref{beta_expression}) we obtain a quadratic equation whose positive root is $\hat{\gamma}$. After estimating $\hat{\beta}$ and $\hat{\gamma}$ from the training data $\boldsymbol{\sigma}$, we obtain the model $P(\boldsymbol{\sigma}|\hat{\beta},\hat{\gamma})$ for the $z$th UAV.
Therefore, given a test RCS data $\boldsymbol{y}$ from an unknown UAV, the log-likelihood that the data is a $z$th UAV is given by $\ln P(\boldsymbol{y}|\hat{\beta},\hat{\gamma})$. Therefore, if we can estimate the statistical model for all the $M$ UAV classes, we can make the UAV classification decision using (\ref{decision_rule_2}).

\begin{figure}[t]
\center{
\begin{subfigure}[]{\includegraphics[width=0.15\linewidth]{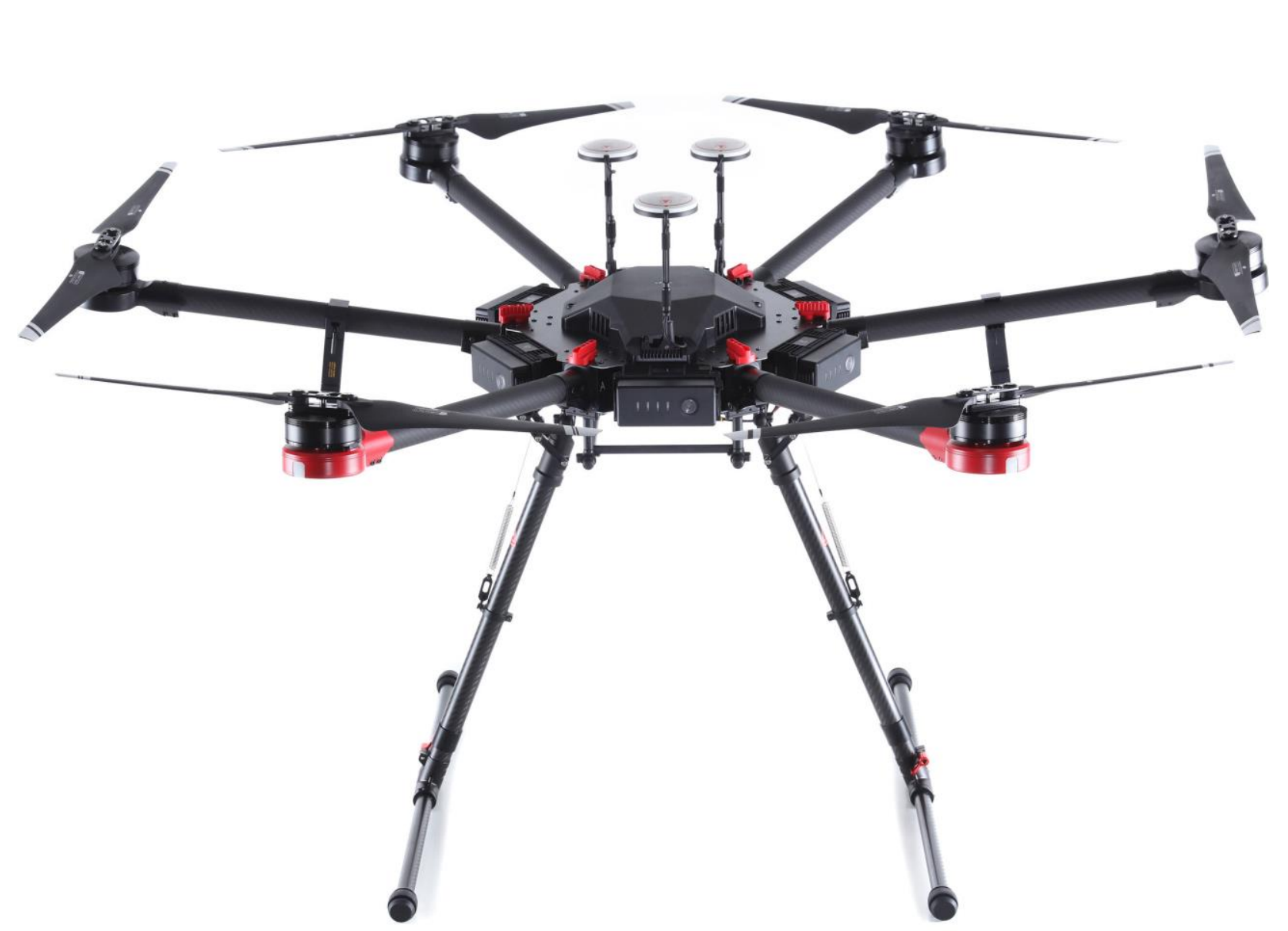}\label{DJI_M600_Pro}}
\end{subfigure}
\begin{subfigure}[]{\includegraphics[width=0.15\linewidth]{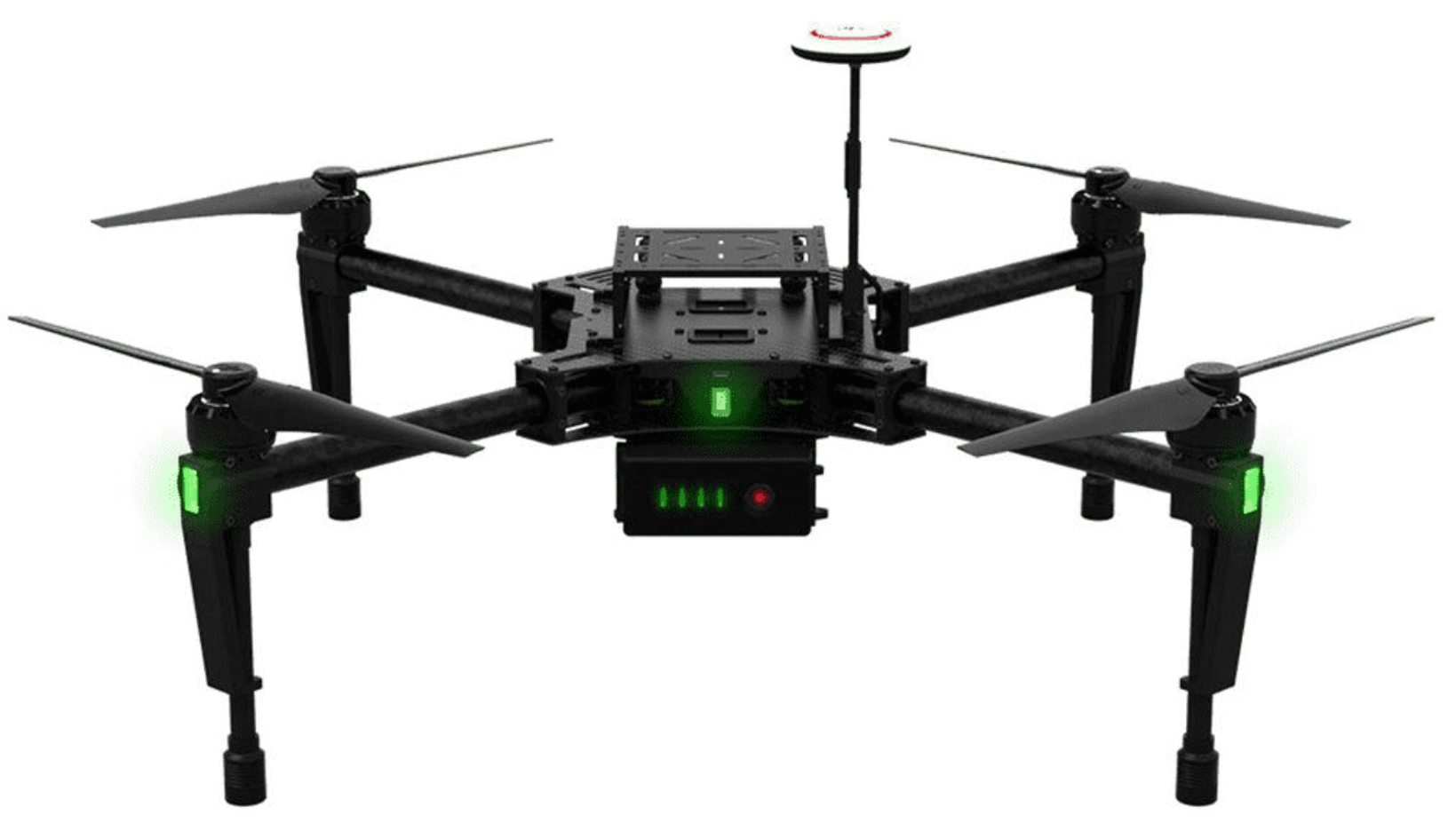}\label{dji_matrice-100}}
\end{subfigure}
\begin{subfigure}[]{\includegraphics[width=0.15\linewidth]{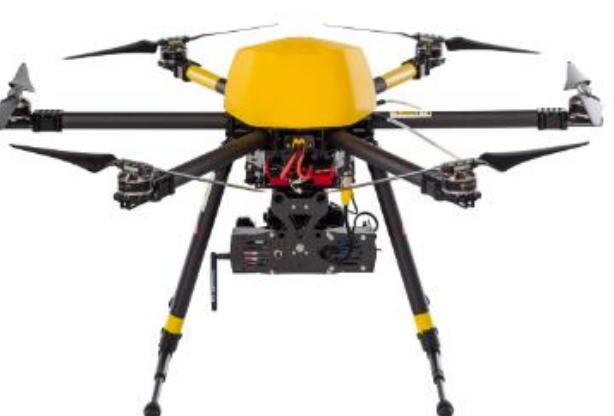}\label{trimble_UAV}}
\end{subfigure}
\begin{subfigure}[]{\includegraphics[width=0.15\linewidth]{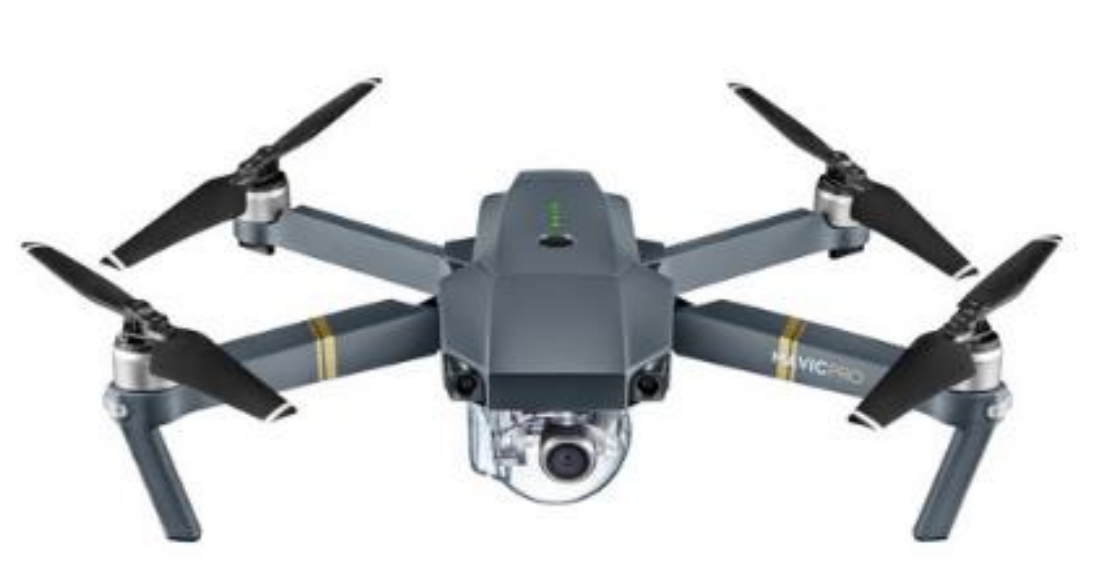}\label{dji_mavic_pro_1}}
\end{subfigure}
\begin{subfigure}[]{\includegraphics[width=0.15\linewidth]{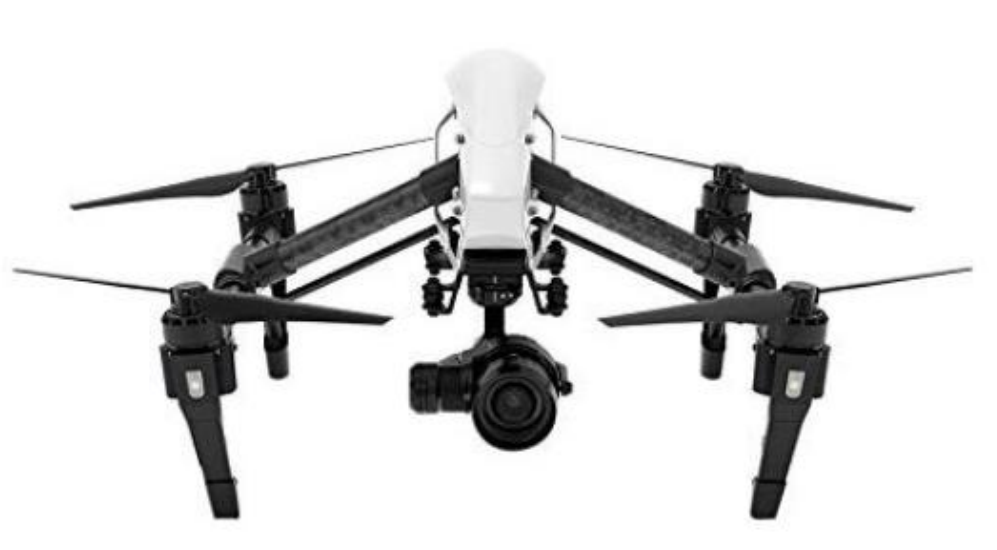}\label{DJI_inspire}}
\end{subfigure}
\begin{subfigure}[]{\includegraphics[width=0.15\linewidth]{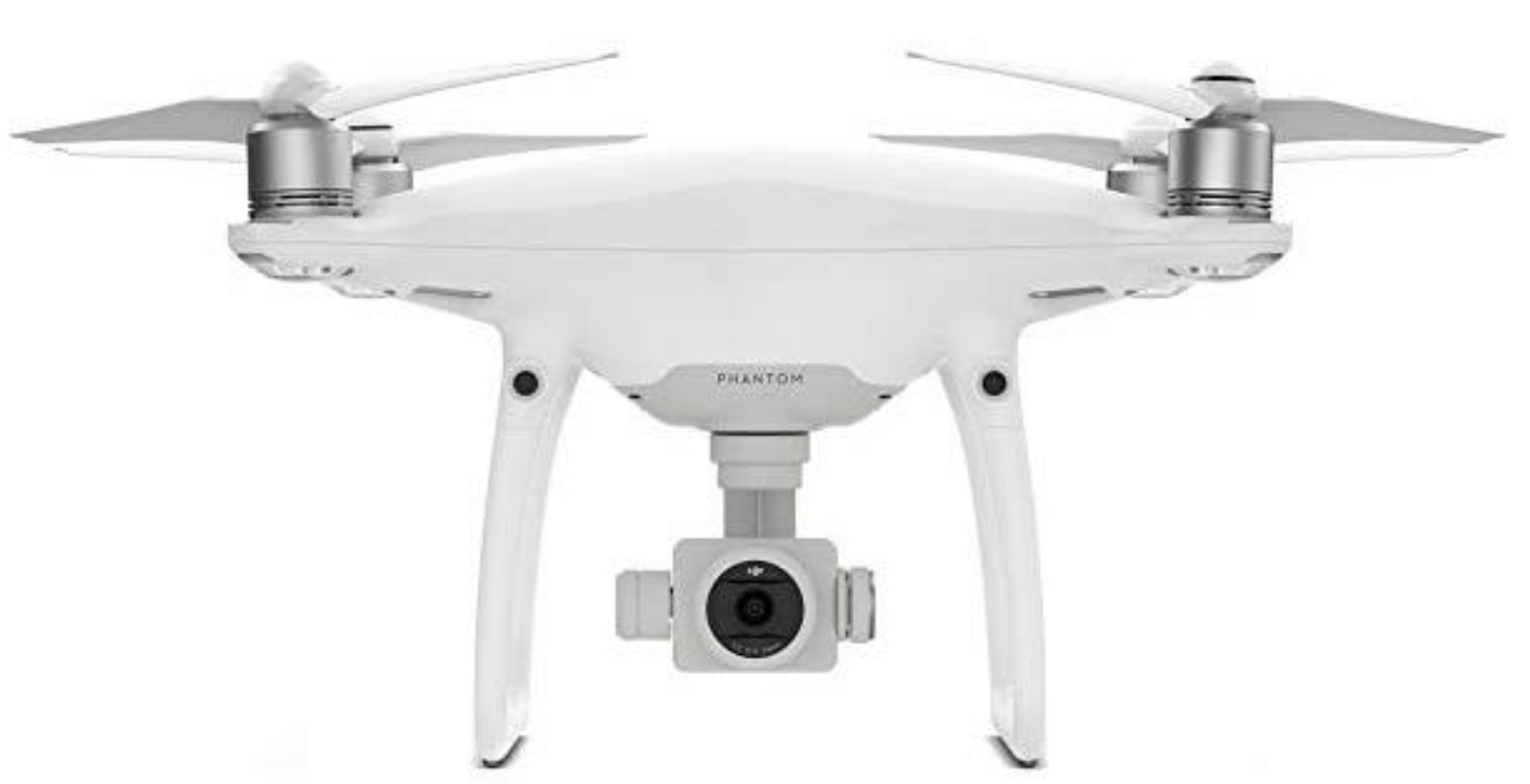}\label{DJI_phantom4pro}}
\end{subfigure}\vspace{-2mm}
 \caption{{Six small UAVs considered: (a) DJI Matrice 600 Pro, (b) DJI Matrice 100, (c) Trimble zx5, (d) DJI Mavic Pro 1, (e) DJI Inspire 1 Pro, (f) DJI Phantom 4 Pro.}}
 \label{SMALL_UAVs}}
 \end{figure}

\subsection{UAV Statistical Model Selection}

While the gamma distribution has been used as an example in the previous subsection to model RCS statistics, it may not be the best parametric model for some UAV classes. If/when that is the case, the chances of misclassification may increase. Therefore, there is a need to select the best statistical model for each UAV type using the RCS training data for each UAV class.
Since the true distribution of the RCS of a given UAV class is unknown, we can only approximate it. To do this, we can define a set of candidate statistical models  $\{P(y|\boldsymbol{\theta})\}$ and develop a statistical criterion for selecting the relatively best statistical model for each UAV type given the RCS training (measurement) data.

\textcolor{black}{In this study, two statistical model selection techniques will be investigated. The first is the Akaike information criterion (AIC) and the second is the Bayesian information criterion (BIC)~\cite{anderson2004model}. For any candidate parametric model, the AIC and BIC are respectively computed from the training data $\boldsymbol{y}=(y_1,\cdots,y_n)$ of the $z^\text{th}$ UAV as}:
\begin{align}
   {\rm AIC(\boldsymbol{y})}&=-2\ln P(\boldsymbol{y}|\boldsymbol{\hat{\theta}}) + 2k~,   \label{AIC}\\
  {\rm BIC(\boldsymbol{y})}&= -2\ln P(\boldsymbol{y}|\boldsymbol{\hat{\theta}}) + k\ln{n}~,   \label{BIC}
\end{align}
\textcolor{black}{where $\theta$ is the parameter of the model, $k$ is the number of parameters in the statistical model and $n$ is the sample size in the training dataset. We apply the AIC and BIC criteria to the training dataset of all the UAVs. The best parametric model for each UAV is stored in the database.}

In general, the best model is the one with the smaller AIC or BIC score. It is pertinent to select the RCS statistical models that do not overfit the training data. Overfitting can be achieved by penalizing the complexity of the models. Models with larger $k$ do well in fitting the data; however, there is a trade-off in the increasing variance~\cite{anderson2004model}. For AIC and BIC, the penalty terms are $2k$ and $k\ln{n}$, respectively. Unlike BIC, AIC is asymptotically efficient. On the other hand, BIC is a more consistent estimator since it has a larger penalty term ($k\ln{n}$). The next section provides the results of measurement, statistical model analysis, and classification.

\begin{table*}[t!]
\setlength{\tabcolsep}{0.5pt} 
\centering
\vspace{-2mm}
\caption{Mean, \text{Std}, and AIC score for \textbf{VV}-polarized RCS data. The models are 1: Log-normal, 2: Generalized extreme value, 3: Gamma, 4: Beta, 5: Generalized Pareto,  6: Weibull, 7: Nakagami, 8: Rayleigh, 9: Rician, 10: Exponential, 11: Normal}
\label{AIC_table_VV}
\begin{tabular}{|c|c|c|c|c|c|c|c|c|c|c|c|c|c|c|}
\hline
Freq & UAV & $\mu$ & Std & \multicolumn{6}{c}{AIC Test Score}\\
\hline
(GHz) & & (dBsm) & (dBsm) & 1 & 2 & 3&  4& 5& 6& 7& 8 & 9 & 10 & 11  \\
\hline
\multirow{6}{*}{\text{15}}
& DJI Matrice 600 &  -11.67 & 1.81 & -773.42 & -770.55 &  \cellcolor{blue!25}-773.45 & -772.72 &-752.08 & -758.20 & -765.92 & -742.09 &  -753.91 & -578.07 & -745.81 \\
&DJI Matrice 100 & -14.69 &  1.69 & \cellcolor{blue!25}-1048.24 & -1047.22 & -1044.92 & -1044.22 & -996.45 & -1021.41 & -1032.95 &-1002.63 & -1017.31 & -833.01 & -1009.84\\
& Trimble zx5& -14.39 & 2.57 & -872.7 & \cellcolor{blue!25}-881.88 &  -842.63 & -838.41 &  -844.95 & -817.41 & -801.34 & -790.48 & -788.48 & -765.57
 & -723.83\\
&  DJI Mavic Pro & -17.06 & 1.51  &  \cellcolor{blue!25}-1287.91 &  -1286.23  & -1287.48 & -1287.33 & -1239.65 &  -1269.53 & -1281.23 & -1225.12 & -1269.71 & -1036.63 & -1266.59\\
& Inspire 1 & -14.24 & 1.56 & \cellcolor{blue!25}-1041.35
 & -1039.38 & -1037.39 & -1036.68 &  -1001.06 &  -1012.32 & -1026.45
 & -981.03 &  -1011.12 & -799.62 &  -1006.31 \\
& DJI Phantom 4 Pro & -15.02 & 1.21 &  -1198.29 & -1198.28 & \cellcolor{blue!25}-1200.24 & -1200.22 & -1120.62 & -1185.36 &-1198.11 & -1085.95 & -1191.81 & -874.72 & -1190.95\\
\hline
\multirow{6}{*}{\text{25}}
&DJI  Matrice 600 & -7.32  &  2.09 & -358.67 & -362.48 & \cellcolor{blue!25}-363.42 & -359.88 & -339.62 & -351.59 & -355.82 & -348.23 &  -347.22 & -207.04 & -331.25 \\
& DJI Matrice 100 &  -11.03 & 2.27 & -637.78 & \cellcolor{blue!25}-639.49 & -623.88 &  -617.68 & -581.15  & -597.57 & -594.91 & -596.91 &  -594.91 & -504.55 & -542.29\\
& Trimble zx5& -9.64 & 2.80&  -445.77 & \cellcolor{blue!25}-473.69 & -392.17 & -350.92 & -423.31 & -367.34 & -326.72  & -268.9 & -266.9 & -346.77 & -208.26 \\
& DJI Mavic Pro & -16.20 &  2.30 &  -1062.91 & \cellcolor{blue!25}-1062.93 & -1049.06 & -1047.52 &  -1021.02 & -1023.99 & -1020.80  & -1022.74 & -1020.74 & -933.78 & -967.88\\
& DJI Inspire 1 &  -11.09 & 2.62 & -589.96 &  \cellcolor{blue!25}-603.86 & -560.19 & -551.33 & -595.7 & -537.58 & -522.88 & -510.72 & -508.72 & -487.83 & -447.18\\
& DJI Phantom 4 Pro & -12.40 & 1.93 & -811.02 & \cellcolor{blue!25}-812.43 & -796.38 & -792.96 & -761.76 & -764.74 & -771.08 & -765.48 & -763.48 & -632.21 & -730.08 \\
\hline
\end{tabular}
\end{table*}

\begin{table*}[t!]
\setlength{\tabcolsep}{0.5pt} 
\centering
\vspace{-2mm}
\caption{Mean, \text{Std}, and BIC score for \textbf{VV}-polarized RCS data. The models are 1: Log-normal, 2: Generalized extreme value, 3: Gamma, 4: Beta, 5: Generalized Pareto,  6: Weibull, 7: Nakagami, 8: Rayleigh, 9: Rician, 10: Exponential, 11: Normal}
\label{BIC_table_VV}
\begin{tabular}{|c|c|c|c|c|c|c|c|c|c|c|c|c|c|c|}
\hline
Freq & UAV & $\mu$ & Std & \multicolumn{2}{c}{} & \multicolumn{6}{c}{BIC Test Score}\\
\hline
(GHz) & & (dBsm) & (dBsm) & 1 & 2 & 3&  4& 5& 6& 7& 8 & 9&10&11  \\
\hline
\multirow{6}{*}{\text{15}}
& DJI Matrice 600 &-11.67 & 1.81 & -767.02 & -760.96 & \cellcolor{blue!25}-767.05 & -766.33 &  -742.48 & -751.81 & -759.51 & -738.89 & -747.52 & -574.87 & -739.5\\
&  DJI Matrice 100 &-14.69& 1.69 & \cellcolor{blue!25}-1041.84& -1037.63 & -1038.53 & -1037.83 &  -986.85 & -1015.01 & -1026.55 & -999.43 & -1010.92 & -829.8 & -1003.44\\
& Trimble zx5& -14.39 & 2.57 & -866.3  & \cellcolor{blue!25}-872.29 & -836.24 & -832.01 & -835.36 & -811.01 & -794.94 & -787.28 & -782.09 & -762.37 &  -717.43\\
& DJI Mavic Pro & -17.06 & 1.51 & \cellcolor{blue!25}-1281.51 & -1276.63 & -1281.08 & -1280.94 & -1230.05 & -1263.13 & -1274.83 & -1221.92 & -1263.31 & -1033.43 & -1260.19\\
& DJI Inspire 1 & -14.24 & 1.56 & \cellcolor{blue!25}-1034.95 & -1029.78 & -1030.99 & -1030.28 & -991.46 & -1005.92 & -1020.05 & -977.83 & -1004.72 & -796.42 &  -999.91\\
& DJI Phantom 4 Pro & -15.02   & 1.21 & -1191.90 & -1188.69 & \cellcolor{blue!25}-1193.84 & -1193.83 & -1111.03 & -1178.97 & -1191.71  & -1082.75 & -1185.41 & -871.52
 & -1184.55 \\
\hline
\multirow{6}{*}{\text{25}}
& DJI Matrice 600 &-7.32 &2.09  & -352.27 & -352.89 & \cellcolor{blue!25} -357.02 &  -353.48 & -330.02 & -345.19 & -349.42 &  -345.03 & -340.82 & -203.84 & -324.86\\
& DJI Matrice 100 &-11.03 & 2.27 &  \cellcolor{blue!25}-631.38
 & -629.9 & -617.48 & -611.28 & -571.56 & -591.17 & -588.52 & -593.71 & -588.52 &-501.36 & -535.89\\
& Trimble zx5& -9.64  & 2.80 & -439.37 &  \cellcolor{blue!25}-464.10 & -385.77 & -344.53 &  -413.71 & -360.94 & -320.32 &  -265.70 & -260.50 & -343.57 &  -201.86\\
& DJI Mavic Pro &-16.20  & 2.30  & \cellcolor{blue!25}-1056.51 & -1053.34 & -1042.66 & -1041.13 & -1011.42 & -1017.59 & -1014.40 & -1019.54 & -1014.35 & -930.58 &  -961.48\\
& DJI Inspire 1 &-11.09 & 2.62 & -583.56 & \cellcolor{blue!25}-594.26 &  -553.79 & -544.93  & -586.10 & -531.19 & -516.48 & -507.52 & -502.33 &  -484.63 &  -440.78 \\
& DJI Phantom 4 Pro & -12.40 & 1.93  & \cellcolor{blue!25}-804.62 & -802.84
 & -789.98 & -786.57 & -752.16 & -758.34 & -764.69 & -762.28 & -757.08 & -629.01 & -723.68 \\
\hline
\end{tabular}
\end{table*}

\begin{table*}[t!]
\setlength{\tabcolsep}{0.5pt} 
\centering
\vspace{-2mm}
\caption{Mean, \text{Std}, and AIC score for \textbf{HH}-polarized RCS data. The models are 1: Log-normal, 2: Generalized extreme value, 3: Gamma, 4: Beta, 5: Generalized Pareto,  6: Weibull, 7: Nakagami, 8: Rayleigh, 9: Rician, 10: Exponential, 11: Normal}
\label{AIC_table_HH}
\begin{tabular}{|c|c|c|c|c|c|c|c|c|c|c|c|c|c|c|}
\hline
Freq & UAV & $\mu$ & Std & \multicolumn{6}{c}{AIC Test Score}\\
\hline
(GHz) & & (dBsm) & (dBsm) & 1 & 2 & 3&  4& 5& 6& 7& 8 & 9 & 10 & 11  \\
\hline
\multirow{6}{*}{\text{15}}
& DJI Matrice 600 &  -12.71 &  2.33   & \cellcolor{blue!25}-768.79 & -764.01 & -762.55 & -761.04 & -753.34 &  -747.13 & -747.22 & -748.94 & -746.94 & -644.67 & -707.83 \\
& DJI Matrice 100 &-14.73 & 2.25 & -948.58 & -948.27 &  -955.08 & \cellcolor{blue!25}-955.36 & -949.13 & -952.00 & -953.29 & -950.33 & -949.80 & -818.86 & -932.27\\
& Trimble zx5& -14.06  & 2.61  & \cellcolor{blue!25}-838.96 & -836.98 & -823.41 & -819.56 & -815.28 & -800.96 & -788.01 & -781.81 & -779.81 &  -740.84 & -718.32 \\
& DJI Mavic Pro & -17.29 & 1.89 & \cellcolor{blue!25}-1226.32 & -1223.21 & -1225.69 & -1225.48 & -1206.99 &-1210.14 & -1216.91 & -1199.86 & -1205.10 &  -1043.91& -1194.16 \\
& DJI Inspire 1 & -14.43 &  2.08 & \cellcolor{blue!25}-953.04 & -951.24 & -941.03 & -939.26 & -928.11 & -915.11 & -918.18 &  -917.00 & 85.00 & -915.00 & -876.39\\
& DJI Phantom 4  Pro&  -14.87 &   1.78   & \cellcolor{blue!25}-1045.89 & -1044.27 & -1037.72 & -1036.84 & -1024.83 & -1013.54 & -1022.86 & -1003.06 & -1006.83 & -844.72 &  -995.17\\
\hline
\multirow{6}{*}{\text{25}}
& DJI Matrice 600 &   -7.07 & 2.25 & \cellcolor{blue!25}-309.72 & -305.35 & -301.24 & -279.44 &  -304.11 &  -279.67 & -279.61 & -281.16 & -279.16 & -176.40 & -235.27 \\
& DJI Matrice 100 & -9.72 & 1.83 & \cellcolor{blue!25}-607.39 & -604.04 & -605.54 & -604.11 &  -574.07 & -590.37 & -597.17 &  -576.1 & -584.89 &  -415.20 & -575.73\\
& Trimble zx5&  -9.69 &  2.83 & \cellcolor{blue!25}-445.56 &  -444.75 & -425.68 &  -411.95 &  -413.99 & -407.86
 & -392.48 & -374.65 & -372.65 & -363.97 &  -315.90\\
& DJI Mavic Pro &  -17.22 & 2.68 & -1092.89 & -1115.53 & -1115.95 & -1116.46 & -1101.2884 & -1120.76 & -1120.81 & \cellcolor{blue!25}-1122.62 & -1120.64 & -1013.82 & -1101.63\\
& DJI Inspire 1 & -12.05 &  3.42 & -574.30 & \cellcolor{blue!25}-587.87 & -535.46 & -525.55 & -574.49 & -524.39 & -494.05 & -403.44 & -401.44& -516.17 &  -374.14\\
& DJI Phantom 4 Pro & -12.24 &  1.69  & \cellcolor{blue!25}-845.30 & -845.08 &  -836.12 & -834.46 & -823.96 & -810.64 & -821.65 & -794.43 & -804.60 & -628.07 & -795.78\\
\hline
\end{tabular}
\end{table*}

\begin{table*}[t!]
\setlength{\tabcolsep}{0.5pt} 
\centering
\vspace{-2mm}
\caption{Mean, \text{Std}, and BIC score for \textbf{HH}-polarized RCS data. The models are 1: Log-normal, 2: Generalized extreme value, 3: Gamma, 4: Beta, 5: Generalized Pareto,  6: Weibull, 7: Nakagami, 8: Rayleigh, 9: Rician, 10: Exponential, 11: Normal}
\label{BIC_table_HH}
\begin{tabular}{|c|c|c|c|c|c|c|c|c|c|c|c|c|c|c|}
\hline
Freq & UAV & $\mu$ & Std & \multicolumn{2}{c}{} & \multicolumn{6}{c}{BIC Test Score}\\
\hline
(GHz) & & (dBsm) & (dBsm) & 1 & 2 & 3&  4& 5& 6& 7& 8 & 9&10&11  \\
\hline
\multirow{6}{*}{\text{15}}
& DJI Matrice 600 & -12.71 & 2.33 & \cellcolor{blue!25}-762.39 & -754.4 & -756.15 & -754.64 & -743.75 & -740.74 & -740.82 & -745.74 & -740.54 & -641.47 &  -701.43 \\
& DJI Matrice 100 & -14.73 & 2.25 & -942.19 & -938.68 & -948.68 & \cellcolor{blue!25}-948.96 &  -939.53 & -945.60 & -946.89 &  -947.13 & -943.41 &  -815.66  & -925.87 \\
& Trimble zx5& -14.06 & 2.61 &  \cellcolor{blue!25}-832.56 & -827.38 & -817.02 & -813.16 &  -805.68 & -794.56 & -781.6 &  -778.61 &  -773.41 & -737.64 &  -711.92 \\
& DJI Mavic Pro & -17.29 & 1.89  & \cellcolor{blue!25}-1219.92 & -1213.61 & -1219.29 & -1219.09 & -1197.39 & -1203.75 & -1210.52 & -1196.66 & -1198.70 & -140.71 & -1187.77\\
& DJI Inspire 1 & -14.43& 2.08  &  \cellcolor{blue!25}-946.64 & -941.65 &  -934.63 & -932.87 & -918.51 & -908.71 & -911.78 & -911.78 & -908.60 &  -793.20 & -870.00\\
& DJI Phantom 4 Pro & -14.87 & 1.78& \cellcolor{blue!25}-1039.50 & -1034.68 & -1031.32 & -1030.44 & -1015.24 & -1007.14 & -1016.46 & -999.86 & -1000.43 & -841.52 & -988.78 \\
\hline
\multirow{6}{*}{\text{25}}
& DJI Matrice 600 & -7.07  & 2.25  & \cellcolor{blue!25}-303.33 & -295.76 & -294.84 & -273.05 & -294.51 & -273.27 &  -273.21 & -277.97 & -272.77 & -173.20 & -228.87 \\
& DJI Matrice 100 & -9.72  & 1.83  & \cellcolor{blue!25}-600.99 & -594.44 & -599.15 & -597.72 & -564.47
 & -583.97 &  -590.78 & -572.90 & -578.49 & -412.01 & -569.34 \\
& Trimble zx5 & -9.69 & 2.83  & \cellcolor{blue!25}-439.16 & -435.15 & -419.28 &  -405.55 & -404.39 & -401.47 & -386.08 & -371.45 &  -366.26 & -360.77 & -309.5\\
& DJI Mavic Pro & -17.22 & 2.68  & -1086.49 & -1105.94 & -1109.55 & -1110.07 & -1091.69 & -1114.37 & -1114.42 & \cellcolor{blue!25}-1119.42 & -1114.24 & -1010.62 & -1095.23\\
& DJI Inspire 1 & -12.05  & 3.42  & -567.91 & \cellcolor{blue!25}-578.28 & -529.06 & -519.16 & -564.89 & -518.00 &  -487.65 & -400.24 & -395.04 & -512.97 & -367.74\\
& DJI Phantom 4 Pro & -12.24& 1.69 & \cellcolor{blue!25}-838.9 & -835.49 & -829.73 & -828.07 & -814.36 & -804.25 & -815.26 & -791.23 & -798.21 & -624.87 & -789.38 \\
\hline
\end{tabular}
\end{table*}
\begin{figure*}{}
\center{
\begin{subfigure}[]{\includegraphics[width=0.3\linewidth]{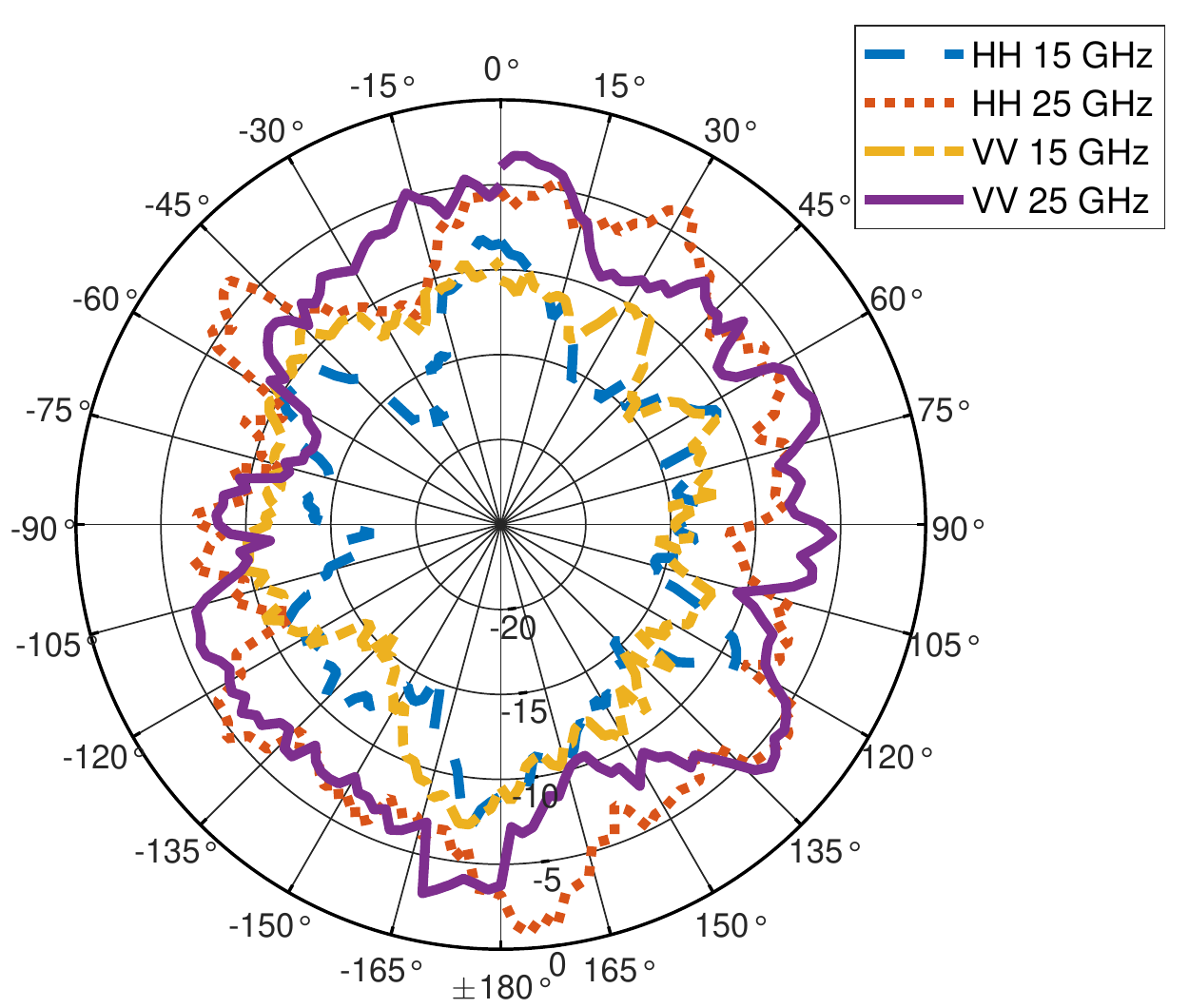}\label{RCS_M600}}
\end{subfigure}
\begin{subfigure}[]{\includegraphics[width=0.3\linewidth]{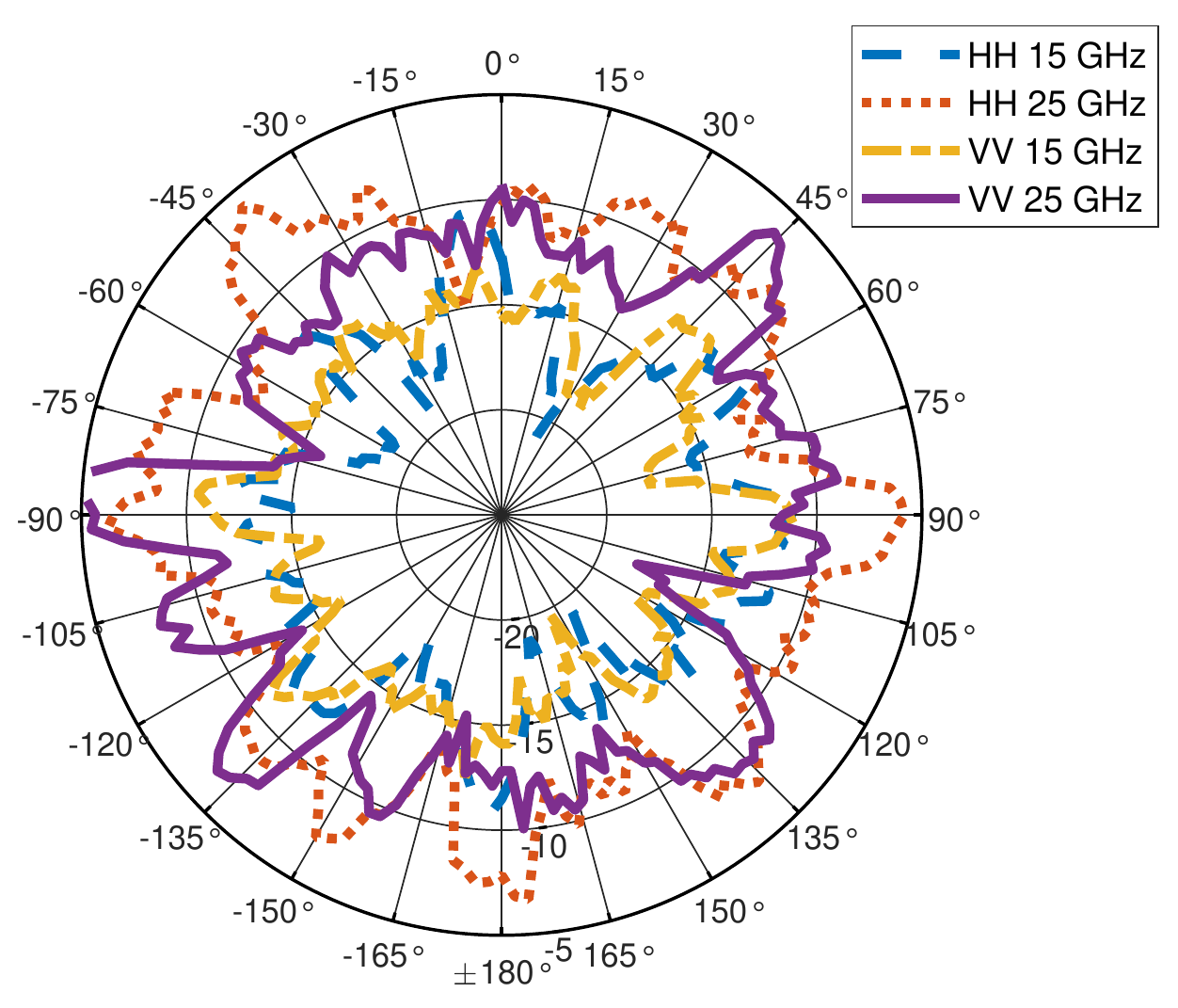}\label{RCS_DJI_M100}}
\end{subfigure}
\begin{subfigure}[]{\includegraphics[width=0.3\linewidth]{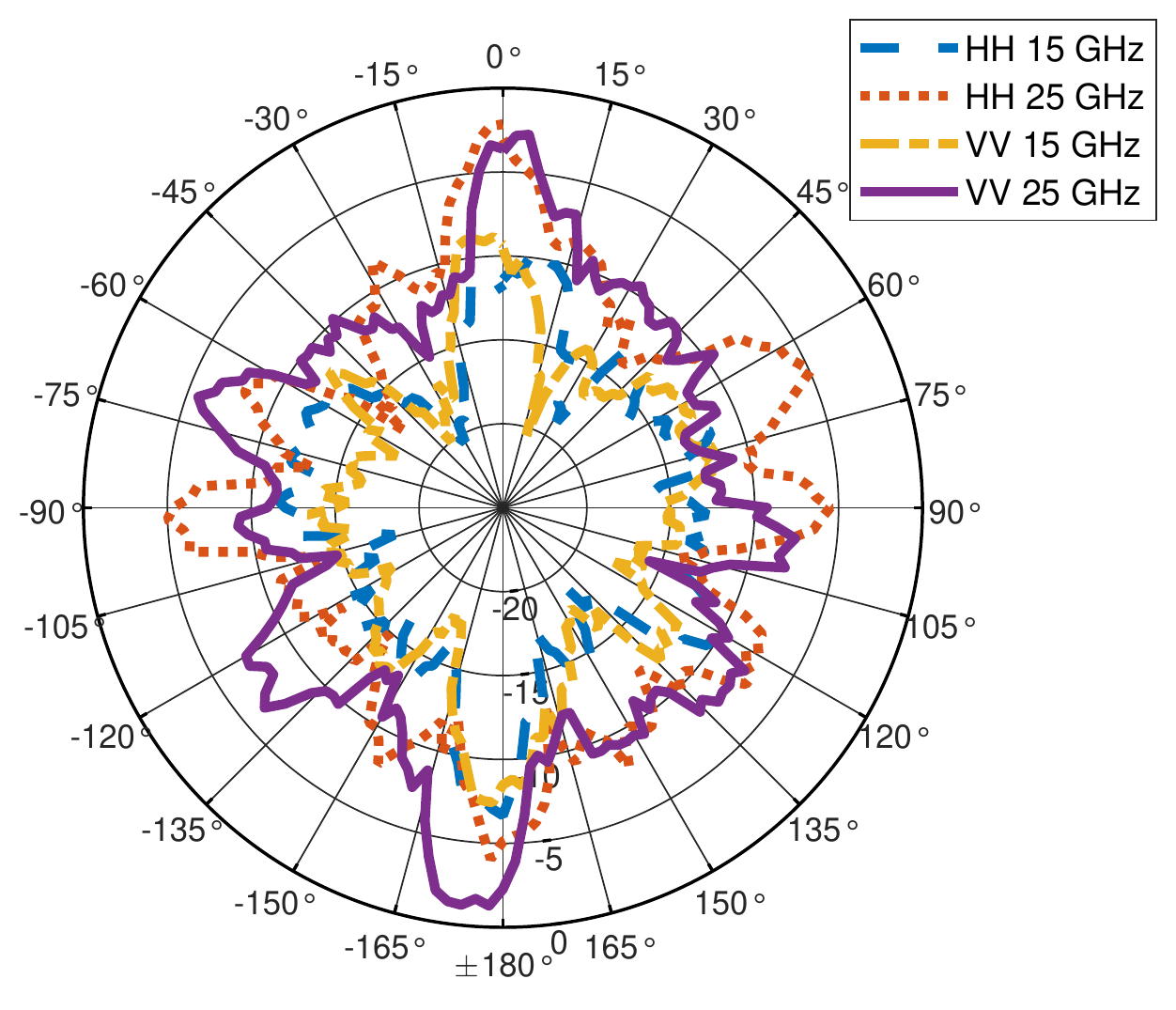}\label{RCS_Trimble}}
\end{subfigure}
\begin{subfigure}[]{\includegraphics[width=0.3\linewidth]{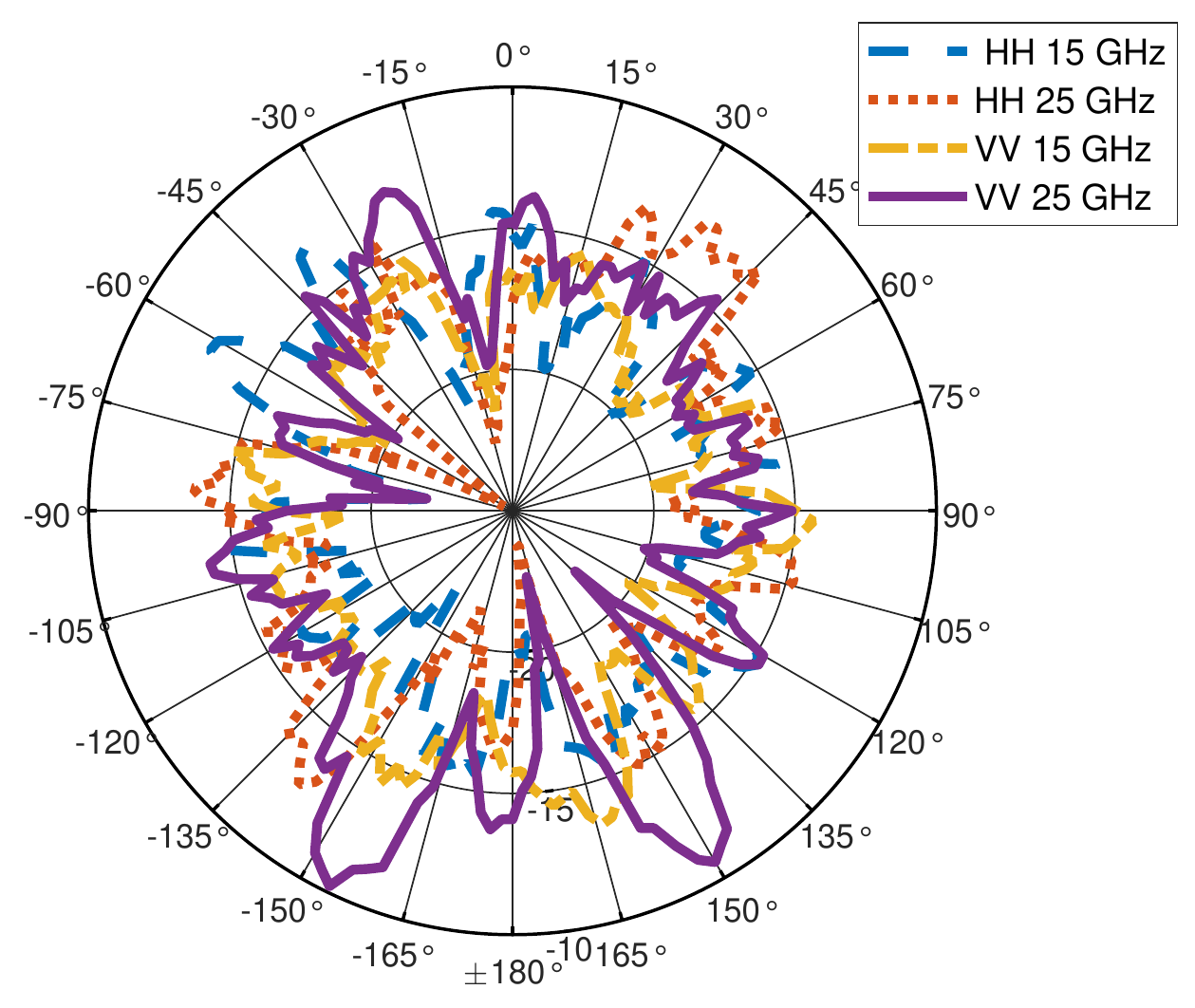}\label{RCS_MAVICPRO}}
\end{subfigure}
\begin{subfigure}[]{\includegraphics[width=0.3\linewidth]{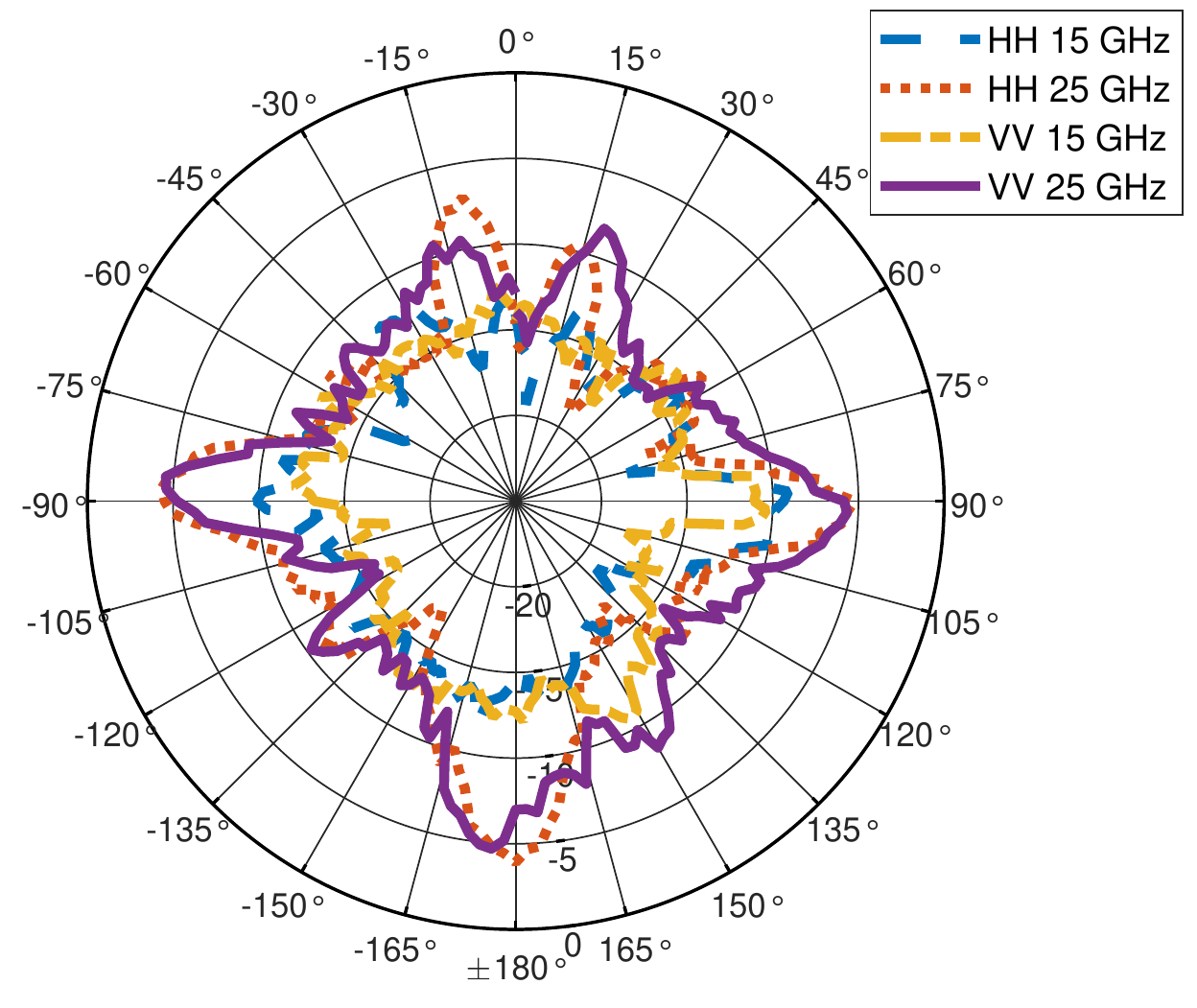}\label{RCS_INSPIRE}}
\end{subfigure}
\begin{subfigure}[]{\includegraphics[width=0.3\linewidth]{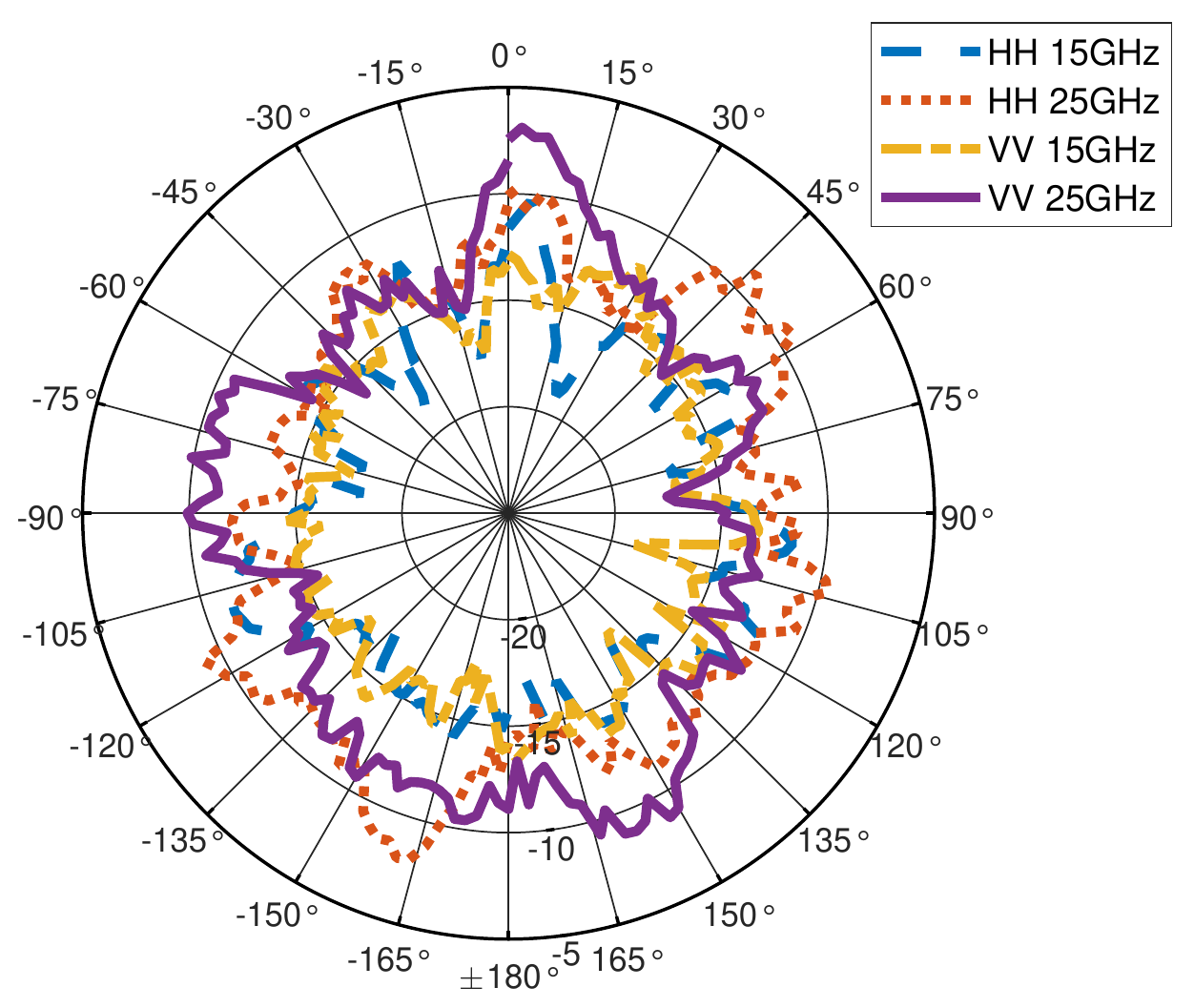}\label{RCS_PHANTOM}}
\end{subfigure}\vspace{-2mm}
\caption{{The measured RCS (dBsm) versus azimuth angles ($\phi\in[0\degree, 360\degree]$) for the small UAVs: (a) DJI Matrice 600 Pro, (b) DJI Matrice 100, (c) Trimble zx5, (d) DJI Mavic Pro 1, (e) DJI Inspire 1 Pro, (f) DJI Phantom 4 Pro.}}
 \label{RCS_POLAR_PLOT_new}}
 \vspace{-4mm}
 \end{figure*}

\section{Measurement and Numerical Results}\label{result_DISCUSSION}
In this section, we present the results of the UAV RCS measurements, model selection analysis, and performance analysis of the UAV statistical recognition system.

\subsection{RCS Measurement Results}
\textcolor{black}{In this section, we present the RCS measurement results of the six UAVs shown in Fig.~\ref{SMALL_UAVs}. For each UAV, the complex transmission coefficient $S_{21}$ is measured in the frequency range 14.5-15.5 GHz and 24.5-25.5 GHz in steps of 5~MHz. This gives 201 frequency points and the measured $S_{21}$ data file has a length of 72762$\times$1. The $S_{21}$ data file can be reshaped to an array 201$\times$181$\times$2 where 2 represent the number of polarization (VV and HH) and 181 is the number of azimuth angles ($\phi\in[0\degree, 360\degree]$ with a 2$\degree$ increment). Therefore, for a specific frequency point and polarization, the measured $S_{21}$ has a length of 181$\times$1. After post-processing and calibration, the RCS of the UAVs is estimated from $S_{21}$ as described in Section III-B. The measured RCS data constitute the training dataset. The database contains the best-fitting statistical model for each UAV class estimated from the training dataset by either the AIC or BIC criteria as described in Section V-B.}

Due to the low reflective materials used in the UAV design, the average RCS values are less than $0$~dBsm. For the 15~GHz VV-polarization measurement, the average RCS of DJI Matrice 600, DJI Matrice 100, Trimble zx5, DJI Mavic Pro, DJI Inspire 1, and DJI Phantom~4 Pro are $-11.67$~dBsm, $-14.69$~dBsm, $-14.39$~dBsm, $-17.06$~dBsm, $-14.24$~dBsm, and $-15.02$~dBsm, respectively. For the same UAV set, the average VV-polarized RCS, measured at 25~GHz are $-7.32$~dBsm, $-11.03$~dBsm, $-9.64$~dBsm, $-16.20$~dBsm, $-11.09$~dBsm, and $-12.40$~dBsm, respectively.  Tables~\ref{AIC_table_VV}-\ref{BIC_table_HH} provide a summary of the mean and the standard deviation (Std) of the RCS for all the cases considered. From these values, we observe that for each of the UAVs, the average VV polarized RCS measured at 25~GHz is greater than the value obtained at 15~GHz. This observation is also true about the average HH-polarized RCS values. This is probably due to the diffraction effects of sharp corners at higher frequencies. Consequently, higher frequency radars are more suitable for ATR and KB radar systems, especially when the target of interest is a small UAV.

Furthermore, from Tables~\ref{AIC_table_VV}-\ref{BIC_table_HH}, we observe that the two bigger UAVs, DJI Matrice 600 and Trimble zx5, have relatively higher average RCS. This is probably because the arms and frames of these UAVs are designed with carbon fiber which can support the weight of large batteries and payload requirements of large UAVs. It is well known that carbon fiber has a higher reflectivity than plastics which is used in the design of the smaller UAVs. Moreover, the measured RCS depends on other factors as well. This observation is depicted by the RCS signature of the six UAVs which is shown in Fig.~\ref{RCS_POLAR_PLOT_new}. The RCS signature is a polar plot of the RCS (dBsm), measured at a specific frequency and polarization, versus azimuth angle. From Fig.~\ref{RCS_POLAR_PLOT_new}, we see that the RCS of each UAV is dependent on the aspect angle, frequency, polarization, and shape of the UAV. For instance, from Fig.~\ref{RCS_POLAR_PLOT_new}, we can recognize some of the physical features (shape) of the UAVs. From these signatures, we can recognize the dominant scattering areas around the arms and the front frame of each UAV. Therefore, through the radar or RCS signatures, UAVs could be distinguished from other airborne objects like manned-aircraft, missiles, and birds. Besides, using the RCS signature we could discriminate between different commercial UAV types by utilizing a statistical recognition system. Moreover, from Fig.~\ref{RCS_POLAR_PLOT_new}, we see that the VV and HH-polarized signature are a little different for each UAV.

\begin{figure*}{}
\center{
\begin{subfigure}[DJI Matrice 600: VV]{\includegraphics[width=0.24\linewidth]{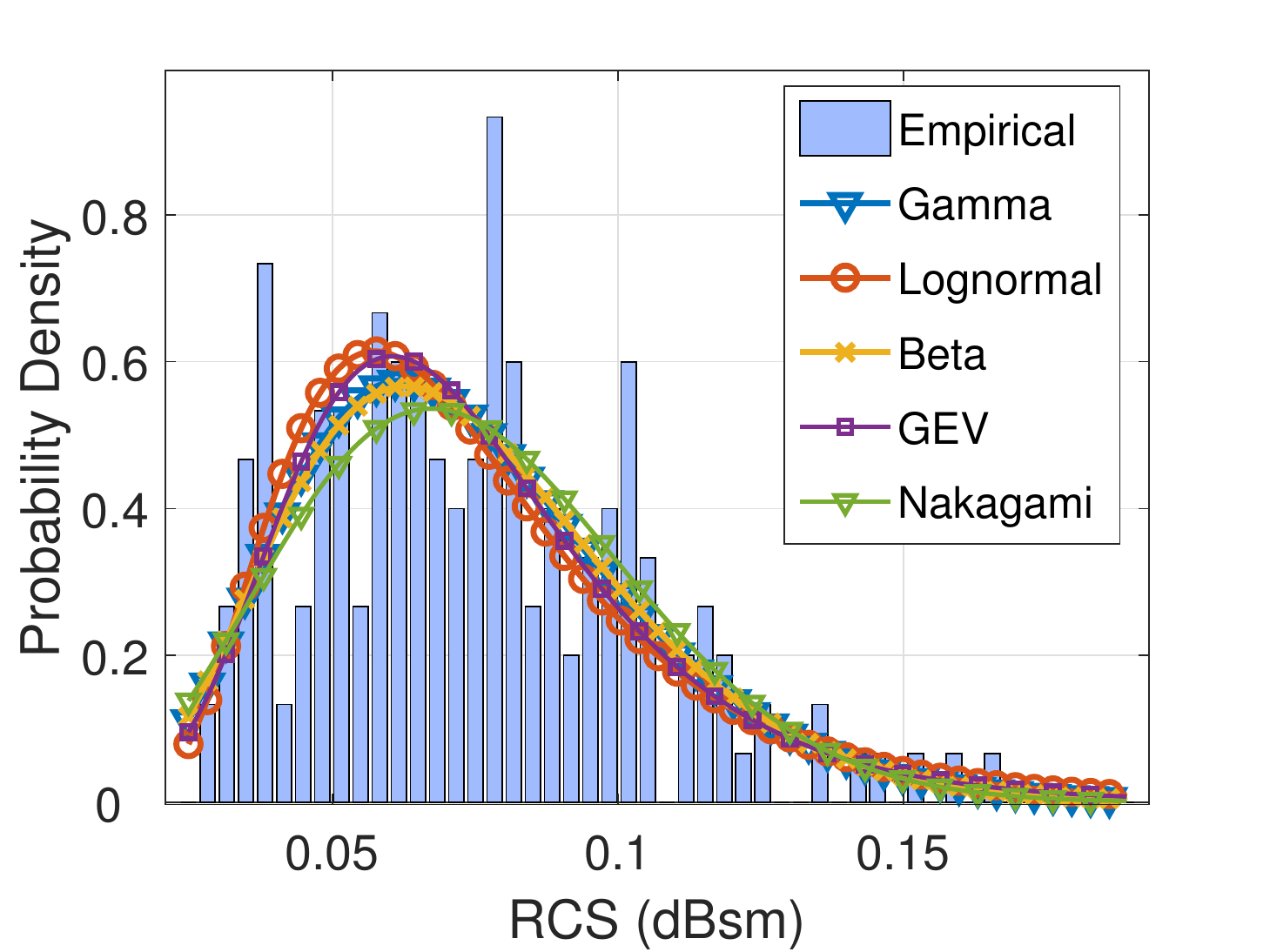}\label{AIC2_Matrice600_15GHz_pdf}}\end{subfigure}
\begin{subfigure}[DJI Matrice 600: HH]{\includegraphics[width=0.24\linewidth]{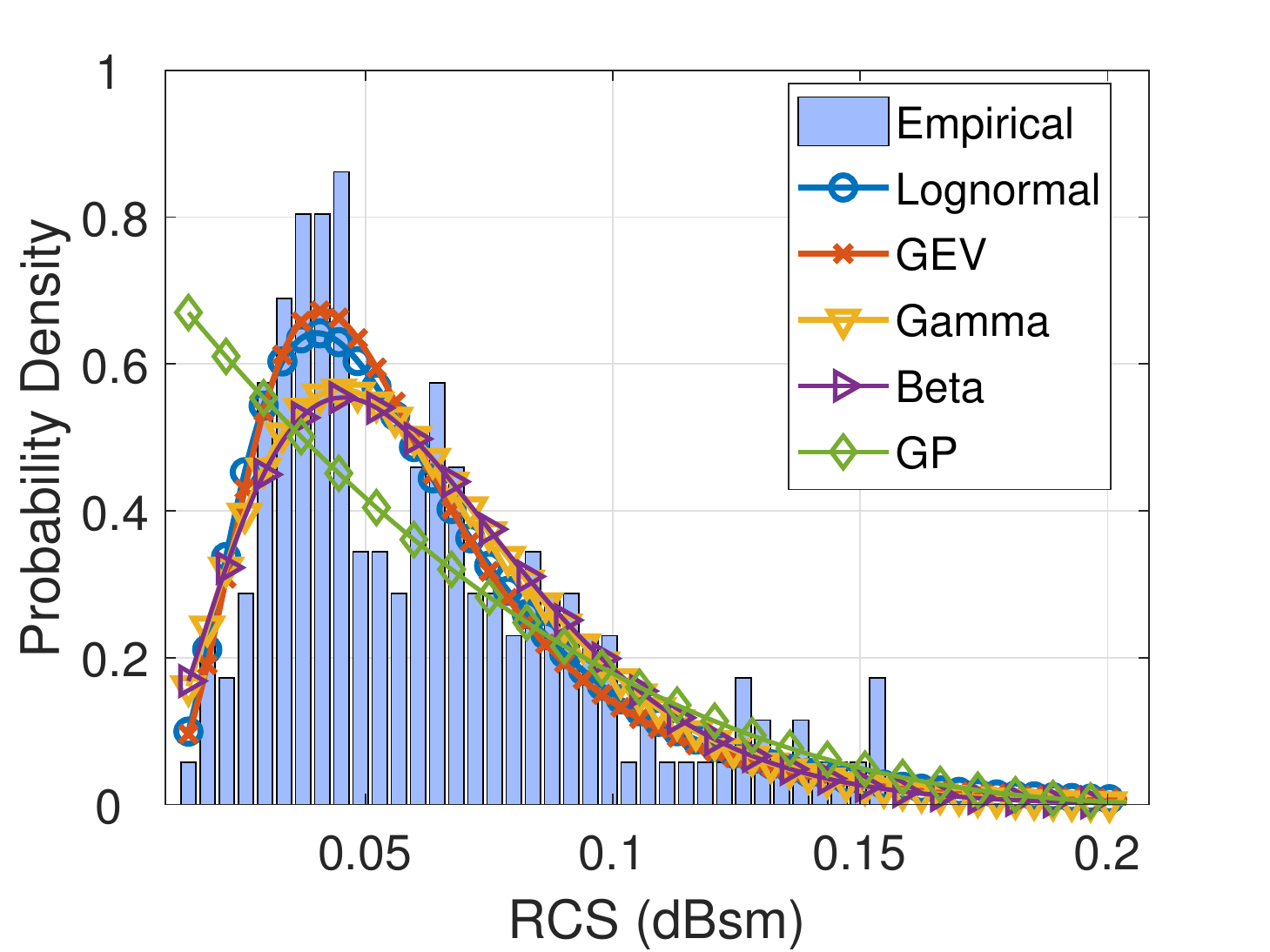}\label{RCS_DJI_M600_15GHz_Hpole_PDF_AIC}}
\end{subfigure}
\begin{subfigure}[DJI Matrice 100: VV]{\includegraphics[width=0.24\linewidth]{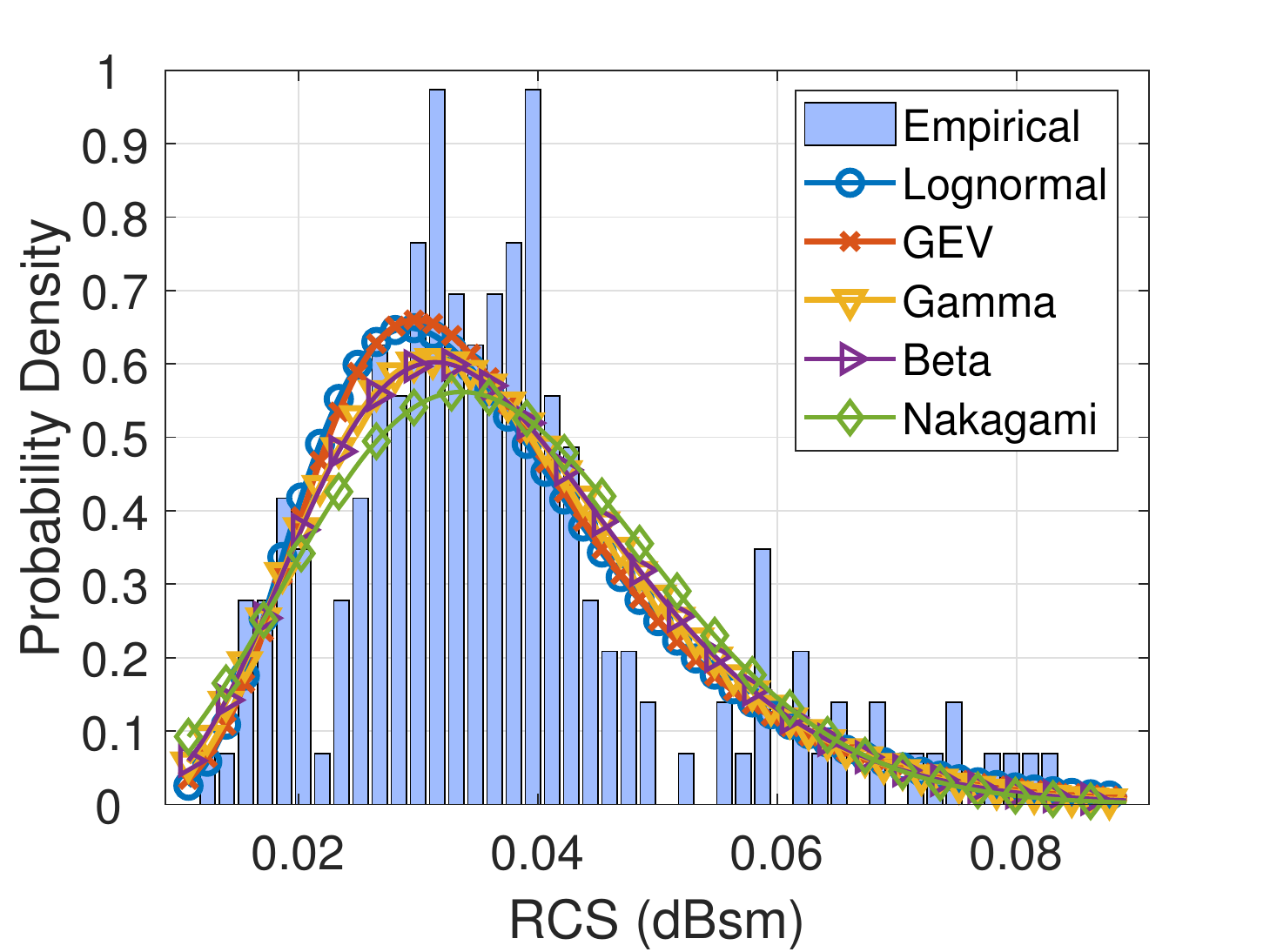}\label{AIC2_DJI_M100_15GHz_pdf}}\end{subfigure}
\begin{subfigure}[DJI Matrice 100: HH]{\includegraphics[width=0.24\linewidth]{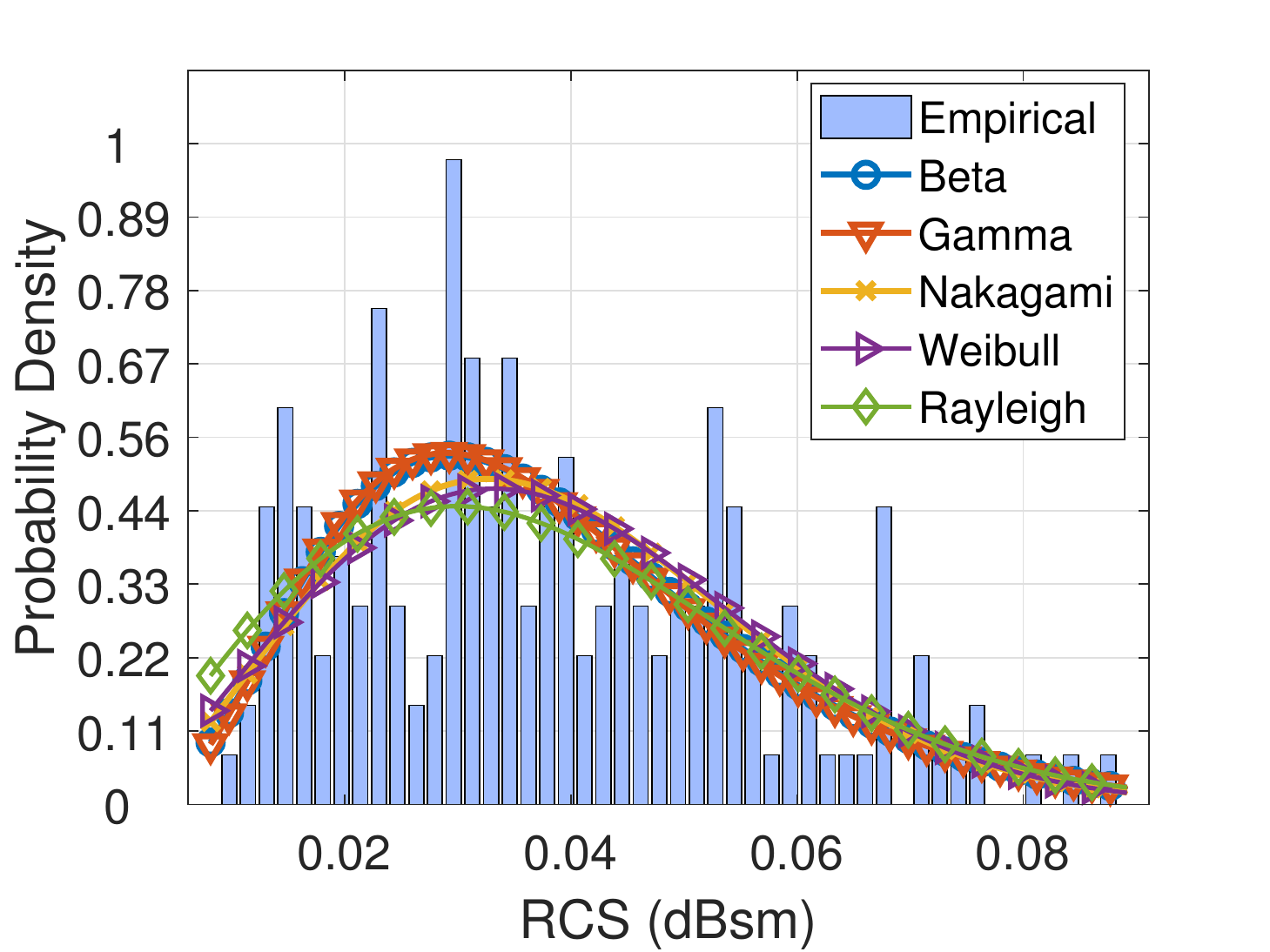}\label{RCS_DJI_M100_15GHz_Hpole_PDF_AIC}}
\end{subfigure}\\
\begin{subfigure}[Trimble zx5: VV]{\includegraphics[width=0.24\linewidth]{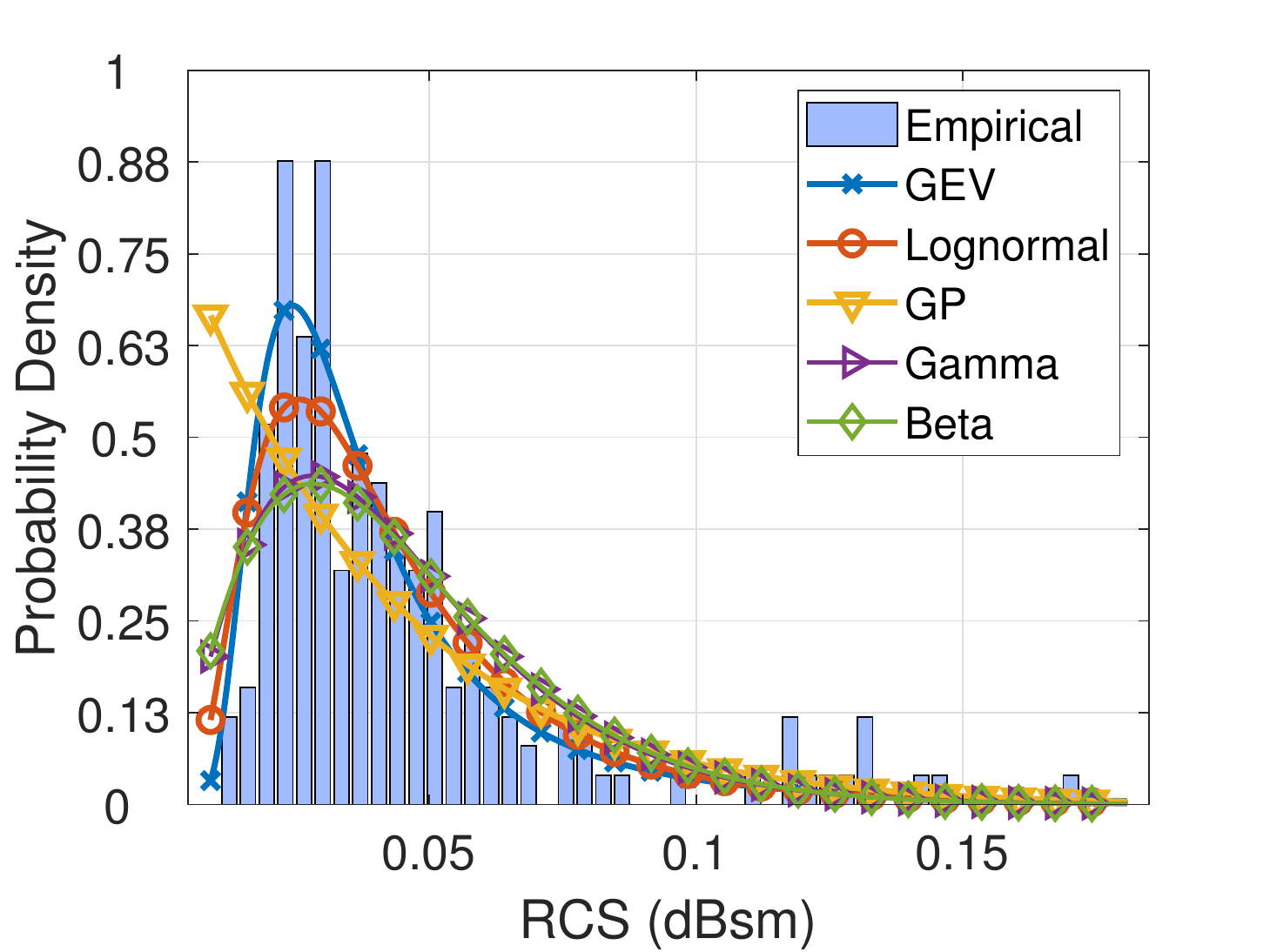}\label{AIC2_Trimble_15GHz_pdf}}
\end{subfigure}
\begin{subfigure}[Trimble zx5: HH]{\includegraphics[width=0.24\linewidth]{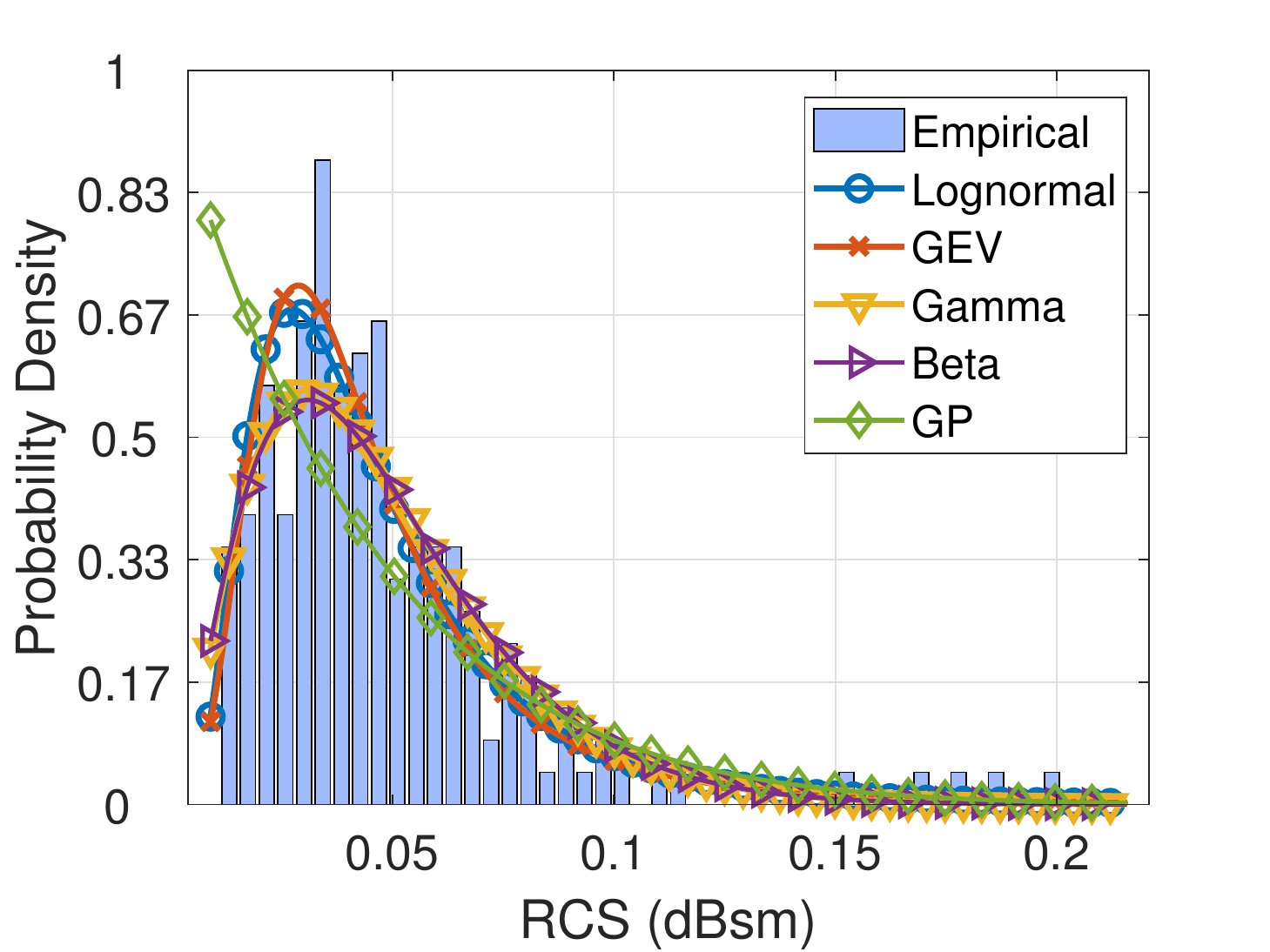}\label{RCS_Trimble_15GHz_HPole_PDF_AIC}}
\end{subfigure}
\begin{subfigure}[DJI Mavic Pro: VV]{\includegraphics[width=0.24\linewidth]{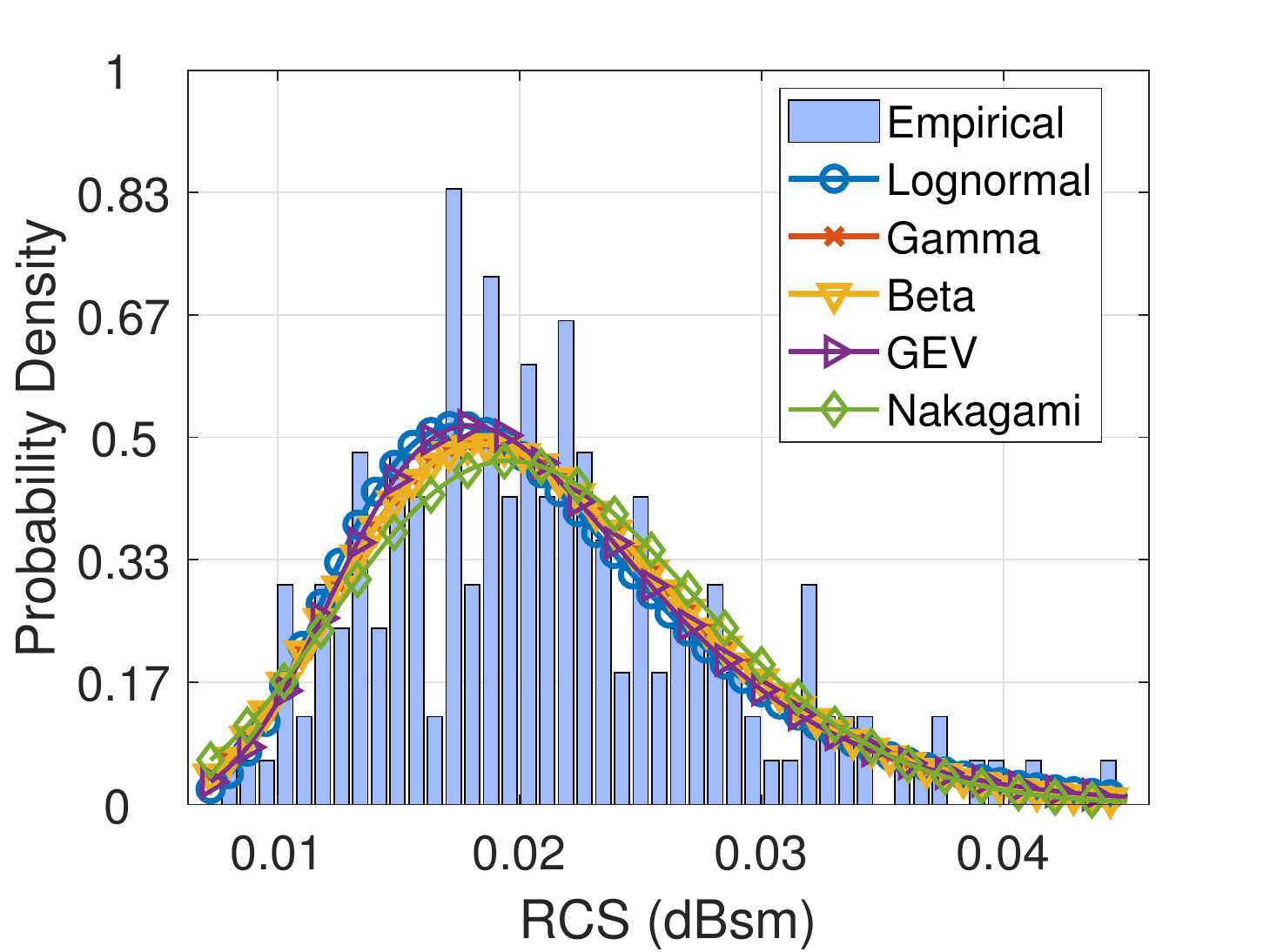}\label{AIC2_mavikPro_15GHz_pdf}}\end{subfigure}
\begin{subfigure}[DJI Mavic Pro: HH]{\includegraphics[width=0.24\linewidth]{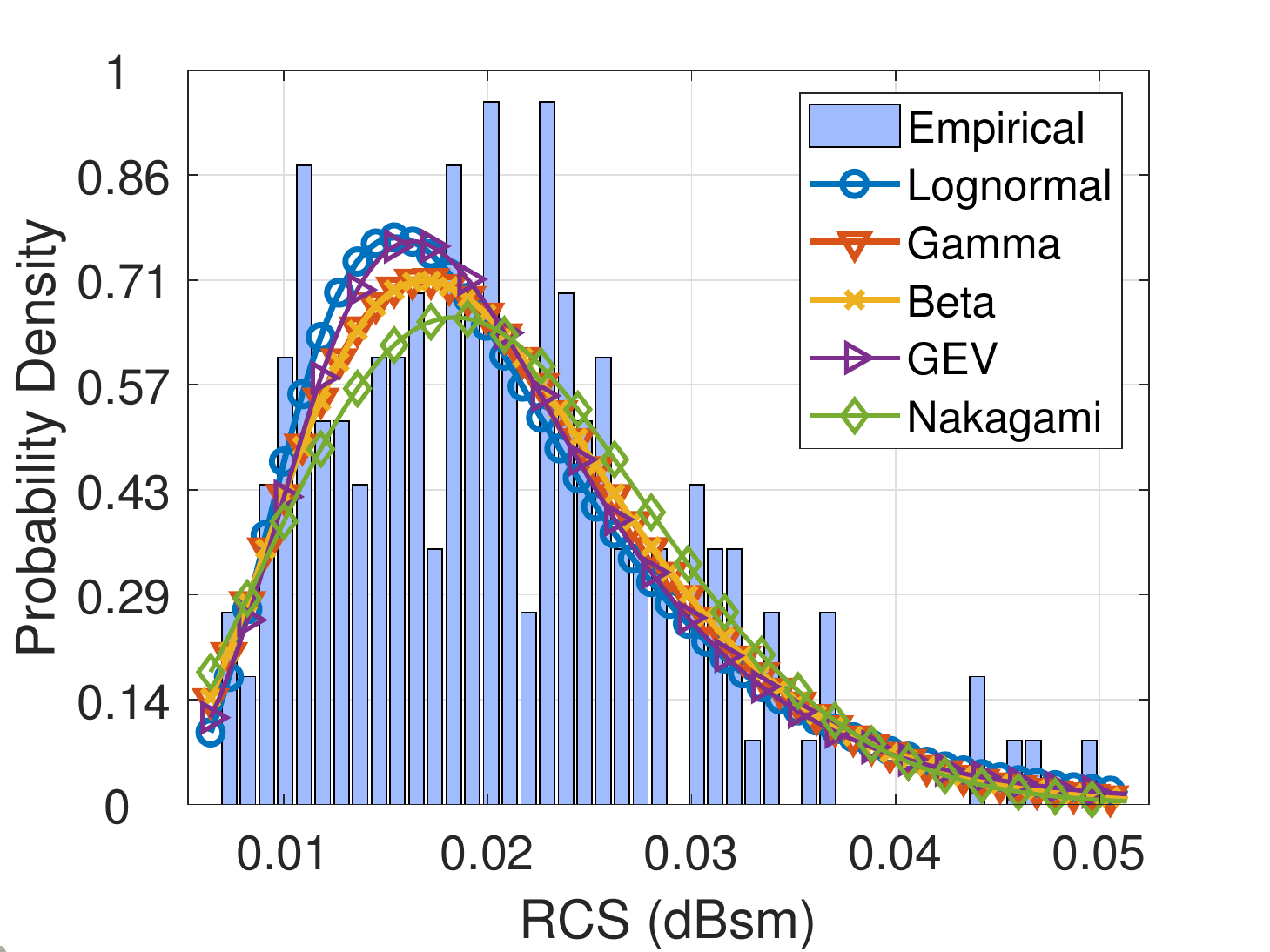}\label{RCS_MavicPro_15GHz_Hpole_PDF_AIC}}
\end{subfigure}\\
\begin{subfigure}[DJI Inspire 1: VV]{\includegraphics[width=0.24\linewidth]{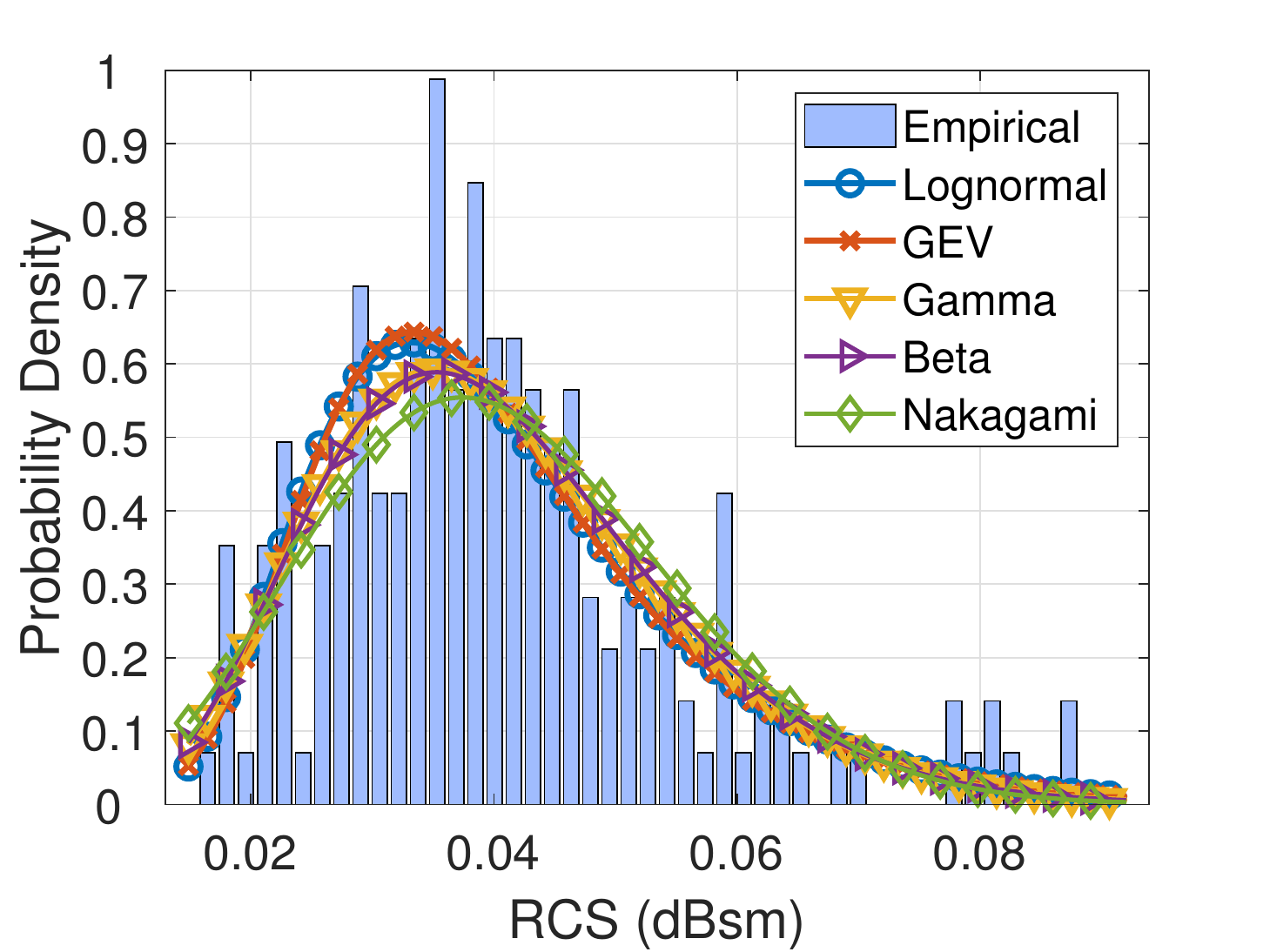}\label{AIC2_inspire_15GHz_pdf}}
\end{subfigure}
\begin{subfigure}[DJI Inspire 1: HH]{\includegraphics[width=0.24\linewidth]{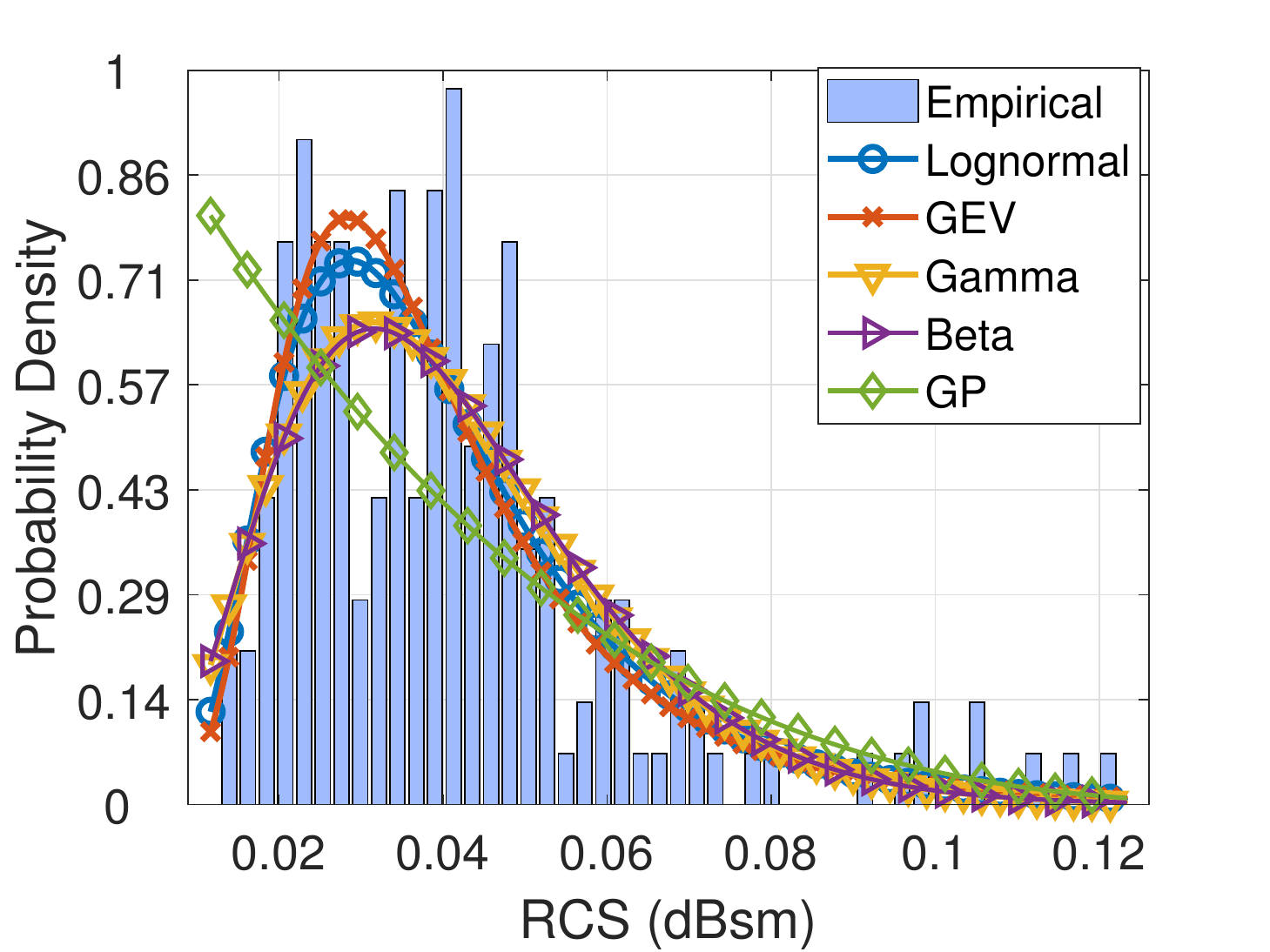}\label{RCS_inspire_15GHz_Hpole_PDF_AIC}}
\end{subfigure}
\begin{subfigure}[DJI Phantom 4 Pro: VV]{\includegraphics[width=0.24\linewidth]{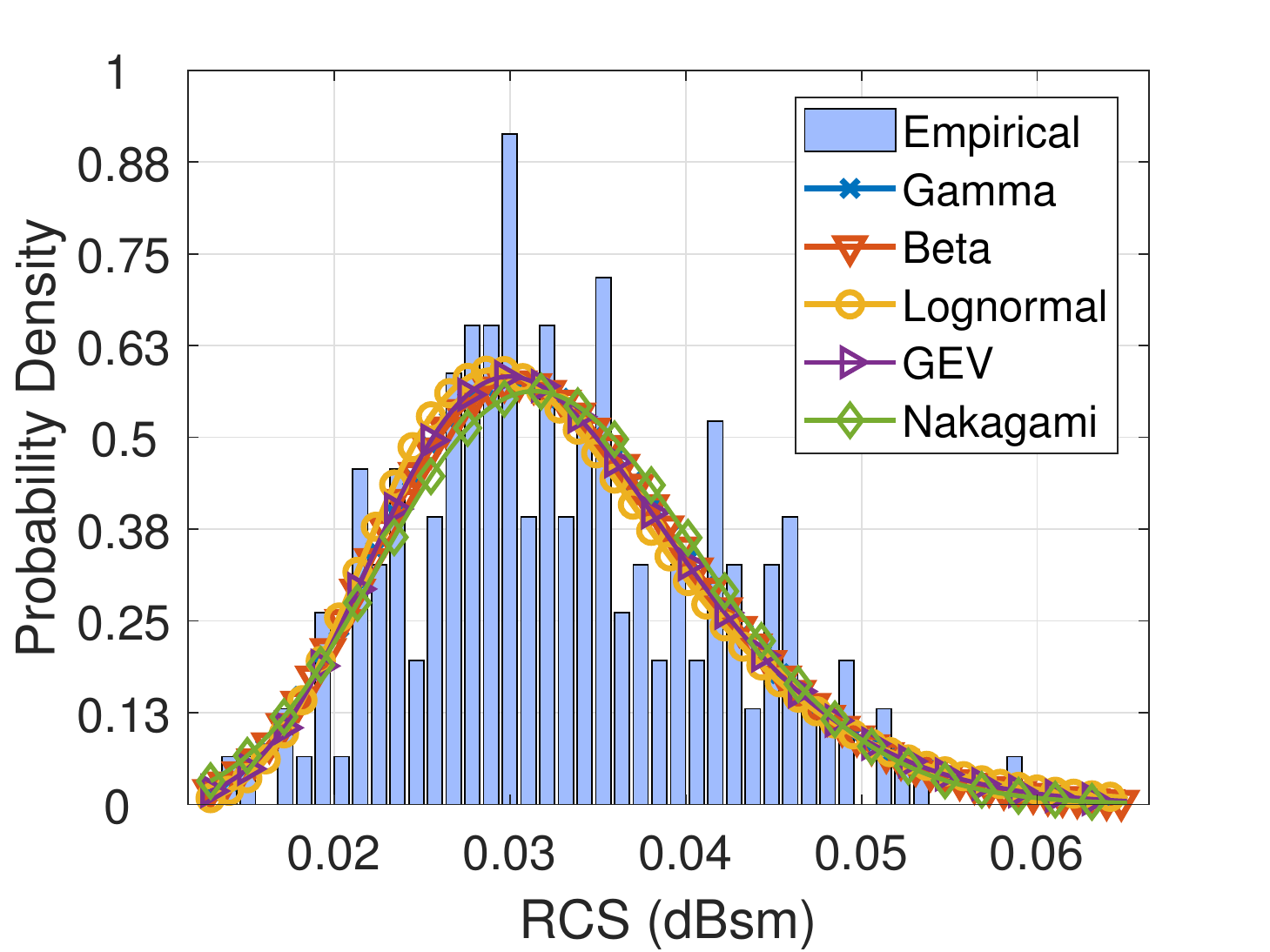}\label{AIC2_Phantom_4_Pro_15GHz_pdf}}
\end{subfigure}
\begin{subfigure}[DJI Phantom 4 Pro: HH]{\includegraphics[width=0.24\linewidth]{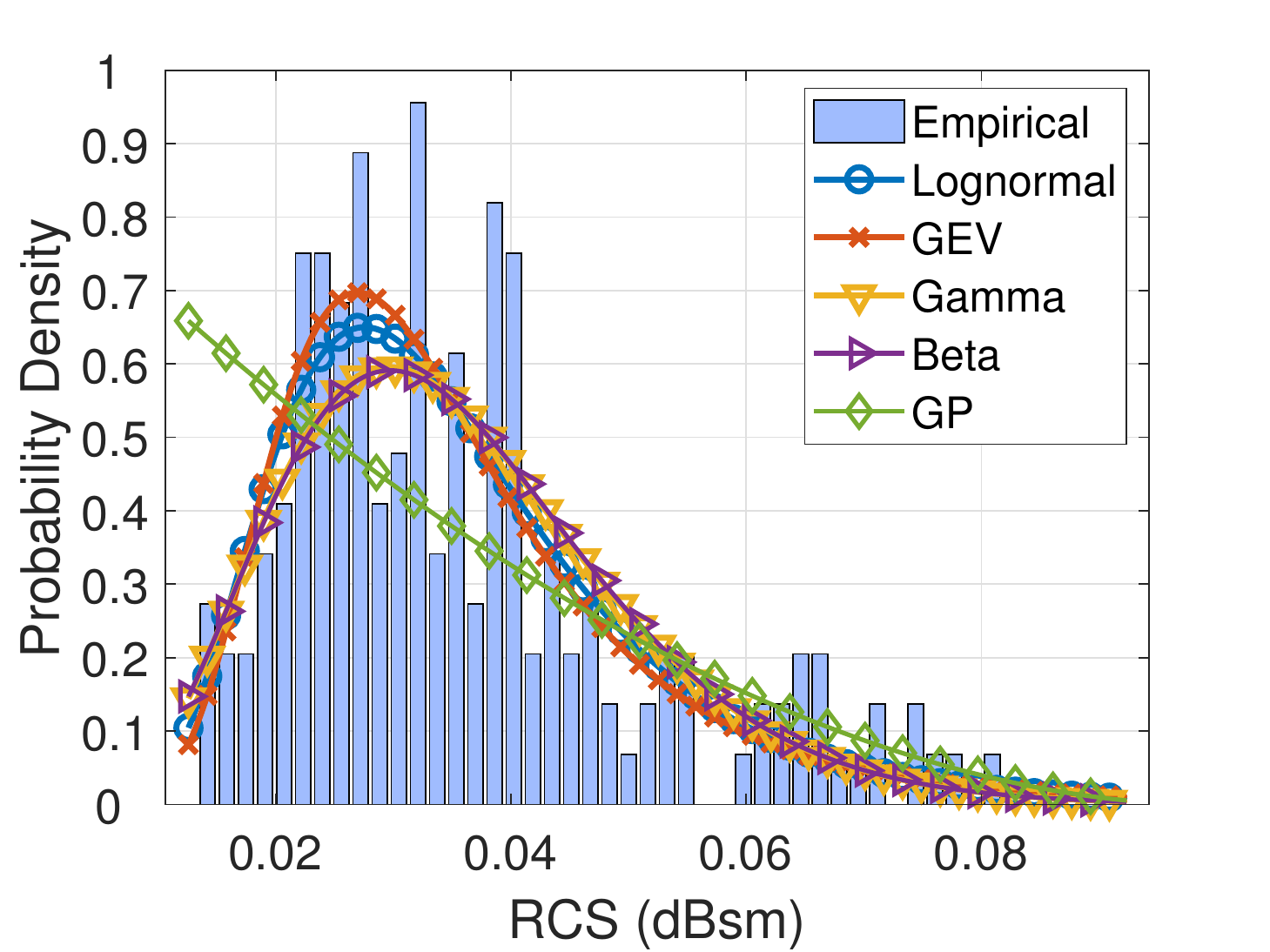}\label{RCS_phantom4pro_15GHz_Hpole_PDF_AIC}}
\end{subfigure}
\caption{{The probability distribution of the best five statistical models fitted to the UAV RCS data measured at 15 GHz (VV and HH-polarization). The histogram is the empirical statistics of the measured data which is fitted to different statistical distributions. The plot legends show in decreasing order (moving down) the relative fit of the statistical models according to the AIC criterion.}}
 \label{pdf_RCS_plot}}
 \vspace{-4mm}
 \end{figure*}

\subsection{Results of RCS Model Selection Analysis}\label{model_selection_results}

In this section, the results of model selection for the measured UAV RCS are presented. In this study, 11 different candidate statistical models will be investigated for all six UAV types. The statistical models are lognormal, generalized extreme value (GEV), gamma, beta, generalized Pareto (GP), Weibull, Nakagami, Rayleigh, Rician, exponential, and normal statistics. The dataset consists of UAV RCS measurement at 15~GHz and 25~GHz  in both the VV-and HH polarization.

For each of the UAVs, Table~\ref{AIC_table_VV} and Table~\ref{BIC_table_VV} present the AIC and BIC test scores, respectively using the VV-polarized RCS measurement data. On the other hand, Table~\ref{AIC_table_HH} and Table~\ref{BIC_table_HH} present the AIC and BIC test scores, respectively using the HH-polarized RCS measurement data. For each UAV type, the best RCS model is selected as the model with the lowest AIC or BIC test score. In the tables, the best model for each UAV is represented by the highlighted cell. From the tables, we observe that except for the 25 GHz VV-polarized RCS data, both the AIC and BIC criteria select the same statistical model as the best. Furthermore, we observe that the lognormal is the most frequently selected statistical model for the UAV RCS measurement data. This is followed by the GEV and gamma statistical models. Interestingly, in~\cite{semkin2020analyzing}, the authors claimed the Gaussian statistics is a good model for the UAV RCS data for measurements at mmWave frequencies. However, our studies have shown there are better statistical models, at least based on our measurements at 15~GHz and 25~GHz.

To better describe the relative performance of all the 11 statistical models, we can analyze their probability distributions. Fig.~\ref{pdf_RCS_plot} provides a plot of the probability distribution of each of the statistical models given the UAV RCS data measured at 15 GHz. From the probability distribution plots, we can verify that both lognormal and GEV models are relatively better in fitting the experimental or empirical RCS data of all six commercial UAVs measured at 15 GHz. This is probably because lognormal and GEV distributions are heavy-tailed and could well fit the extremities of the skewed measurement data. On the contrary, the Gaussian distribution is often suitable for symmetric data. However, from the histograms in Fig.~\ref{pdf_RCS_plot}, we see that the empirical data is not symmetric.
Besides, for each UAV type, the parameters of the statistical models will vary depending on the radar system parameters such as frequency and polarization. Thus, for a given radar system, knowing the most appropriate RCS statistics of a specific UAV type could help discriminate/classify the UAV from other objects.
The results of statistical-based UAV recognition/classification will be discussed next.

\subsection{Analysis of UAV Statistical Recognition System Using all Azimuth RCS Data}\label{classification_subsection}
In this subsection, we describe the results of the UAV recognition/classification at different SNR using all azimuth RCS data from each of the UAV (($\phi\in[0^{\circ}, 360^{\circ}$) with a 2$^{\circ}$ increment). First, we need to estimate the Gaussian noise power ($\sigma_N^2$) for the different SNR considered in the analysis. For instance, given a specific SNR and the time-domain radar response (scattered data) $a_k (t)$ from an unknown $k$-th UAV, the noise power $\sigma_{\rm N}^2$ is estimated as:
\begin{align}
\label{SNR_EQU}
  \sigma_N^2 =P_{k}10^{-{\rm SNR}/10},
\end{align}
where
\begin{equation}
\label{power_measurement}
  P_k =\frac{1}{T}\int_{0}^{T} a_k (t)^2 {\rm d}t =\\
  \begin{cases}
    \frac{\sum_{i=1}^{N} \big|\sqrt{\sigma_{\rm VV}}\big|^2}{N},~\text{VV-polarization} \\
    \frac{\sum_{i=1}^{N} \big|\sqrt{\sigma_{\rm HH}}\big|^2}{N},~\text{HH-polarization}\\
  \end{cases}
\end{equation}
is the average power of the time domain return $a_k (t)$ from the unknown UAV, $T$ is the response interval, and $N$ is the number of discrete samples of the frequency domain RCS data. The outcome in (\ref{power_measurement}) is a consequence of the Parseval's theorem.

\begin{figure}[t!]
 \center
 \includegraphics[scale=0.85]{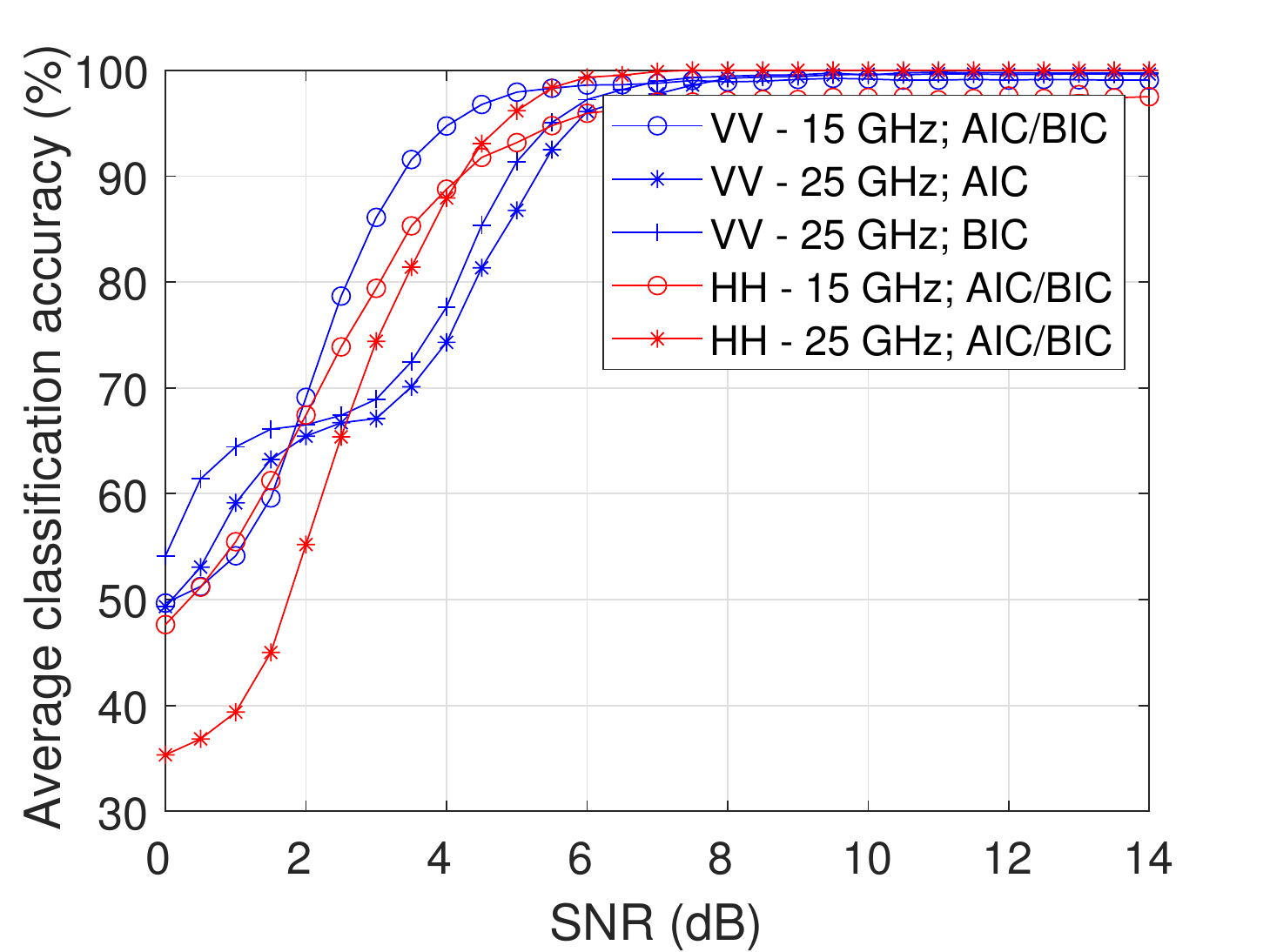}
\caption{Average classification vs SNR for six drones evaluated at 15 GHz and 25 GHz measured using both the VV and HH-polarized RCS data. This result is based on all the azimuth RCS data from each of the UAV (($\phi\in[0^{\circ}, 360^{\circ}$) with a 2$^{\circ}$ increment)}
\label{Fig:classification_plot}\vspace{-5mm}
\end{figure}

Once $\sigma_{\rm N}^2$ has been estimated, we can add the appropriate noise signal to the test signals and perform classification. \textcolor{black}{To evaluate the performance of the UAV recognition system, we generate a test dataset by running the Monte Carlo simulation over a range of SNRs. For each SNR, 500 noisy RCS test data are generated by adding appropriate Gaussian noise to the measurement data. The test data are used for classification. The test samples are passed through the statistical recognition system which computes the likelihood that each of the test data is a member of one of the UAV classes in the database.}

Fig.~\ref{Fig:classification_plot} shows the average classification accuracy versus SNR at different frequencies and polarizations. Also, Fig.~\ref{Fig:classification_plot} highlights the performance of the different model selection criteria used for the training database. We see that the average UAV classification accuracy is the same in the case of the 15 GHz VV-polarized test data, irrespective of which criteria was used for selecting the best statistical models for the UAV in the database. A similar observation can be seen in the case of the 15~GHz and 25 GHz HH-polarized test data. To explain this observation, we look at Tables~\ref{AIC_table_VV}-\ref{BIC_table_HH}. From these tables, we see that both AIC and BIC selects the same statistical models for the database. Therefore, the classification results are similar in these cases. On the contrary, for the 25 GHz VV-polarized test data, AIC and BIC criteria select a different statistical model for some of the UAVs in the database. Therefore, the classification accuracy slightly differs as shown in Fig.~\ref{Fig:classification_plot}.

\begin{figure}[t!]
\center{
 \begin{subfigure}[]{\includegraphics[width=0.4\linewidth]{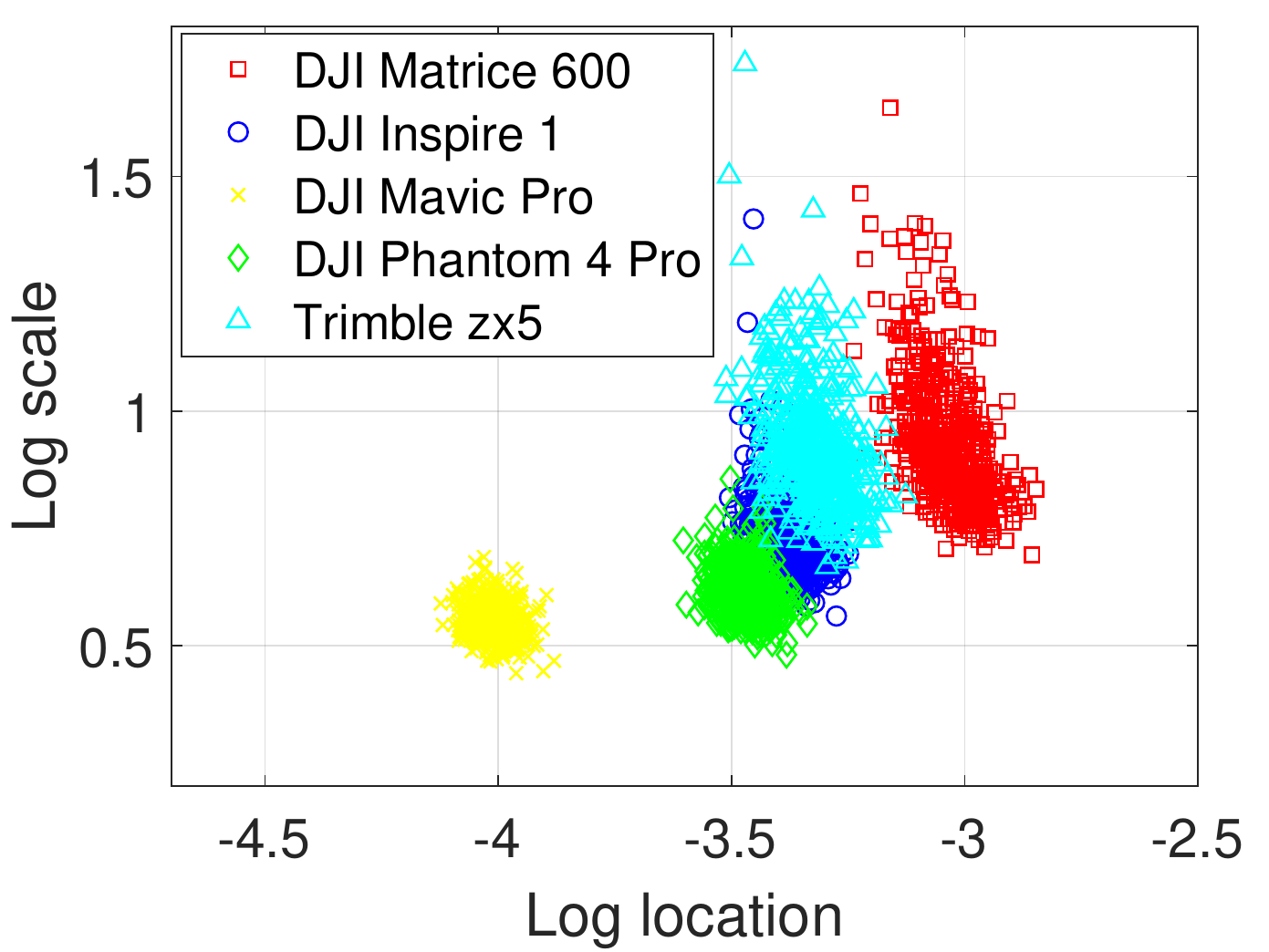}
 \label{uav_scatterplot_0dB}}
\end{subfigure}
 \begin{subfigure}[]{\includegraphics[width=0.4\linewidth]{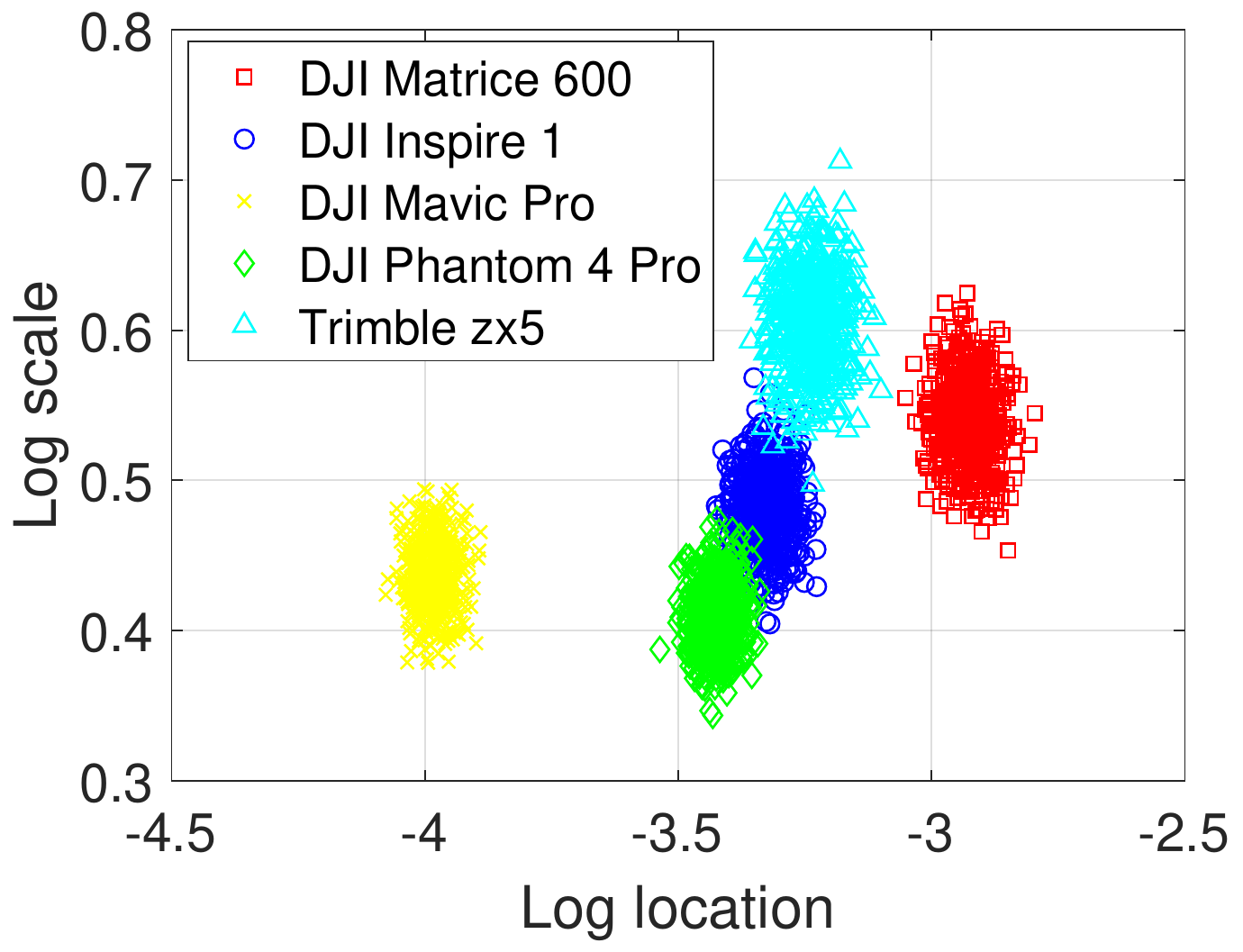}\label{uav_scatterplot_10dB}}
\end{subfigure}
\caption{{Scatter plots for parameters of the lognormal statistical model fitted to the 15~GHz HH polarized RCS test data at: (a) 0~dB SNR, (b) 10~dB SNR. }\label{scatter_plot}}}\vspace{-5mm}
\end{figure}

Also, from Fig.~\ref{Fig:classification_plot}, we see that the UAV classification accuracy increases with SNR. For the 15~GHz and 25~GHz VV-polarized RCS test data, the average classification accuracy is 49.67\% and 49.33\% at 0~dB SNR and 99.17\% and 99.53\% at 10~dB SNR.
Similarly, for the 15~GHz and 25~GHz HH-polarized RCS test data, the average classification accuracy is 47.63\% and 35.53\% at 0~dB SNR and 97.43\% and 100\% at 10~dB SNR. Besides, Fig.~\ref{Fig:classification_plot} also shows that at SNR below 3~dB, the classification accuracy for the HH-polarized RCS test data is relatively lower than the VV-polarized RCS test data. This is probably because AIC and BIC mostly select the same statistical model (lognormal) for all the HH-polarized UAV RCS data in the database. Even if each UAV type in the database is described by a unique lognormal statistics, having the same model type for the different UAVs could lead to some confusion at low SNR. This is because at low SNR, the additive Gaussian noise cause the parameter values of the lognormal statistical models of the HH-polarized RCS test data to come closer together which can lead to confusion in the UAV classification. This observation is depicted by the scatter plots shown in Fig.~\ref{scatter_plot}. In Fig.~\ref{uav_scatterplot_0dB}, we see a significant overlap between the lognormal parameters of DJI Inspire 1 and Trimble zx5 on one hand, and between DJI Inspire 1 and DJI Phantom 4 Pro on the other hand. Also, there is a little overlap, between the parameters of the DJI Matrice 600 and Trimble zx5. The overlapping of the lognormal parameter at of 0~dB SNR will affect the classification accuracy of the UAV statistical recognition system. However, at an SNR of 10~dB, Fig.~\ref{uav_scatterplot_10dB} shows a reduced overlap between the lognormal parameters. Therefore, we would expect a higher recognition rate as SNR increases.

\begin{figure*}[t!]
    \centering
         \begin{subfigure}[VV 15 GHz, AIC]{\includegraphics[width=0.45\linewidth]{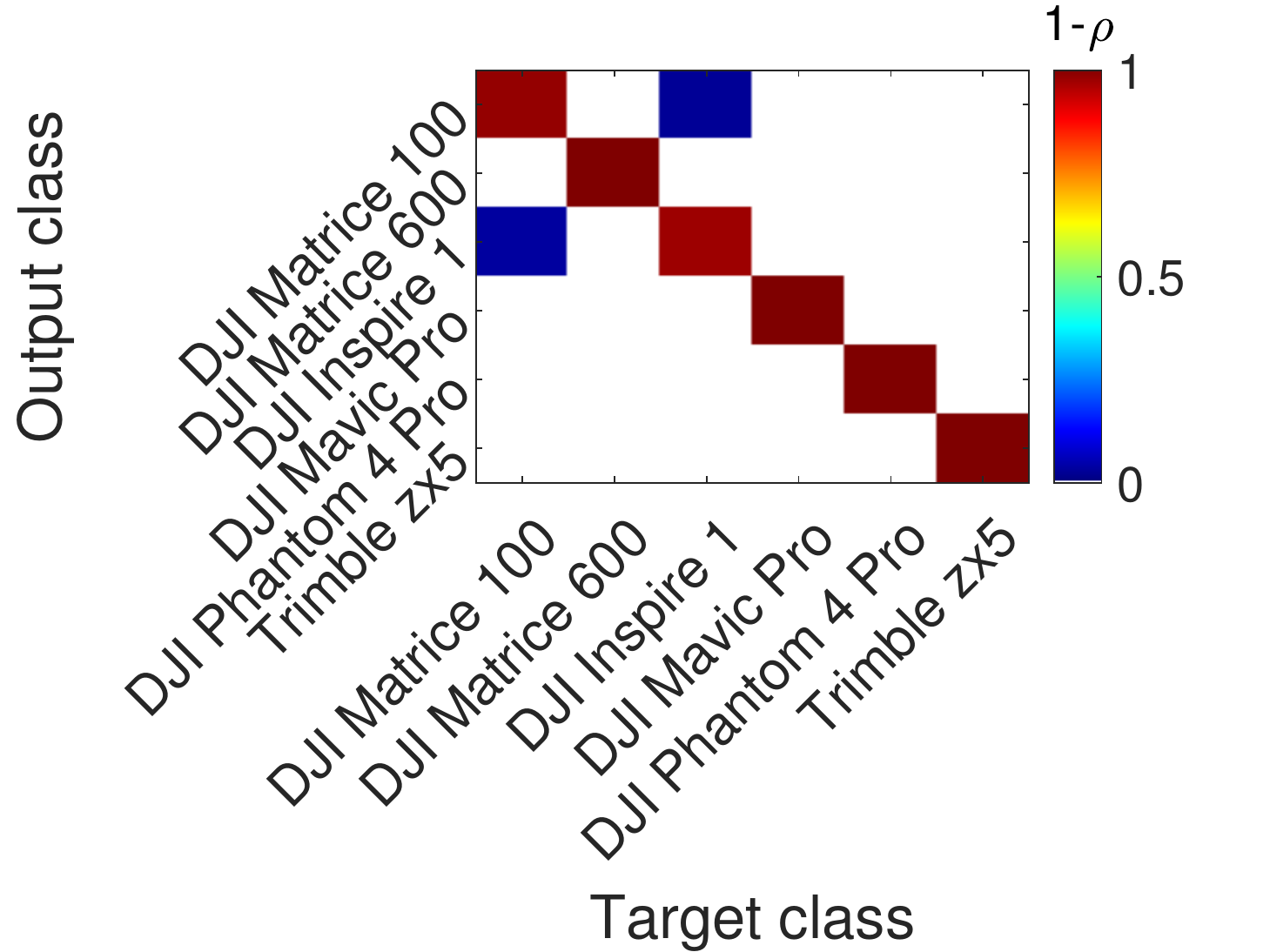}\label{VV_15_GHz_AIC}}
\end{subfigure}
     \begin{subfigure}[VV 25 GHz, AIC]{\includegraphics[width=0.45\linewidth]{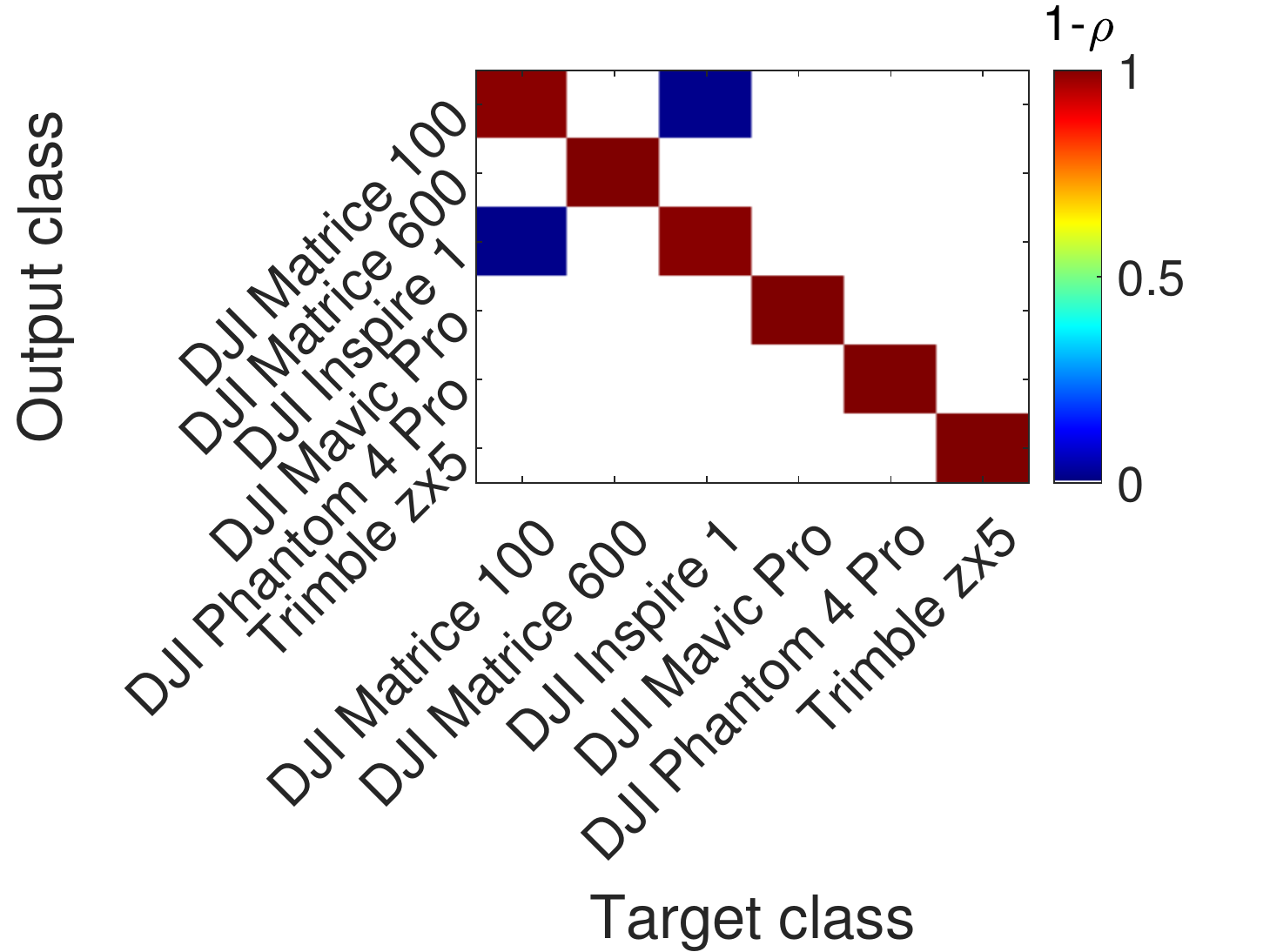}\label{VV_25GHz_AIC}}
\end{subfigure}\\
\begin{subfigure}[HH 15 GHz, AIC]{\includegraphics[width=0.45\linewidth]{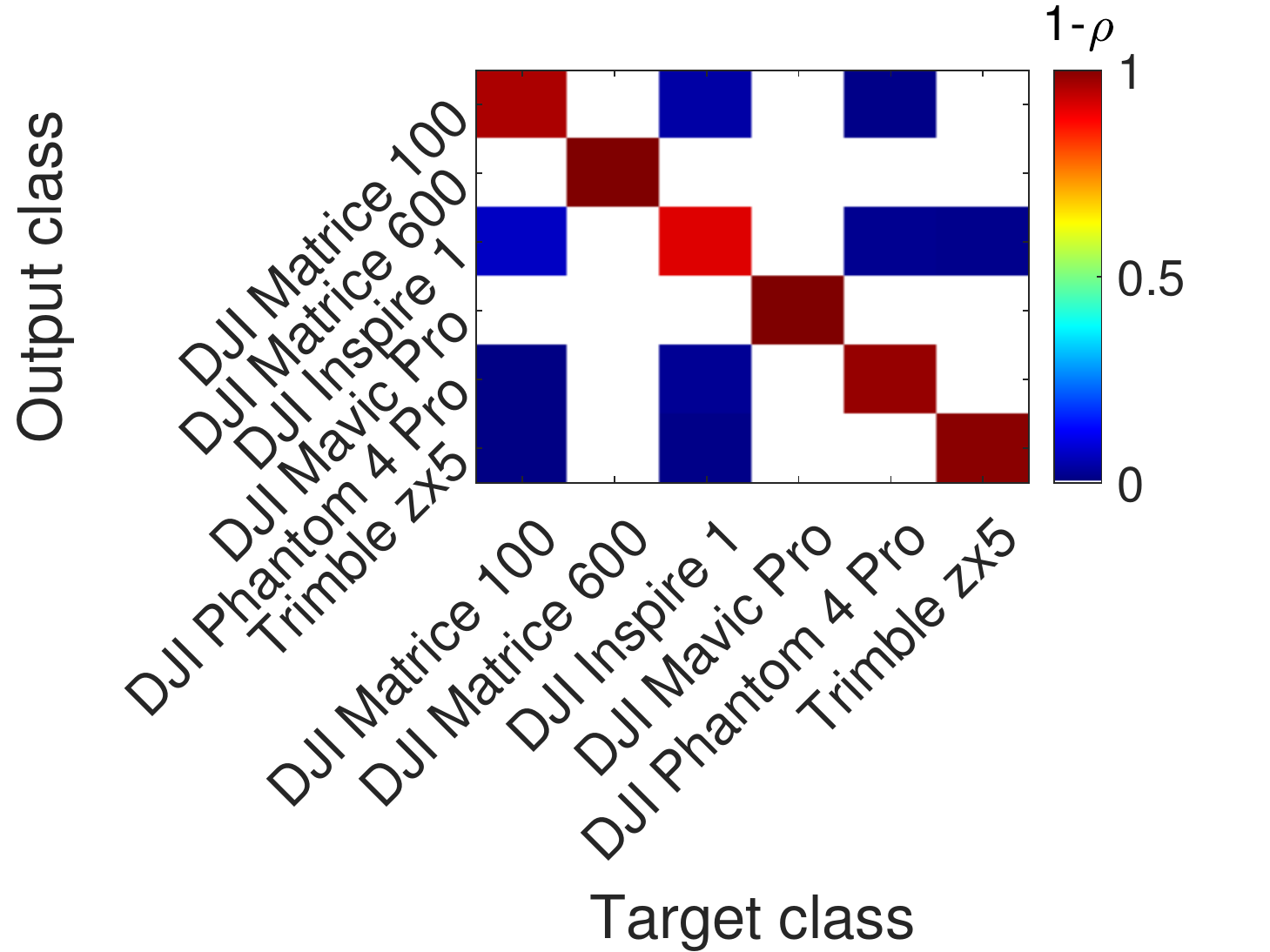}\label{HH_15GHz_AIC}}
\end{subfigure}
 \begin{subfigure}[HH 25 GHz, AIC]{\includegraphics[width=0.45\linewidth]{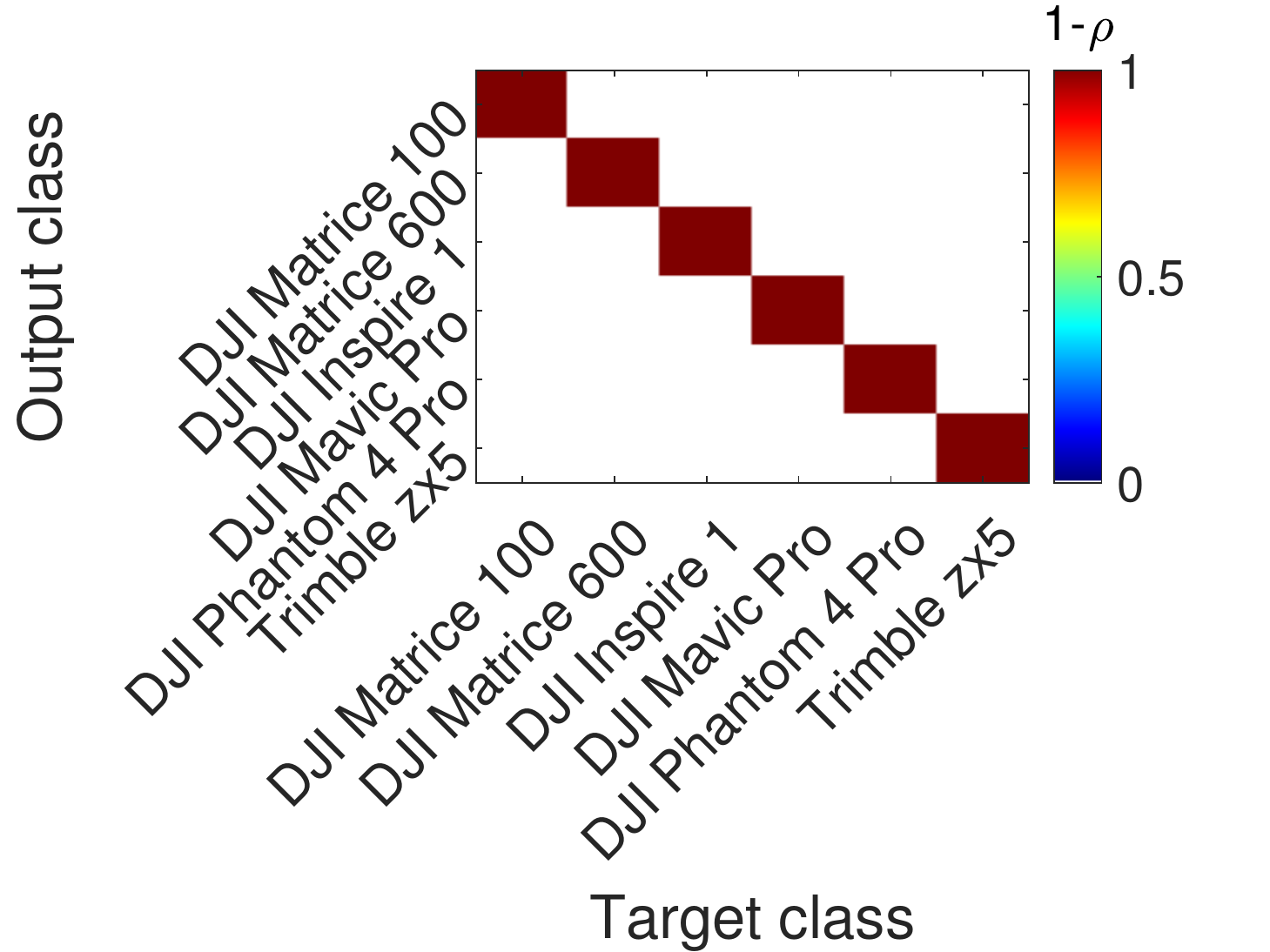}\label{HH_15GHz_BIC}}
\end{subfigure}
\caption{The confusion matrices of the UAV statistical recognition system at 10 dB SNR. These matrices are obtained when we use the AIC criterion for the database. In each matrix, the degree of confusion is specified by the colorbar in terms of the probability of confusion $\rho$. Moving down the colorbar, the value of $\rho$ increases, signifying increasing confusion.}
\label{confusion_matrix}
\end{figure*}



To further analyze the classification  performance of the UAV statistical recognition system, we generate the confusion matrices for the UAV statistical recognition system using the VV and HH-polarized test data and AIC model selection criterion for the database. The confusion matrix gives us an idea of what the UAV recognition system is getting right and what kind of misclassification occurs. In the confusion matrix, the diagonal elements represent the instances of correct classification while the off-diagonal elements represent the misclassifications. Fig.~\ref{confusion_matrix} shows the confusion matrices generated at 10 dB SNR by averaging the results of the Monte Carlo simulation. For the 15~GHz and 25~GHz VV-polarized RCS test data, the DJI Matrice 100 is sometimes confused (misclassified) as the DJI Inspire 1 and vice-versa. This accounts for all the misclassification when we use the VV-polarized RCS test data to perform UAV statistical recognition. On the other hand, using the 15~GHz HH-polarized RCS test data, there are several instances of misclassification between DJI Matrice 100, DJI Inspire 1, Trimble zx5, and Phantom 4 Pro UAVs. However, the percentage of the misclassification (sum of the off-diagonal rates) is not significant at 10 dB. This observation is consistent with the scatter plot in Fig.~\ref{uav_scatterplot_10dB}. For instance, using the 15 GHz HH-polarized, only 0.4\% of the DJI Matrice 100 test data are misclassified as either Trimble Zx5 or DJI Phantom 4 Pro; only 0.8\% of the DJI Inspire 1 test data are misclassified as Trimble zx5; only 0.8\% of DJI Phantom 4 Pro are misclassified as DJI Matrice 100. Besides, the percentage of correct classification (average of all diagonal elements) is about 97.43\%. Furthermore, Fig.~\ref{confusion_matrix} shows that there are no misclassifications for the 25 GHz HH-polarized RCS test data at 10 dB SNR. Therefore, for improved UAV recognition, we suggest the use of the 25~GHz HH-polarized RCS test data and the lognormal statistics for the database.
\begin{figure}[t!]
 \center
 \includegraphics[scale=0.85]{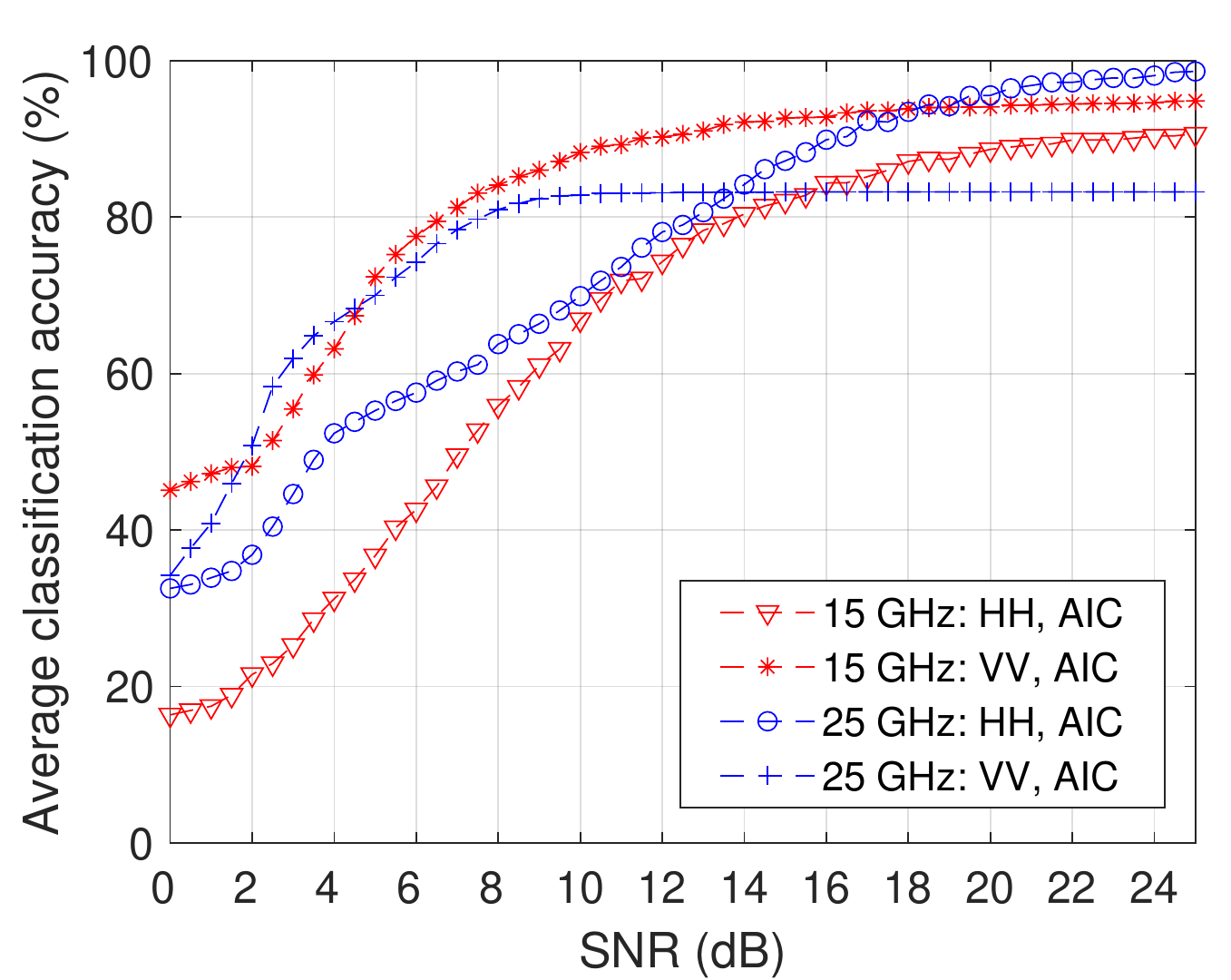}
\caption{The average classification of the six drones in the case where the radar sees only 120 $^{\circ}$ sector RCS reflections from the approaching targets. The classification is evaluated at 15 GHz and 25 GHz measured using both the VV and HH-polarized RCS data.}
\label{limited_azimuth}
\end{figure}
\subsection{Analysis of UAV Statistical Recognition System Using Limited Azimuth RCS Data}\label{classification_subsection_2}
\textcolor{black}{
In Section~\ref{classification_subsection}, we evaluated the performance of the UAV statistical recognition system using the RCS data measured from all around the UAVs (($\phi\in[0^{\circ}, 360^{\circ}$) with a 2$^{\circ}$ increment). However, in many instances, the field of view (FOV) is limited. That is, the radar only sees a sector of the approaching target. For instance, some practical radars, like the Fortem radar~\cite{fortem_radar1}, can only detect an incoming UAV within a 120$^{\circ}$ sector. In that case, UAV recognition system has to make a classification using the radar reflection data (RCS) captured from within the limited FOV of the radar. Fig.~\ref{limited_azimuth} shows the average classification of the six drones when the FOV of the radar is 120$^{\circ}$ (centered around 0$^{\circ}$ or ($\phi\in[-60^{\circ}, 60^{\circ}$] with a 2$^{\circ}$ increment) in the azimuth plane. In this case, we observe the average classification accuracy is relatively smaller than what is reported in Fig.~\ref{Fig:classification_plot}. At 14 dB SNR, Fig.~\ref{limited_azimuth} shows that the classification algorithm achieves an accuracy of 80.4\%, 92.1667\%, 84.2\%, and 83.2\% for 15 GHz HH, 15 GHz VV, 25 GHz HH, and  25 GHz VV respectively. In comparison, at 14 dB, Fig.~\ref{Fig:classification_plot} shows that the classification algorithm achieves 97.5\%, 99.07\%, 100\%, and 100\% for 15 GHz HH,  15 GHz VV, 25 GHz HH, and 25 GHz VV respectively. However, Fig.~\ref{limited_azimuth} shows that as SNR increases, the average classification accuracy gets better even with limited azimuth angles. In general, the wider the azimuth beamwidth of the radar and the higher the SNR, the better the performance of the UAV statistical recognition system.}
\begin{figure}[t!]
\center{
 \begin{subfigure}[]{\includegraphics[width=0.45\textwidth]{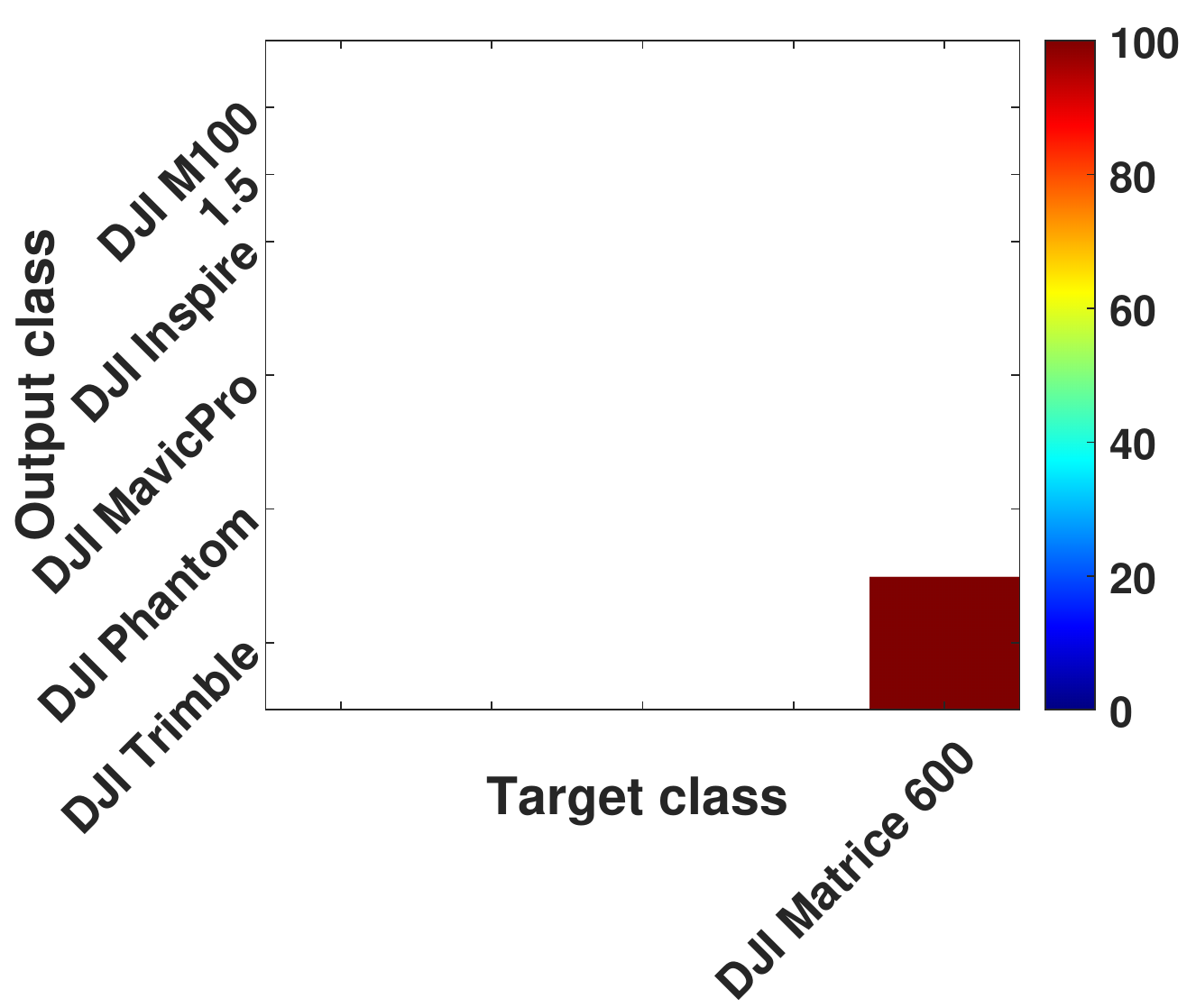}
 \label{M600_as_Trimble}}
\end{subfigure}
 \begin{subfigure}[]{\includegraphics[width=0.51\textwidth]{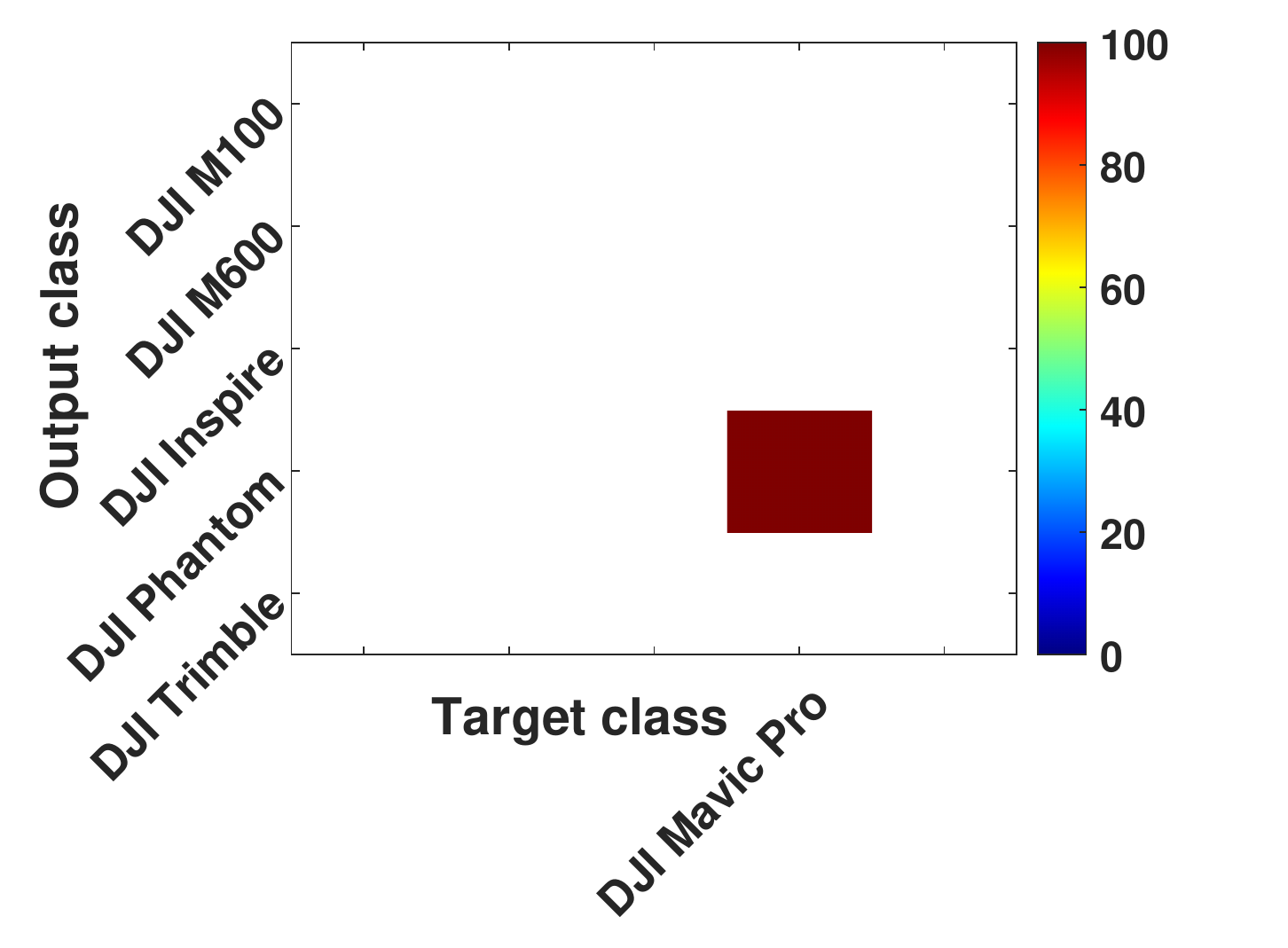}\label{Mavic_as_phantom}}
\end{subfigure}
\caption{Classifying UAVs of the same family at 14 dB SNR and 25 GHz HH : (a) DJI Matrice 600 classified as Trimble zx5, and (b) DJI Mavic Pro classified as DJI Phantom 4 Pro. }\label{UAV_same_family}\label{UAV_same_family}}

 \end{figure}
\subsection{Classifying an Unknown UAV from the Same Family as One of the UAVs in the Database}\label{classification_subsection_2}

\textcolor{black}{
 In this subsection, we are interested in investigating the performance of the UAV statistical recognition system when an unknown UAV is observed by the radar. Suppose the shape/material properties of the unknown UAV is similar to one of the UAVs in the database, then for the purpose of this experiment, we will assume both belong to the same family. Therefore, we are interested in knowing if the UAV statistical recognition system will identify the similarity in the database and appropriately classify the unknown UAV. For example, by observing the UAVs in Fig.~\ref{SMALL_UAVs}, we notice that two pairs of UAVs look alike. The first pair of UAVs is DJI Matrice 600 Pro and Trimble zx5. The second pair of UAVs is DJI Mavic Pro and DJI Phantom 4 Pro. }

 \textcolor{black}{To perform this classification experiment, we train the UAV statistical recognition system with five drones instead of six. Then, we take the sixth UAV (which is excluded from the training database) and use it for classification. For example, we use the unknown UAV as the DJI Matrice 600, which belongs to the same family as the Trimble zx5 UAV in our training database. To perform a classification experiment, we run a Monte Carlo simulation with the 100 DJI Matrice 600 RCS test data, average the classification results, and then plot the confusion matrix at 14 dB SNR. The result is shown in Fig.~\ref{M600_as_Trimble}. We noticed all 100 RCS test data of DJI Mavic Pro are classified as the Trimble zx5 drone. We performed a similar experiment, this time our database includes DJI Matrice 600 but excludes DJI Mavic Pro (the unknown UAV). We performed classification using a similar Monte Carlo analysis. Fig.~\ref{Mavic_as_phantom} shows that all 100 DJI Mavic Pro test data are classified as DJI Phantom 4 Pro at 14 dB SNR.}
 \begin{figure}[t!]
 \center
 \includegraphics[scale=0.75]{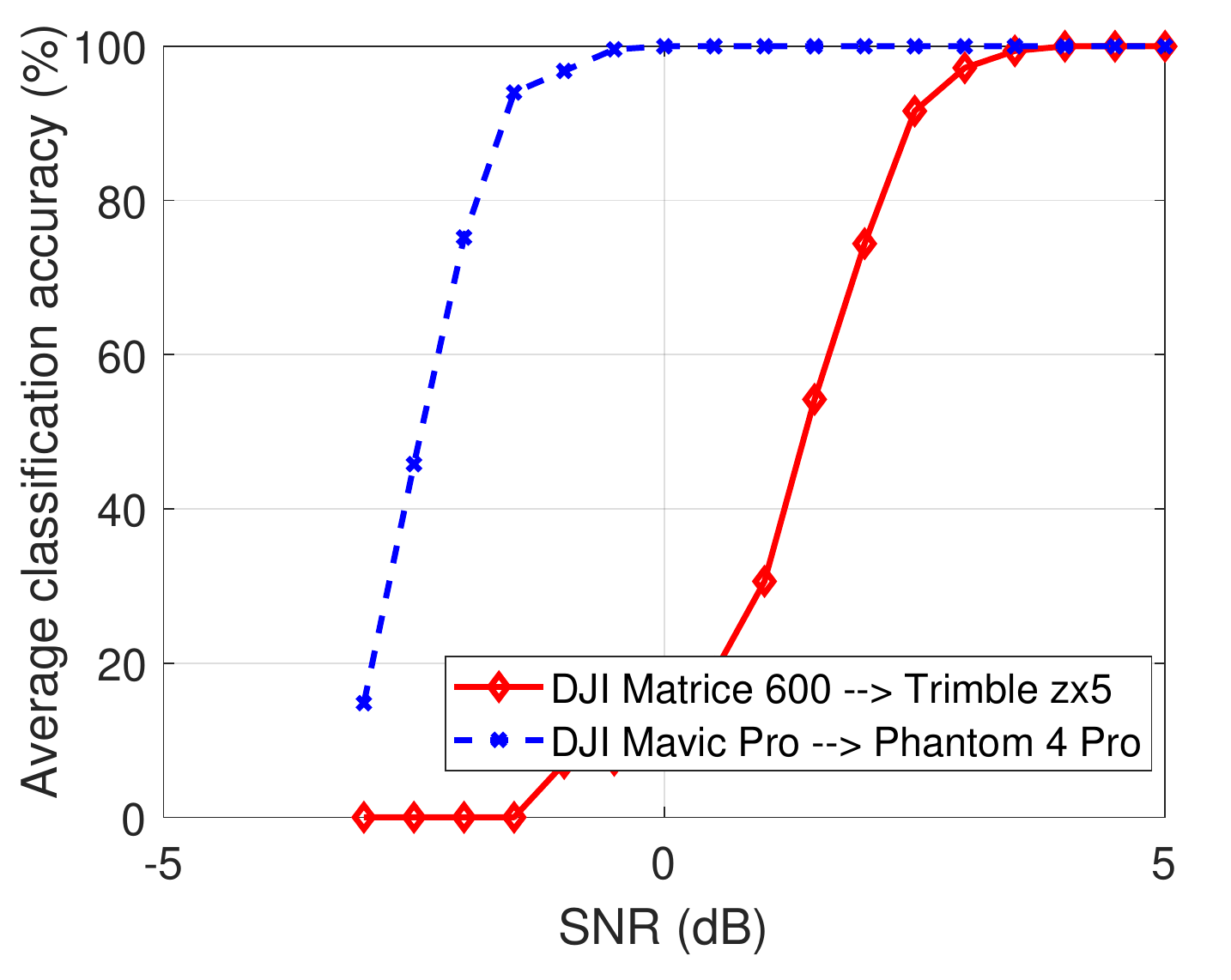}
\caption{Average classification accuracy versus SNR for the unknown UAV classification experiment. In this experiment, the unknown UAV is similar to one of the UAVs in the database}
\label{unknown_UAV_classification}
\end{figure}
  \textcolor{black}{Fig.~\ref{unknown_UAV_classification} shows the result of classifying the unknown UAVs versus SNR. Form Fig.~\ref{unknown_UAV_classification}, we see that even at 5 dB SNR, the UAV statistical recognition system achieves an accuracy of 100$\%$ when classifying the unknown UAV into the appropriate family in the training database. However, as the SNR reduces below 3 dB, the average classification accuracy falls sharply. Therefore, we conclude that the UAV statistical recognition system can identify the similarity between the RCS statistics of an unknown UAV and a known UAV in the training database.}

 \textcolor{black}{On the contrary, if the unknown UAV or radar target is not similar to any of the classes or family in the database, then any attempt to classify the object would yield an error. This is a fundamental limitation of all supervised learning systems (statistical and machine learning algorithms). However, several studies have proposed the use of a ``rejection algorithm"~\cite{copsey2003bayesian,fischer2015efficient,fischer2016optimal} when classifying an unknown target or object. A rejection algorithm basically would prevent any attempt to classify an unknown target if the target statistics are not similar or close to the known classes/families in the training database. However, to measure similarity or closeness, a rejection algorithm has to estimate a rejection threshold, which is a non-trivial task. Often times, there is no closed-form analytical expression to use when estimating the rejection threshold. One has to try classifying all possible ``unknown object" and use that information to decide the value of the rejection threshold. This is an exhaustive process since there is no end to the list of possible unknown objects that could confuse a supervised learning system. Besides, the design of such a rejection algorithm is beyond the scope of the current study.}

\section{Conclusion}\label{conclusion}
 The paper presented a detailed experimental procedure for accurately measuring the RCS of commercial UAVs in a compact range anechoic chamber at 15 GHz and 25 GHz in both VV and HH polarization. The paper describes how the target UAVs can be modeled by fitting their RCS data to a set of 11 statistical models whose parameters are estimated using maximum likelihood techniques. The best statistical model that represents a given UAV type is selected with the aid of AIC and BIC model selection criteria. From the model selection analysis, we observe that the lognormal and GEV distributions are very good at modeling the RCS of the commercial UAVs. This is probably because the data are heavy-tailed and skewed. We also observe that the Gaussian statistics did not perform so well since the radar target returns from the UAVs are not symmetric. Using the best statistical model for each UAV type in the database, we propose a UAV statistical recognition system whose performance is evaluated at different SNRs. We provided the average classification plot which shows that at 10~dB, an average accuracy of 97.43\% or more is achievable. Also, the confusion matrices are provided to analyze the strength and weaknesses of the recognition system. The confusion matrices show that at 10~dB SNR, the HH-polarized RCS data is best for identifying the UAVs. \textcolor{black}{Furthermore, we evaluated the performance of the UAV statistical recognition system in a scenario where the radar can only capture limited RCS aspect reflections from the target object. We showed that at high SNR, the UAV statistical recognition system still achieves good average classification accuracy. Also, we evaluated the performance of the UAV statistical recognition system when an unknown UAV, similar to one of the UAVs in the training database, is to be classified.}


In the future, we are planning to extend our work to include statistical recognition of UAVs using inverse synthetic aperture radar (ISAR) image data. The ISAR image is a 2-dimensional radar imaging technique that can provide rich spatial features for target recognition. Moreover, to enhance UAV recognition rates in narrow field of view (FOV) scenarios, we plan to explore ways to harness features from both ISAR images and micro-Doppler signature. Fusing multi-domain radar statistical features could greatly improve UAV recognition, especially at far distances. Also, we are interested in investigating the use of mixture statistical models for UAV RCS modeling, which could provide better UAV classification at low SNR scenarios.

\section*{Acknowledgment}
The authors would like to thank Mr. Kenneth Ayotte and the management of the Ohio State University Electroscience Laboratory for their help  with the RCS measurements.

\bibliography{IEEEabrv,references}
\bibliographystyle{IEEEtran}
\end{document}